\documentclass[aps,prd,amsmath,amssymb,eqsecnum,nofootinbib,notitlepage,superscriptaddress,floatfix]{revtex4-1}

\usepackage{hyperref}
\usepackage[retainorgcmds]{IEEEtrantools}
\usepackage{graphicx}
\usepackage{graphics}
\usepackage{color}

\newcommand{\f}{f_{\ell}}
\newcommand{\g}{g_{\ell}}
\newcommand{\W}[1]{W\left(#1\right)}
\newcommand{\Aout}{A^{out}_{\ell,\omega}}
\newcommand{\Ain}{A^{in}_{\ell,\omega}}

\newcommand{\taum}{\tau_m}

\newcommand{\rc}{r_0}
\newcommand{\lcut}{\ell_{\rm cut}}
\newcommand{\Lcut}{L_{\rm cut}}

\newcommand{\Gret}{G_{\text{ret}}}
\newcommand{\Glret}{G^{\text{ret}}_{\ell}}
\newcommand{\GlQNM}{G^{QN}_{\ell}}
\newcommand{\GlnQNM}{G^{QN}_{\ell,n}}
\newcommand{\GlBC}{G^{BC}_{\ell}}
\newcommand{\FullGlret}{\mathcal{G}^{ret}_{\ell}}
\newcommand{\FullGlQNM}{\mathcal{G}^{QN}_{\ell}}
\newcommand{\FullGlBC}[1]{\mathcal{G}^{BC}_{#1}}
\newcommand{\DGl}{\Delta G_{\ell}}
\newcommand{\DGw}[3]{\DGl(#1,#2;#3)}

\newcommand{\GQNM}{G_{QN}}
\newcommand{\GBC}{G_{BC}}

\newcommand{\wQNM}{\omega_{\ell n}}

\newcommand{\AoutQNM}{A^{out}_{\ell n}}

\begin{document}
\title{Self-Force and Green Function in Schwarzschild spacetime via Quasinormal Modes and Branch Cut}

\author{Marc Casals}
\email{marc.casals@ucd.ie}
\affiliation{School of Mathematical Sciences and Complex \& Adaptive Systems Laboratory, University College Dublin, Belfield, Dublin 4, Ireland}

\author{Sam Dolan}
\email{s.dolan@shef.ac.uk}
\affiliation{Consortium for Fundamental Physics, School of Mathematics and Statistics, University of Sheffield, S3 7RH, United Kingdom}

\author{Adrian C.~Ottewill}
\email{adrian.ottewill@ucd.ie}
\affiliation{School of Mathematical Sciences and Complex \& Adaptive Systems Laboratory, University College Dublin, Belfield, Dublin 4, Ireland}

\author{Barry Wardell}
\email{barry.wardell@gmail.com}
\affiliation{School of Mathematical Sciences and Complex \& Adaptive Systems Laboratory, University College Dublin, Belfield, Dublin 4, Ireland}

\begin{abstract}
The motion of a small compact object in a curved background spacetime deviates from a geodesic due
to the action of its own field, giving rise to a self-force. This self-force may be calculated by
integrating the Green function for the wave equation over the past worldline of the small object.
We compute the self-force in this way for the case of a scalar charge in Schwarzschild spacetime,
making use of the semi-analytic method of matched Green function expansions.
 Inside a local neighbourhood of the compact object, this method uses the Hadamard form for the Green function in order to render regularization trivial.
 Outside this local neighbourhood, we
calculate the Green function using a spectral decomposition into poles (quasinormal modes) and a
branch cut integral in the complex-frequency plane. We show that both expansions overlap in a
sufficiently large matching region for an accurate calculation of the self-force to be possible.
The scalar case studied here is a useful and illustrative toy-model for the gravitational case,
which serves to model astrophysical binary systems in the extreme mass-ratio limit.
\end{abstract}

\maketitle

\section{Introduction}
\label{sec:Introduction}

The motion of a point particle in curved spacetime can be modelled as the particle deviating
from a geodesic of the background spacetime due to the action of its own field, which gives rise to
the self-force (see~\cite{Poisson:2011nh,Barack:2009ux} for reviews). The particle may be a scalar
charge, an electric charge or a point mass. In the case of an electric charge moving in flat
spacetime, the self-force is the celebrated Abraham-Lorentz-Dirac force~\cite{Dirac:1938nz}. The case
of a point mass serves to model, for example, a small compact object moving in the background of a
massive black hole, i.e., an Extreme Mass-Ratio Inspiral. Such a binary system is of major
astrophysical importance due to the expected ubiquity of massive black holes in our Universe
\cite{Magorrian:1997hw}.

Astrophysical black holes are expected to be rotating and as such their gravitational field is described by the Kerr metric.
However, even in the simpler case of the static and spherically-symmetric
Schwarzschild black hole background, the calculation of the self-force (SF)
is technically challenging. 
This is partly due to the fact that the field of the small object diverges towards the position of the object itself and so regularization procedures are required.
Various methods for the calculation of the SF are already in place and have been used by different research groups to achieve
 landmark results in the Schwarzschild background spacetime in the recent years. 
Among them, there is concordance between gravitational SF, post-Newtonian and Numerical Relativity calculations~\cite{Blanchet:2009sd,LeTiec:2011bk},
a gravitational SF correction to the frequency of the Innermost Stable Circular Orbit~\cite{Barack:2009ey},
a correction to the precession effect~\cite{Barack:2010ny},
evolution of a `geodesic' SF orbit (i.e., where the motion is evolved using the value of the SF corresponding to an instantaneously tangent geodesic) in the gravitational case~\cite{Warburton:2011fk} and evolution of a self-consistent orbit (i.e.,  where the coupled system of equations for the motion and for the SF are simultaneously evolved) in the scalar case~\cite{Diener:2011cc}.

These results were obtained via two leading methods for SF calculations:
 the mode-sum regularization scheme \cite{Barack:2001gx,Barack:1999wf}, and the 
effective source method \cite{Barack:2007jh,Vega:2007mc}.
Despite first appearances, these approaches share a common foundation: the SF is related to the gradient of a regularized (`R') field (or metric perturbation), which is obtained by subtracting a certain singular (`S') field from the physical field \cite{Detweiler:2002mi}. A key point is that, since the gradient is evaluated on the worldline, only \emph{local} knowledge of the S field is needed. 


Typically, these schemes employ numerical methods to calculate the field (or metric perturbation)
directly by solving the differential equation that it satisfies. One drawback of these methods is that, by their nature, they yield relatively little insight into the origin of the SF, except of course through the highly accurate numerical results that they provide. To seek a complementary approach, we may return to an approach which pre-dates both the mode-sum regularization and effective source methods. This is a method built on the so-called MiSaTaQuWa equation \cite{Mino:1996nk,Quinn:1996am}.

In the 1960s, DeWitt \cite{DeWitt:1960fc} obtained an expression for the self-force experienced by an electromagnetic charge in curved spacetime, by writing the SF in terms of the retarded Green function. The MiSaTaQuWa equation may be viewed as a generalization of DeWitt's formula~\cite{DeWitt:1960fc} to the case of a gravitational point source (within perturbation theory). 
In this paper we focus on the case of a scalar charge moving on a curved background \cite{Quinn:2000wa} since it is technically easier than the gravitational case (e.g., there are no issues of gauge freedom in the scalar case) and yet it shares the key calculational and conceptual features.

The SF on a scalar charge $q$ moving on a vacuum background spacetime may be written as
\begin{equation}\label{eq:S-F}
F_{\mu}(\tau)=q^2
\int_{-\infty}^{\tau^-}d\tau'\ \nabla_{\mu}\Gret(z(\tau),z(\tau')),
\end{equation}
where $z(\tau)$ is the worldline of the particle, $\tau$ is its proper time, $u^{\mu}$ its four-velocity
and $\Gret$ is the retarded Green function (GF) of the Klein-Gordon equation obeyed by the scalar field.
The integral over the past worldline of the derivative of the GF in Eq.~(\ref{eq:S-F}) is referred to as the tail integral. An equivalent of this tail integral occurs in both the electromagnetic and gravitational cases.
 Equation (\ref{eq:S-F}) 
 features an upper limit in the integral which excludes $\tau'=\tau$,
thus removing the known divergence that the Green function $\Gret(x,x')$ possesses at coincidence $x'=x$.
Eq.~(\ref{eq:S-F}) allows one to view the SF as arising from the propagation of field waves emitted in the past by the point particle: these waves are generally `scattered' 
or refocused by the background spacetime and return to the particle to affect its motion.

\begin{figure}[htb!]
\begin{center}
\includegraphics[height=6.5cm]{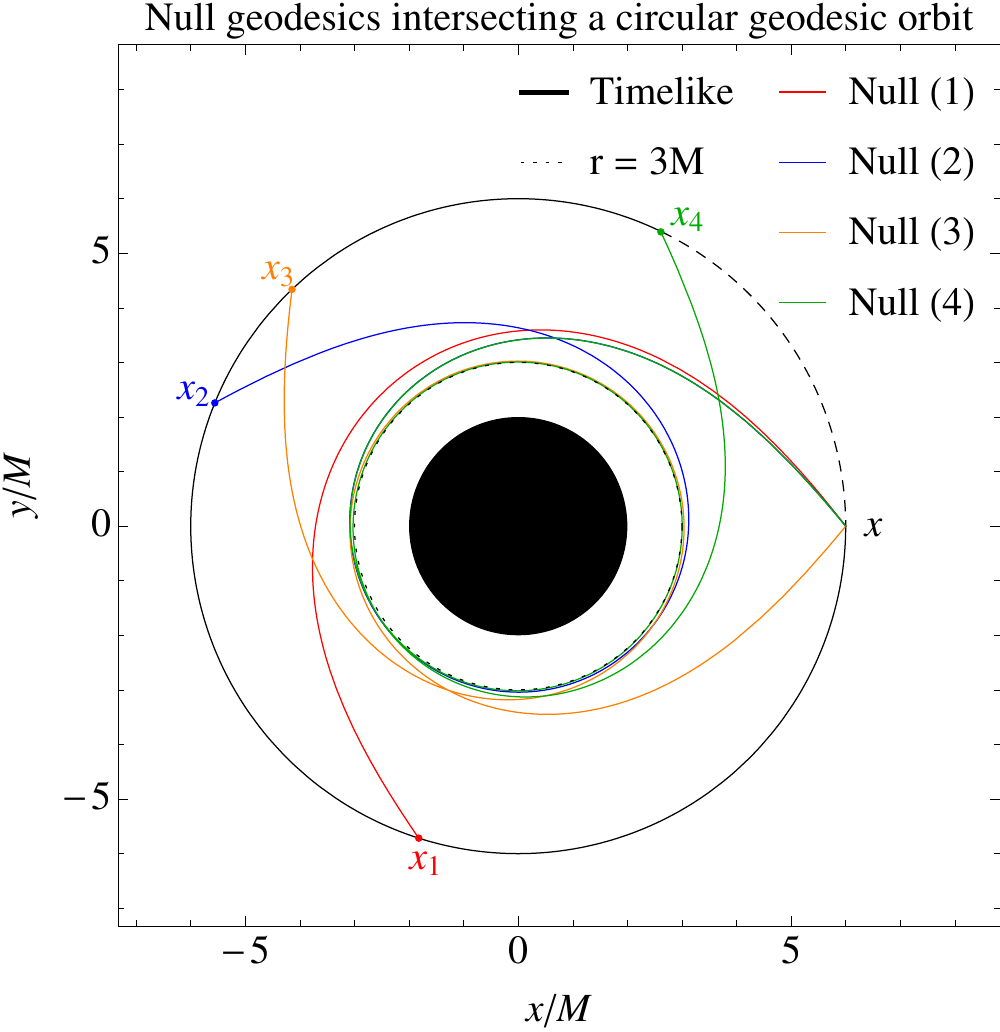}
\includegraphics[height=6.625cm]{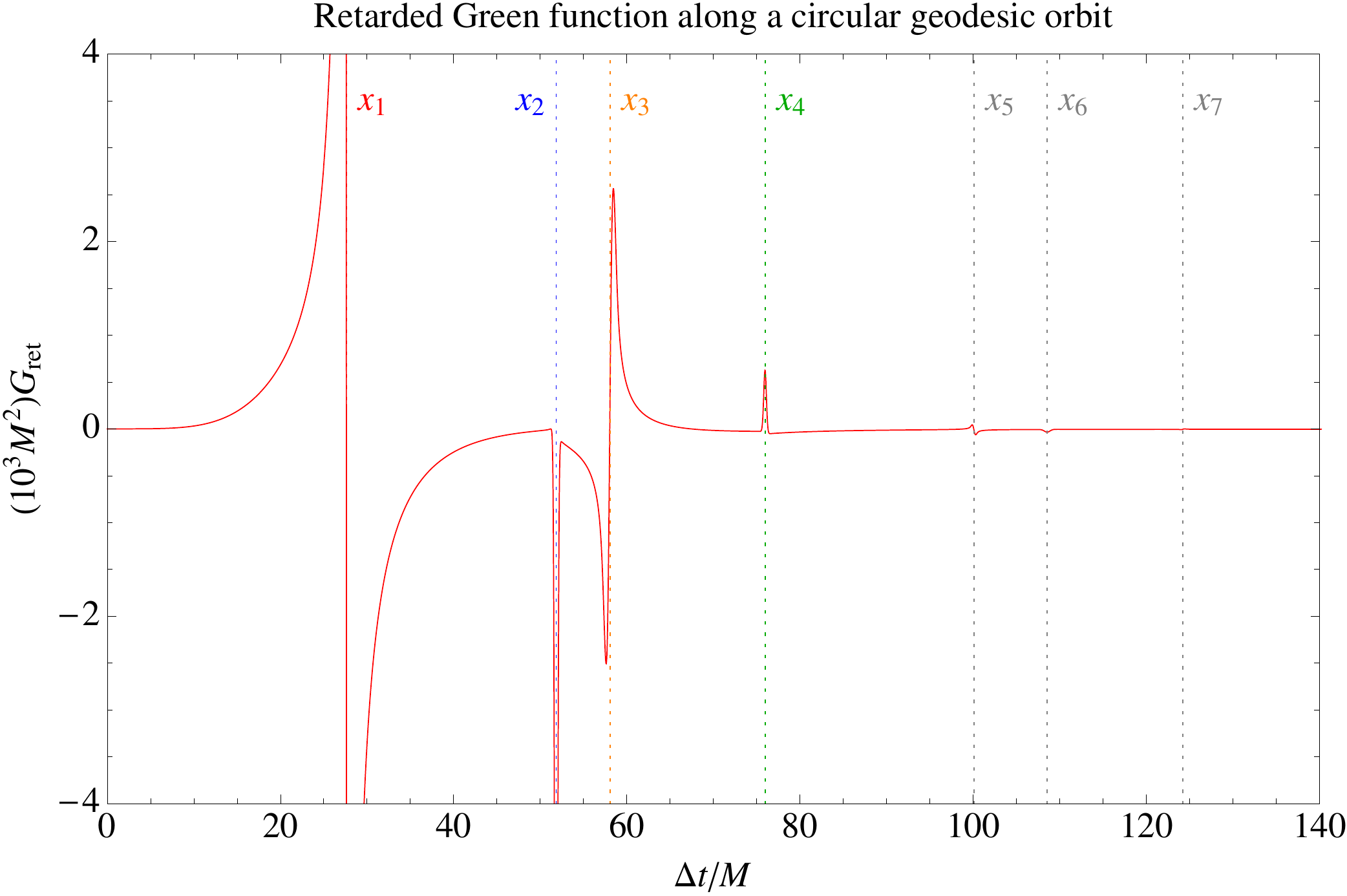}
\includegraphics[height=6.5cm]{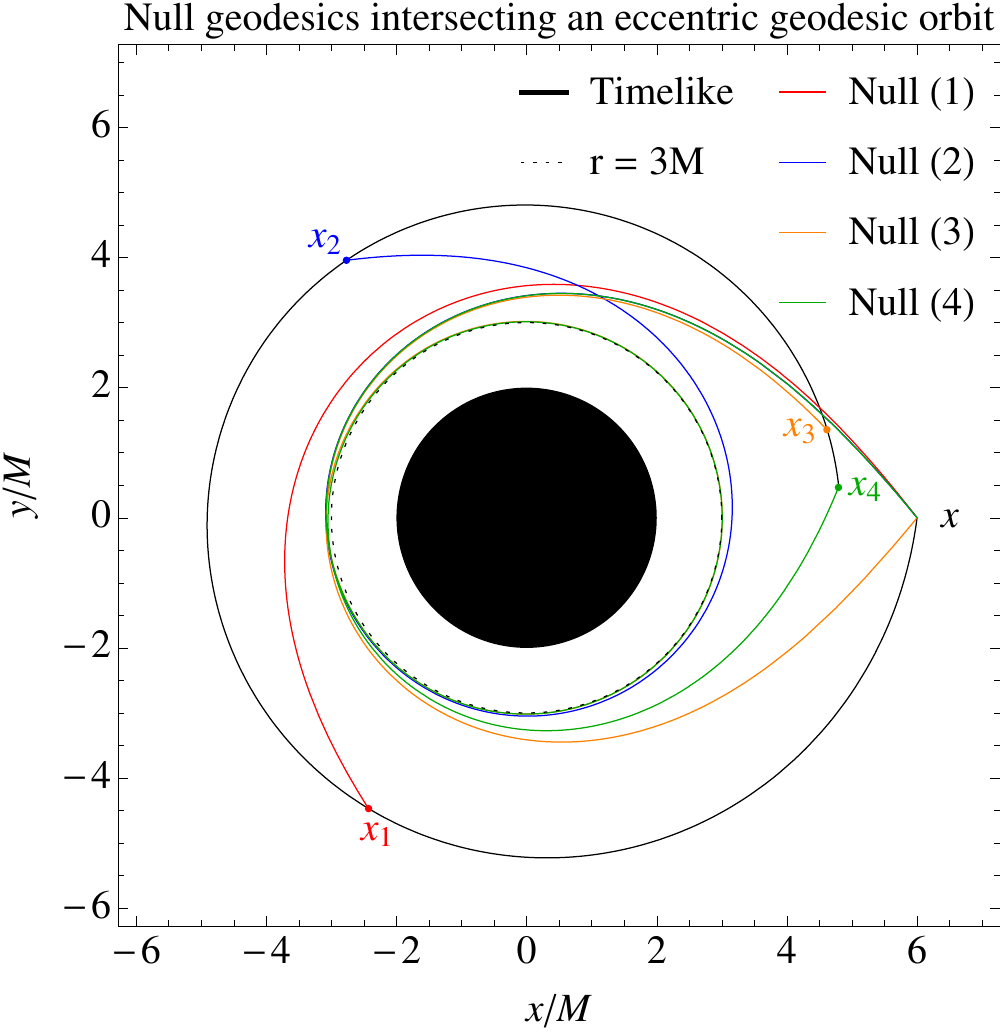}
\includegraphics[height=6.625cm]{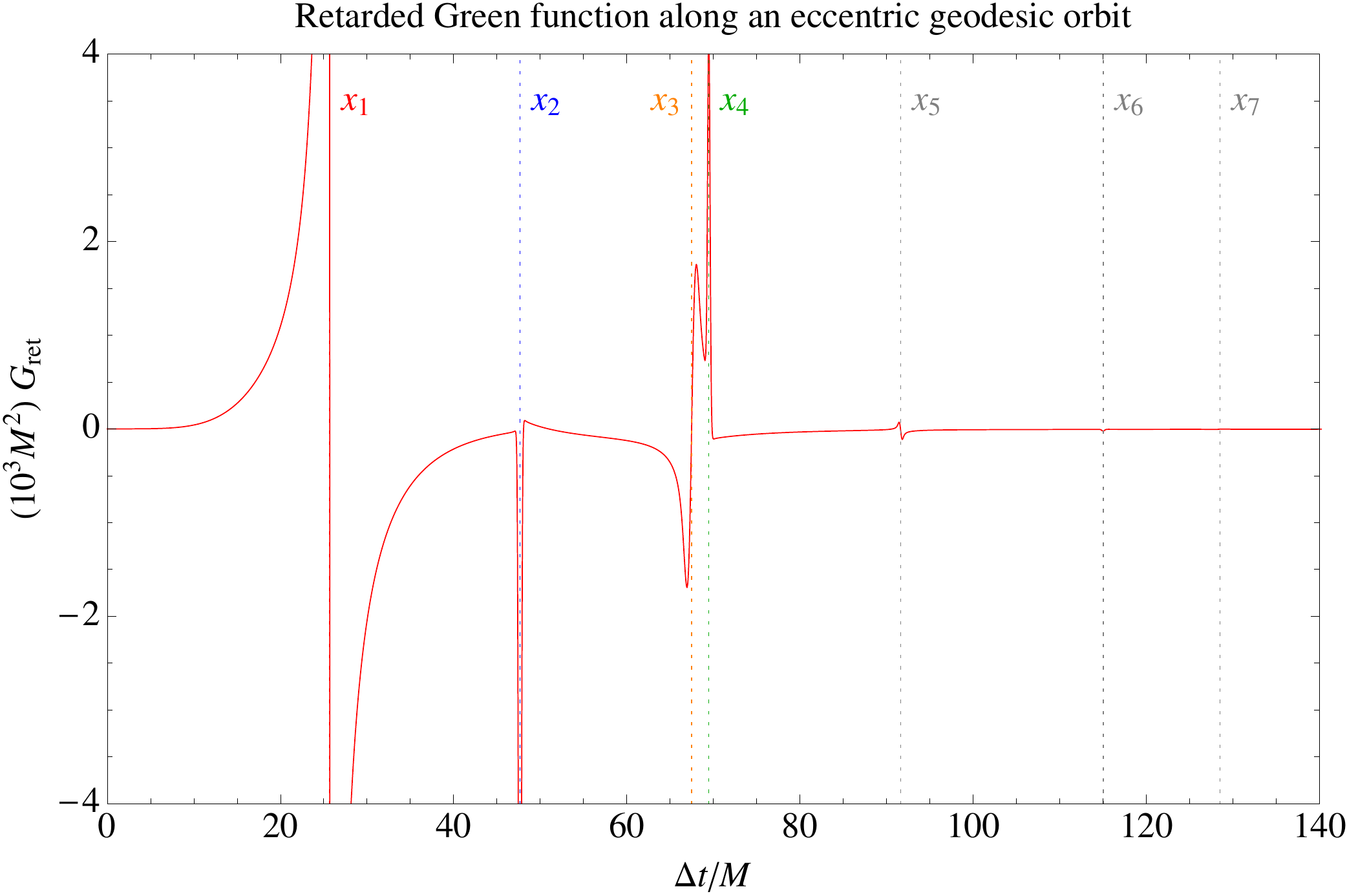}
\end{center}
\caption{Green function along the innermost stable circular geodesic orbit of radius $\rc = 6M$
(top) and along the eccentric geodesic orbit with $p=7.2, e=0.5$ at the point in the orbit where
$r = 6M$ and 
$\frac{d r(\tau)}{d\tau}>0$
(bottom).
The left plots show the orbit and the first four null geodesics that start on the orbit, pass
around the Schwarzschild black hole, and re-intersect the orbit. The right plot illustrates
the retarded Green function $G_{\text{ret}}(x, x')$ where $x'$ is a point on the orbit a time $\Delta t$
in the past of $x$. The singular features at points $x_1, x_2, x_3, \cdots$ correspond to the
arrival of the null geodesics $(1), (2), (3), \cdots$ shown in the left plots. The singular
features follow a four-fold repeating pattern, arising from the passage of the null wavefront
through caustics. 
}
\label{fig:intro}
\end{figure}

If the background spacetime possesses an unstable photon orbit, also known as a light-ring, then a null geodesic emitted from a point on the particle's worldline may circle around the light-ring and return to intersect the worldline. Figure \ref{fig:intro} illustrates this feature in Schwarzschild black hole spacetime, where the light-ring is located at the radial Schwarzschild coordinate $r=r_{\text{orb}}\equiv 3M$.  
It has long been known that the retarded Green function $\Gret(x,x')$ is singular whenever $x$ and $x'$ are connected by a null geodesic~\cite{Garabedian,Ikawa}. More recently, it has been appreciated that the  singular structure of $\Gret(x,x')$ is affected by the formation of caustics~\cite{Ori1short,Casals:2009zh,Dolan:2011fh,Harte:2012uw,Zenginoglu:2012xe,Casals:2012px}. Caustics are the set of focal points associated with a continuous family of null geodesics which originate from a single spacetime point; in principle, in the Schwarzschild spacetime an infinite number of caustics will eventually form along antipodal lines, as the null wavefront orbits at $r=r_{\text{orb}}$. The singular part of the Green function undergoes a transition each time the null wavefront encounters a caustic, exhibiting a repeating four-fold cycle, as shown in Fig.~\ref{fig:intro}. In principle, the change in the singular form of $\Gret$ may have a direct impact on the value of the SF via Eq.~(\ref{eq:S-F}).

In 1998, Poisson and Wiseman~\cite{Poisson:Wiseman:1998} suggested a method\footnote{We note that,
in the same year, Capon~\cite{Capon-1998} calculated a regularized Green function using mode sum
expansions with various approximations (e.g., large radius).}
for calculating the SF via Eq.~(\ref{eq:S-F}) by using a method of matched expansions. 
Essentially, this method consists of splitting the tail integral in two different time regimes, as shown in Fig.~\ref{fig:matched exps}. For times $\tau'$ `close' to $\tau$, 
in a region demarcated as the quasilocal (QL) region,
the so-called Hadamard form for the GF is used.
An advantage of this form is that the divergence of the GF at $\tau'=\tau$ is explicit and so it is straightforward to remove it. The main disadvantage of this form is that it is only valid in a local neighbourhood of $z(\tau)$ and so a different method, typically a mode-sum expansion, must be used for
times  $\tau'$ `far' from $\tau$, in a region denoted as the distant past region (DP).
\emph{A priori}, it is not clear whether the two regions (QL region and DP) have a shared domain in which the expansions may be matched together. 
In~\cite{Anderson:2005gb}, the authors investigated the feasibility of the method of matched expansions for a scalar charge on Schwarzschild spacetime
by using an approximation to the mode-sum expansion in the weak-field regime.
In particular, they carried out a calculation of the SF in the QL region  in the cases of a static charge and of a charge in a circular geodesic at radius $r=6M$ and, in the former case, of (a weak-field approximation to) the SF in the DP.
They found that the convergence of the mode-sum method
in the DP was ``poor", making matching between the calculations 
in the QL region and in the DP difficult to achieve. They concluded that this method is
``promising but possessing some technical challenges". 

In Ref.~\cite{Casals:2009zh} 
we overcame some of these technical challenges to achieve the first successful calculation of SF with the method of matched expansions. This calculation was carried out on a static region of the Nariai spacetime ($dS_2\times \mathbb{S}^2$), which served as a toy-model for Schwarzschild spacetime.
In this paper we extend the calculation in Ref.~\cite{Casals:2009zh} to the case of Schwarzschild spacetime. 
We calculate the SF on a scalar charge on Schwarzschild spacetime using a
high-order expansion for the Hadamard form 
in the QL region, and a `full' mode-sum expansion in the DP. We achieve good matching between the
two calculations.
The mode-sum expansion method consists of a spectral decomposition of the GF~\cite{Leaver:1986,*PhysRevD.38.725} into its two main contributions: a sum over poles (the so-called quasinormal modes) and an integral around a branch cut on the negative-imaginary axis of the complex-frequency plane, as shown in Fig.~\ref{fig:contour}. In other words, we calculate the full GF in Schwarzschild spacetime outside a local region by adding a quasinormal mode (QNM) contribution and a branch cut  (BC) contribution.
The BC contribution in Schwarzschild is a new addition with respect to the Nariai case, where 
the exponential fall-off of the radial potential ensures there is no BC and so the GF in the DP is fully given by the QNM contribution only.

We carry out the calculations using the method of matched expansions via the use of QL, QNM and BC contributions in the case of a scalar charge on Schwarzschild spacetime at $r=\rc\equiv 6M$ which is moving on (1) a circular geodesic (with angular velocity $\Omega=\sqrt{M}/\rc^{3/2}\approx 0.068/M$)
 and (2) an eccentric geodesic with $p=7.2$ and $e=0.5$, where $M$ is the mass of the black hole, $p$ is the semi-latus rectum and $e$ is the eccentricity. Figure \ref{fig:intro} shows these two timelike geodesics; the null geodesics which re-intersect these orbits; and the GF along these orbits, calculated by the method of matched expansions presented. 


The layout of the paper is as follows. 
In Sec.~\ref{sec:matched exp} we recap the method of matched expansions.
In Secs.~\ref{sec:QL} and \ref{sec:DP} we describe the methods for calculating the GF 
in the QL region and DP, respectively. 
In Sec.~\ref{sec:matching} we show the matching between the calculation 
in the QL region and in the DP. In Sec.~\ref{sec:SF} we calculate the SF.
We conclude the paper with a discussion in Sec.~\ref{sec:Discussion}. Throughout, we use geometrized units, with $c = G = 1$.

\section{The Method of Matched Expansions}\label{sec:matched exp}

As outlined in the Introduction, the method of matched expansions is a scheme for the direct evaluation of tail integrals such as Eq.~(\ref{eq:S-F}). It was proposed fifteen years ago \cite{Poisson:Wiseman:1998}, but until now it has not been developed into a practical scheme for SF calculation on a black hole spacetime. In this approach, the integral in Eq.~(\ref{eq:S-F}) is split into two different time regimes, QL region and DP. In each regime the GF is obtained via a suitable expansion. If the expansions may be shown to share a common region of validity, then the tail integral may be  evaluated. An overview of the method is given below, with technical details of the QL and DP expansions following in Secs.~\ref{sec:QL} and \ref{sec:DP}, respectively.

\begin{figure}[htb!]
\begin{center}
\includegraphics[width=7cm]{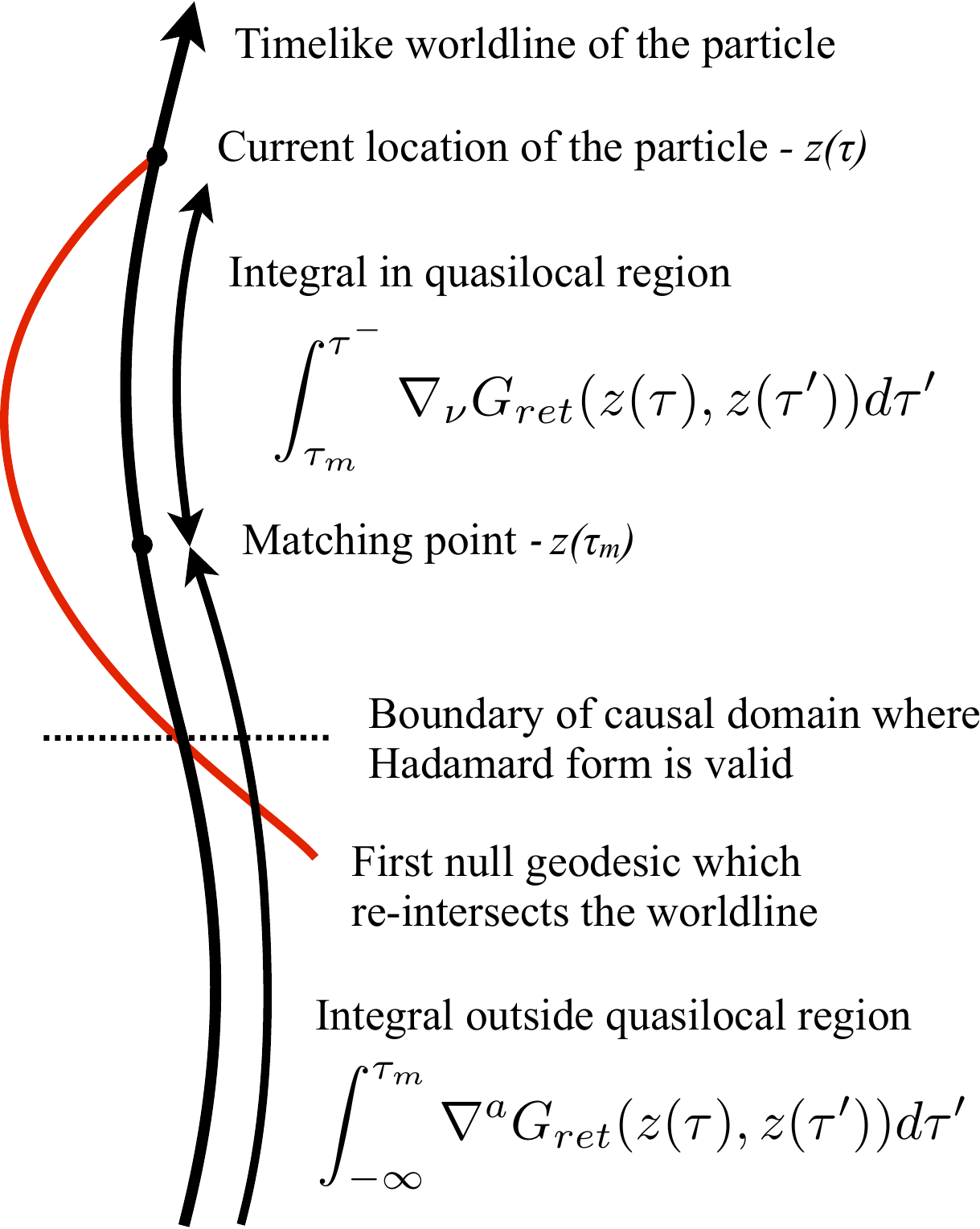}
\end{center}
\caption{Schematic representation of the method of matched expansions (see Eq.~(\ref{eq:S-F match})).}
\label{fig:matched exps}
\end{figure}

In summary, the method of matched expansions consists on calculating the SF via the expression
\begin{equation}\label{eq:S-F match}
f_{\mu}(\tau)=
q^2
\nabla_{\mu}
\left\{\int_{\taum}^{\tau^-}d\tau'\ V(z(\tau),z(\tau'))+\int_{-\infty}^{\taum}d\tau'\ 
\Gret(z(\tau),z(\tau'))\right\},
\end{equation}
where 
$\taum$
 is a properly chosen value of the proper time such that matching is possible between the QL (first term in
Eq.~(\ref{eq:S-F match})) and DP (second term in Eq.~(\ref{eq:S-F match})) calculations.
The method is schematically represented in Fig.~\ref{fig:matched exps}.  

\subsubsection{Green function in Hadamard form}
The tail integral in Eq.~(\ref{eq:S-F}) is taken along the entire past worldline, up to \emph{but not including} the point of coincidence at $x = x'$. 
The divergence of the GF at coincidence is explicitly manifest in the Hadamard form~\cite{Friedlander,Hadamard,DeWitt:1960fc} for the GF:
\begin{equation}\label{eq:Hadamard}
\Gret(x,x')=\theta_-(x,x')\left[U(x,x')\delta(\sigma(x,x'))-V(x,x')\theta\left(-\sigma(x,x')\right)\right],
\end{equation}
where $U(x,x')$ and $V(x,x')$ are regular bitensors, $\delta$ and $\theta$ are the usual Dirac-delta and Heaviside
distributions, respectively, and $\theta_-(x,x')$ is equal to $1$ if $x$ lies to the future of $x'$ and it is equal to $0$ otherwise.
The two-point function $\sigma(x,x')$
is Synge's world-function, i.e., one-half of the square of the geodesic distance along the unique geodesic connecting the points $x$ and $x'$. Because the Hadamard form exhibits an explicit form for  the singularity of the GF at coincidence, it is trivial to remove this singular point from the integral (namely by subtracting the term with $U\delta(\sigma)$ from Eq.~(\ref{eq:Hadamard})).

The limitation of the Hadamard form Eq.~(\ref{eq:Hadamard}) is that it is only valid in a convex normal neighbourhood of the base point $x$, that is, 
in a region $\mathcal{N}(x)$ containing $x$ with the property that every $x'\in\mathcal{N}(x)$ is
connected to $x$ by a unique geodesic which lies in $\mathcal{N}(x)$.
This effectively requires that, in order to use the Hadamard form Eq.~(\ref{eq:Hadamard}) for the GF in the SF expression Eq.~(\ref{eq:S-F}),  the worldline points $z(\tau)$ and $z(\tau')$ must be connected by a unique
nonspacelike geodesic which stays within $\mathcal{N}(z(\tau))$.

There are known uniformly convergent series which can be used to calculate $V(x,x')$~\cite{DeWitt:1960fc,Friedlander} throughout the domain of validity (i.e.~for all $x'\in \mathcal{N}(x)$), and 
methods for taking these series to very high order have been developed (see Sec.~\ref{sec:QL}). It remains to be shown that the convergence of such series is sufficiently rapid that an accurate calculation of $\Gret$ may be extended towards the boundary of $\mathcal{N}(x)$.

In~\cite{Casals:2012px} a method was proposed for calculating the GF globally by `propagating' the Hadamard form to points $x'$ outside $\mathcal{N}(x)$ via the use of Kirchhoff's integral representation for the field. This method, however, requires a deep knowledge of the properties of $\sigma$, $U(x,x')$ and $V(x,x')$ which, in practice, makes the method difficult to apply to Schwarzschild spacetime.

\subsubsection{Singular structure of Green function}
Equation (\ref{eq:Hadamard}) shows that the singularities of $\Gret(x,x')$
within 
$\mathcal{N}(x)$
 occur not only at $x=x'$ but also 
along $\sigma(x,x')=0$, i.e., 
when the two spacetime
points are connected by a null geodesic.
It is known~\cite{Garabedian,Ikawa} that  for $x'$ outside
$\mathcal{N}(x)$,
the singularities of the GF continue to occur when 
the two spacetime points are connected by a null geodesic. 
The singularity of the GF for $x'$ outside 
$\mathcal{N}(x)$
develops a structure which is richer than the solo Dirac-delta distribution exhibited
for $x'\in \mathcal{N}(x)$.
It was recently shown~\cite{Ori1short,Casals:2009zh,Dolan:2011fh,Harte:2012uw,Zenginoglu:2012xe,Casals:2012px} 
that in various 
spacetimes 
the singularity of the GF when the spacetime points are null-separated exhibits a four-fold
structure: $\delta(\sigma)$, $\text{PV}\left(1/\pi\sigma\right)$, $-\delta(\sigma)$, $-\text{PV}\left(1/\pi\sigma\right)$, $\delta(\sigma)$, \dots, 
where \text{PV} is the principal value distribution.
The character of the singularity changes as the null geodesic crosses through a caustic point
of the background spacetime.
Such four-fold singularity structure was initially noted in General Relativity in~\cite{Ori1short}, first proven
in~\cite{Casals:2009zh} in a static region of Nariai spacetime ($dS_2\times \mathbb{S}^2$), 
then shown in Schwarzschild spacetime in~\cite{Dolan:2011fh},
generalized in~\cite{Harte:2012uw}, proven in Pleba\'nski-Hacyan spacetime ($\mathbb{M}_2\times \mathbb{S}^2$) in~\cite{Casals:2012px} and
physically explained as well as
 beautifully
illustrated via numerical simulations in  Schwarzschild spacetime in~\cite{Zenginoglu:2012xe}.
We note that certain points in these spacetimes escape the previous singularity structure, such as
points lying along a caustic line~\cite{Dolan:2011fh,Casals:2012px,Zenginoglu:2012xe}.

\subsubsection{Multipole expansion}
To move beyond the normal neighbourhood, one may expand the GF in multipoles and take a Fourier-mode decomposition. It was found in \cite{Anderson:2005gb} that a lack of convergence of the sum over multipoles ($\ell$) and integral over frequency ($\omega$) presented a serious impediment to progress (convergence issues are somewhat inevitable, as the GF is a distribution with singular features). As described in the Introduction, a step forward came with writing the Fourier integral in terms of a quasinormal mode (QNM) sum (a sum of the residues of the GF at poles in the complex-frequency plane) \cite{Casals:2009zh}, whose convergence properties are now well-understood (see Sec.~\ref{sec:QNM}). On a black hole spacetime, the expansion in QNMs must be supplemented by an integral along a branch cut, which has been the focus of some recent work (see Sec.~\ref{sec:BC}).

\section{Quasilocal Expansion of the Green Function}\label{sec:QL}
In the QL region, the spacetime points $x$ and $x'$ are assumed to be
sufficiently close together that the GF is uniquely given
by the Hadamard parametrix, Eq.~\eqref{eq:Hadamard}. The term involving
$U(x,x')$ does not contribute to the integral in Eq.~\eqref{eq:S-F} since it
has support only when $x=x'$, and the integral excludes this point. We will
therefore only concern ourselves in the QL region with the calculation
of the function $V(x,x')$.

The fact that $x$ and $x'$ are close together suggests that an expansion of
$V(x,x')$ in powers of the separation of the points may give a good
approximation within the QL region. Reference~\cite{Casals:2009xa} used a WKB
method to derive such a coordinate expansion for spherically symmetric spacetimes
and gave estimates of its range of
validity. Referring to the results therein, we have $V(x,x')$ as a power series
in $(t-t')$, $(1-\cos \gamma)$ and $(r-r')$,
\begin{equation} \label{eq:CoordGreen}
V(x,x') =
 \sum_{i,j,k=0}^{\infty} v_{ijk}(r) ~ (t-t')^{2i} (1-\cos\gamma)^j (r-r')^k,
\end{equation}
where $\gamma$ is the angular separation of the points
and $v_{ijk}$ are dimensionful coefficients. Note that ---
since $V(x,x')$ is symmetric and the Schwarzschild spacetime is static ---
time reversal invariance ensures that only even powers of $(t-t')$ can appear in this expansion.
Up to an overall minus
sign, Eq.~\eqref{eq:CoordGreen} therefore gives the QL contribution to
GF as required in the present context. It is also
straightforward to take partial derivatives of these expressions at either
spacetime point to obtain the derivative of the GF.

As proposed in Ref.~\cite{Casals:2009xa}, we have not used the QL expansion in the specific form of
Eq.~\eqref{eq:CoordGreen}. Instead, we begin with an expansion of that form to 52-nd order in the
coordinate separation and make two key modifications to improve both its accuracy and its domain of
validity. Firstly, we factor out the leading form of the singularity at the first light crossing
time. In the GF this singularity has the form $(t-t_c)^{-1}$ and for the derivative of the
GF it has the form $(t-t_c)^{-2}$, where $t_c$ is the first light-crossing time.
Secondly, we compute diagonal Pad\'e approximants from the residual series expansions. In the circular
orbit case, this second step is straightforward as all coordinates may be written in closed form in
terms of the radius of the orbit and the time separation of the points. In the eccentric orbit case,
we require the additional step of using the geodesic equations to rewrite the coordinates
as expansions in $(t-t')$ along the orbit, and then using the resulting series expansion for $V(z(\tau),z(\tau'))$
in $(t-t')$ alone as a starting point for computing a Pad\'e approximant. The final result is an
approximation for $V(z(\tau),z(\tau'))$ in the form of a ratio of two polynomials in $(t-t')$, with coefficients
which are functions of $r$ only,
\begin{equation} \label{eq:CoordGreenPade}
V(z(\tau),z(\tau')) =
 \frac{\sum_{i=4}^{26} a_{i}(r) ~ (t-t')^{i}}{(t-t_c)\sum_{j=0}^{26} b_{j}(r) ~ (t-t')^{j}}.
\end{equation}
This expansion is an accurate representation of the GF throughout
$\tau '\in (0,\tau_m]$.

\section{Spectral decomposition of the Green Function in the Distant Past}
\label{sec:DP}

After a multipole-$\ell$ decomposition in the angular distance $\gamma$, 
the GF in Schwarzschild spacetime can be expressed as 
\begin{align} \label{eq:Green}
&
\Gret(x,x')=\sum_{\ell=0}^{\infty}\FullGlret,\quad
\FullGlret\equiv 
\frac{1}{r\, r'}
(2\ell+1)P_{\ell}(\cos\gamma)\Glret(r,r'; \Delta t),
\end{align}
where $r$ and $t$ are the radial and time Schwarzschild coordinates of 
the spacetime point $x$, respectively, and similarly $r'$ and $t'$ of the spacetime point  $x'$ and $\Delta t\equiv t-t'$.
By next carrying out  a Fourier-mode decomposition in time,
one obtains
\begin{align} \label{eq:Green ell}
\Glret(r,r';\Delta t)\equiv
\frac{1}{2\pi}
\int_{-\infty+ic}^{\infty+ic} d\omega\ G_{\ell}(r,r';\omega)e^{-i\omega  \Delta t},
\end{align}
where $c>0$. The Fourier modes $G_{\ell}$ of the Green function satisfy
the following second order radial ODE:
\begin{align} \label{eq:radial ODE}
&\left[\frac{d^2}{dr_*^2}+\omega^2-V(r)\right]G_{\ell}(r,r';\omega)=-\delta(r_*-r'_*),
\\
&V(r)\equiv 
\left(1-\frac{2M}{r}\right)\left[\frac{\ell(\ell+1)}{r^2}+\frac{2M}{r^3}\right],
\nonumber
\end{align}
together with appropriate retarded boundary conditions.
Here,
 $r_*\equiv r+2M\ln\left(\frac{r}{2M}-1\right)$
 is the radial `tortoise coordinate'.
 The  Fourier modes of the GF can be calculated as
 \begin{align}\label{eq:Gl}
G_{\ell}(r,r';\omega)=\frac{\f(r_<,\omega)\g(r_>,\omega)}{\W{\omega}},
\end{align}
where  $r_>\equiv \max(r,r'),\ r_<\equiv \min(r,r')$.
The radial functions $\f(r,\omega)$ and $\g(r,\omega)$ are
two linearly-independent solutions of the homogeneous version of the ODE (\ref{eq:radial ODE}).
For $\text{Im}(\omega)\ge 0$ and $r_*/M\in\mathbb{R}$ they satisfy the boundary conditions:
\begin{align} \label{eq:f,near hor}
&
\f(r,\omega) \sim 
e^{-i\omega r_*}, & r_*/M\to -\infty,
\\&
\f(r,\omega) \sim \Aout e^{+i\omega r_*}+\Ain e^{-i\omega r_*}, & r_*/M\to +\infty,
\nonumber
\end{align}
where $\Ain$ and $\Aout$ are complex-valued coefficients,
and 
\begin{equation} \label{eq:g}
\g(r,\omega)\sim  e^{+i\omega r_*}, \quad \ r_*/M\to +\infty.
\end{equation}
For $\text{Im}(\omega) < 0$, with $r_*/M\in\mathbb{R}$, the solutions $\f$ and $\g$ must 
be defined by analytic continuation.
The function 
\begin{align}  \label{eq:Wronskian}
\W{\omega}\equiv 
\g\f'-\f\g'=2i\omega \Ain,
\end{align}
where a prime indicates a derivative with respect to $r_*$,
is the Wronskian of the two radial solutions.

\begin{figure}[htb!]
\begin{center}
\includegraphics[width=9cm]{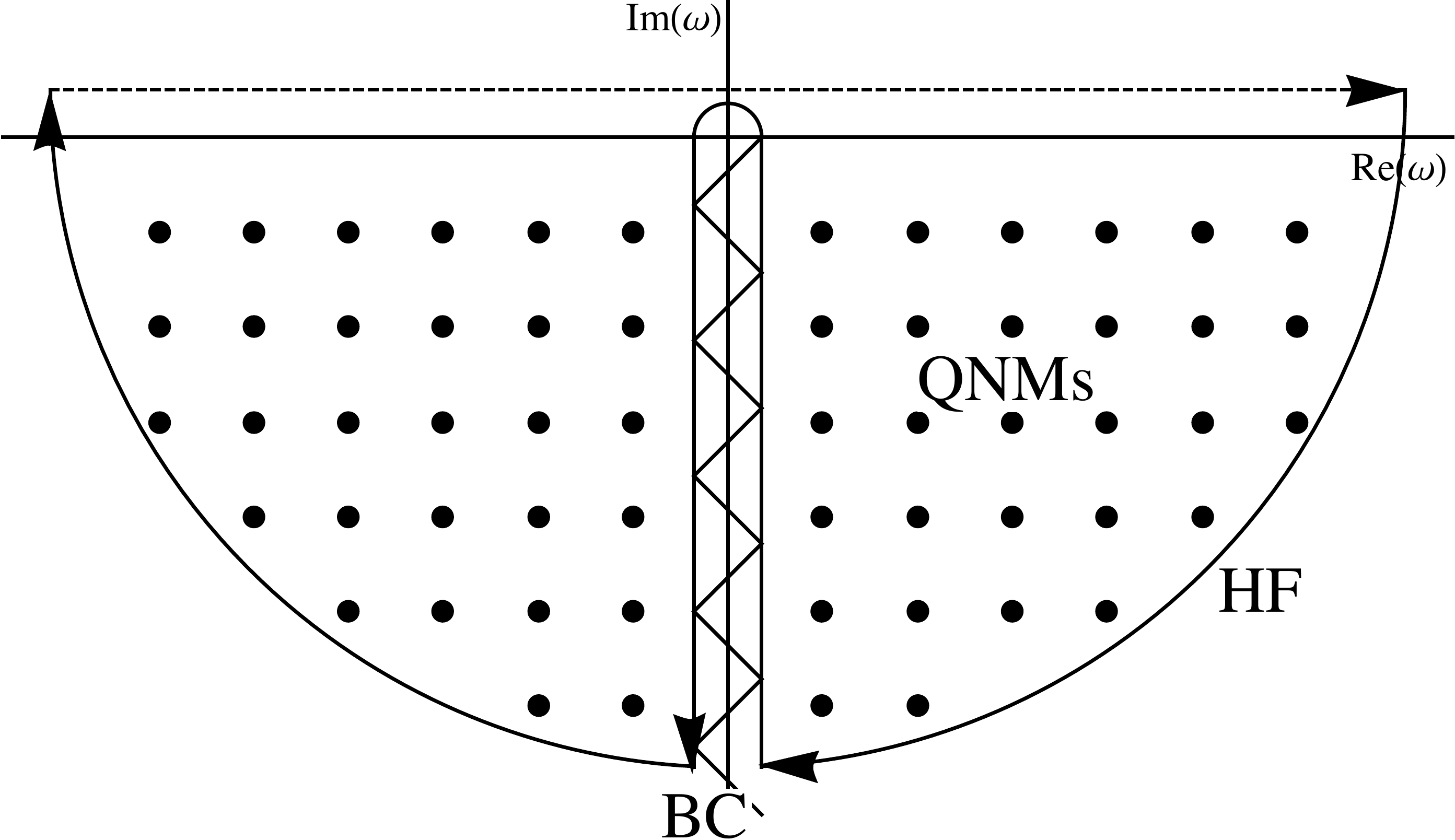}
\end{center}
\caption{Contour deformation in the complex-frequency plane.
The residue theorem of complex analysis allows one to re-express the integral over (just above) the real line of the Fourier modes
$\Glret$ of the GF as an integral over a high-frequency arc (HF) plus an integral around a branch cut (BC) of the Fourier modes
plus an integral over the residues at the poles (QNMs) of the Fourier modes.}
\label{fig:contour}
\end{figure}

In an influential paper, Leaver~\cite{Leaver:1986,*PhysRevD.38.725} deformed the integral in Eq.~(\ref{eq:Green ell})  over (just above)
the real axis on the complex-frequency plane to a contour along a high-frequency
arc (HF) on the lower semiplane, as in Fig.~\ref{fig:contour}. In doing so, the residue theorem of complex analysis dictates that one must take into account the singularities of the integrand. The Fourier modes $G_{\ell}$ possess simple poles (QNM frequencies) on the lower
semiplane and a branch cut (BC) down the negative-imaginary axis on the complex-frequency plane. 
Therefore, the integral in Eq.~(\ref{eq:Green ell}) can be rewritten as a sum of the integral $G^{HF}_{\ell}$ along the high-frequency arc, a sum $\GlQNM$ over the residues at the poles, and an integral $\GlBC$ around the branch cut:

\begin{equation}
\Glret=G^{HF}_{\ell}+\GlQNM+\GlBC.
\end{equation}
The QNM contribution $\FullGlQNM$ and the BC contribution $\FullGlBC{\ell}$ to $\FullGlret$ are given by
replacing $\Glret$ by, respectively, $\GlQNM$ and $\GlBC$ in Eq.~(\ref{eq:Green}).
The full QNM contribution $\GQNM$ and the full BC contribution $\GBC$ to $\Gret$ are then respectively given by summing over
all modes $\GlQNM$ and $\GlBC$. 

The integral along the
HF arc
yields a `direct' contribution which is expected to vanish after a
certain finite time
corresponding to a point $x'$ not lying beyond the boundary of $\mathcal{N}(x)$~\cite{Ching:1994bd,Ching:1993gt,Ching:1995tj,Capon-1998}.
Furthermore,
the quality of the match to the
QL expansion in the region between $\Delta t \approx r_* + r'_*$ and the edge of the normal neighbourhood
is strong empirical evidence that it also doesn't contribute there (at least in the specific cases
investigated). We therefore will not be considering the HF contribution any further under the assumption
that its contribution is either negligible or zero in the DP region.

In order to calculate the QNM and BC contributions $\GQNM$ and $\GBC$, we mainly
adopt the method introduced by Mano, Suzuki and Takasugi (MST) in Refs.~\cite{Mano:1996vt,Mano:1996mf,Sasaki:2003xr}
(for the BC we also use other methods described below). In this approach, solutions to the radial equation are expressed via two complementary infinite series, using (i) Gaussian hypergeometric functions, and (ii) Coloumb wave functions. Formally, the radius of convergence of these series 
respectively extends towards (i) spatial infinity, and (ii) the event horizon. MST established the key relationships between these representations, which enable one to compute radial wave functions and derived quantities, such as the Wronskian, to high accuracy. MST's method is particularly well-suited to calculations at low frequency. 

In the next two subsections we give more details of the calculations of the QNM and BC contributions  $\GQNM$ and $\GBC$ and present the results.
Further details on the calculational methods can be found in a series of 
papers~\cite{Dolan:2011fh,Casals:2012tb,Casals:2012ng,Casals:Ottewill:2011smallBC,Dolan:Ottewill:QNMMST}.
We note that the plots in this section which depend on the orbit of the particle correspond only to the circular case with radius $\rc=6M$ since their features are essentially shared by the eccentric orbit case at the same radius with $p=7.2$ and $e=0.5$.

\subsection{Quasinormal Mode Sum} \label{sec:QNM}

The Fourier modes $G_{\ell}$ of the GF have simple poles on the complex-frequency plane at the frequencies $\wQNM$ where the Wronskian passes through zero. The condition $W(\wQNM)=0$ defines a discrete spectrum of quasi-normal modes, i.e.~an infinite (but countable) set of complex frequencies $\wQNM$ which are labelled by angular momentum 
$\ell$ and overtone $n$ numbers. The spectrum is symmetric under reflection in the imaginary axis, i.e. under $\wQNM \rightarrow -\wQNM^\ast$. 

The key properties of the QNM spectrum can be understood by considering two asymptotic regimes. In the large angular momentum regime ($\ell \gg 1$ and $\ell \gg n$),
 the asymptotic spectrum is
\begin{equation}
\omega_{\ell n} \sim (\ell+1/2) \Omega - i (n+1/2) \Lambda ,   \label{qn-freq-l}
\end{equation}
where $\Omega$ and $\Lambda$ are, respectively, the orbital frequency and Lyapunov exponent (instability timescale) of the photon orbit at $r = r_{\text{orb}}$ \cite{1972ApJ...172L..95G,PhysRevD.31.290} (known as the light-ring). For the Schwarzschild black hole, $\Omega = \Lambda = 1/(\sqrt{27}M)$ and $r_{\text{orb}} = 3M$. In the large overtone regime ($n \gg 1$ and $n \gg \ell$), the asymptotic spectrum is
\begin{equation}
\omega_{\ell n} \sim k - i \kappa (n+1/2).   \label{qn-freq-n}
\end{equation}
Here $k$ is an $\ell$-independent constant which depends on the spin of the field, and $\kappa$ is the surface gravity at the horizon, $\kappa = 1/(4M)$. For the scalar field on the Schwarzschild spacetime, $k = \ln(3) / (8 \pi M)$ \cite{Andersson:2003fh}.

We define $\GlQNM$ to be the sum of residues of the Fourier mode $G_{\ell}$ at its (simple) poles
 \footnote{Note that due to a typographical error there is an extra factor of $2$ in Eq.62 of Ref.~\cite{Casals:2011aa}
 and a factor $1/(r\cdot r')$ missing inside the $\ell$-sum in Eq.1 of Ref.~\cite{Casals:2011aa}.}:
\begin{equation}\label{eq:Gl QNM}
\GlQNM(r,r';\Delta t)=
 \sum_{n=0}^{\infty}\GlnQNM(r,r';\Delta t),\quad
\GlnQNM(r,r';\Delta t)\equiv 
2 \, \text{Re}\left(
\frac{\mathcal{B}_{\ell n}}{\left(A^{out}_{\ell,n}\right)^2} 
\f(r,\wQNM)\f(r',\wQNM)
e^{-i \wQNM \Delta t}
\right),
\end{equation}
where 
$A^{in/out}_{\ell,n} \equiv A^{in/out}_{\ell,\omega=\wQNM}$.
The  QNM `excitation factors' are defined by $\mathcal{B}_{\ell  n}\equiv \AoutQNM/(2 \wQNM \alpha_{\ell n})$ and $\alpha_{\ell n}$ is defined via $\Ain\sim (\omega-\wQNM)\alpha_{\ell n}$ as $\omega\to \wQNM $.
 We note that the radial function $\g$ does not appear in Eq.~(\ref{eq:Gl QNM}) because it may be replaced by $\f/\Aout$ at a QNM frequency: the two quantities are equal when $\omega=\wQNM$, as follows from the fact that $A^{in}_{\ell,n}=0$ and from the boundary conditions  (\ref{eq:f,near hor}) and (\ref{eq:g}).

\subsubsection{QNM frequencies and excitation factors}
\begin{figure}[htb!]
\begin{center}
\includegraphics[width=10cm]{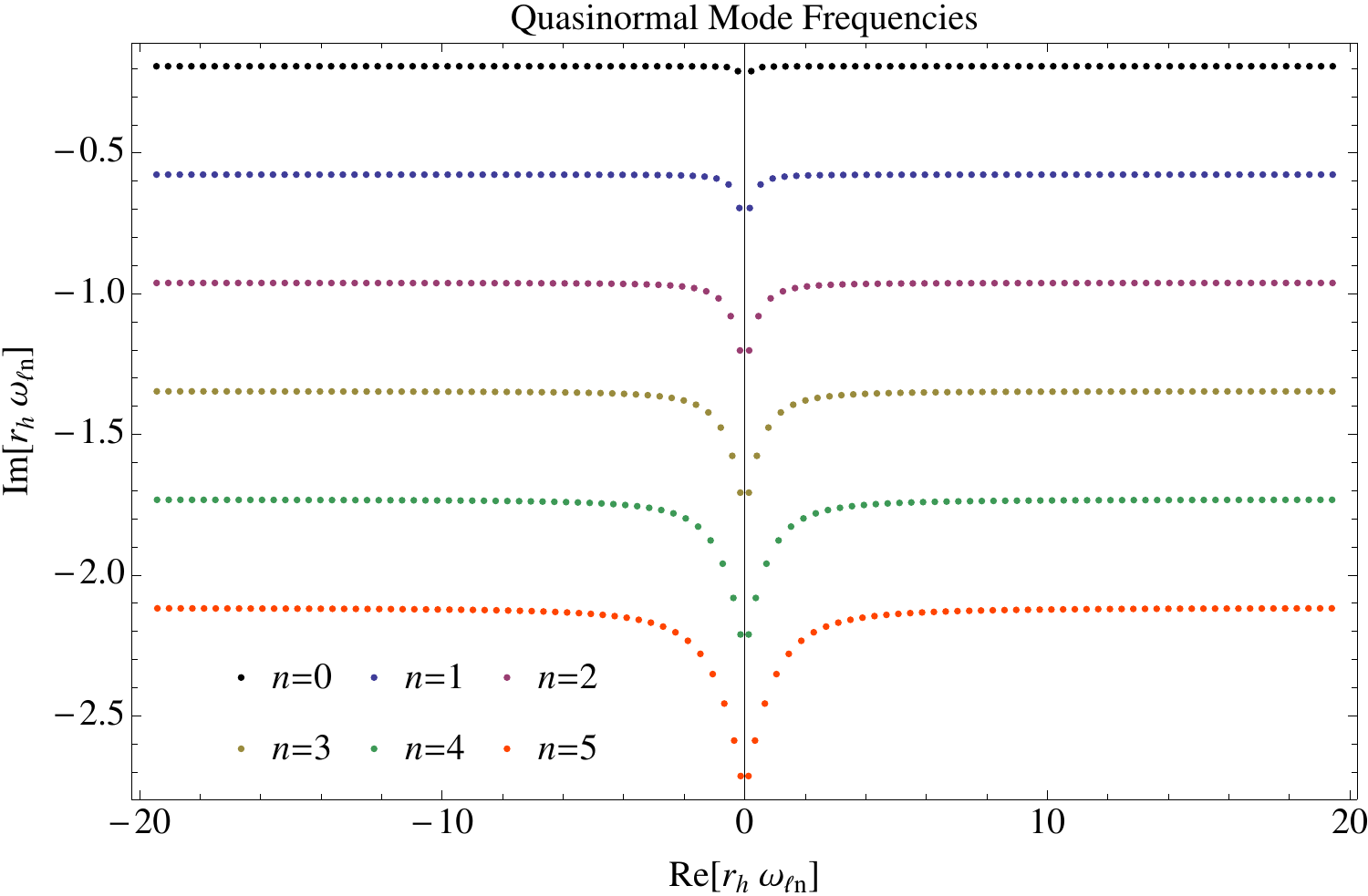}
\end{center}
\caption{Scalar QNM frequencies in the complex-frequency plane in Schwarzschild spacetime for $n:0\to 5$ and $\ell: 0\to 100$. The frequencies are symmetric with respect to the negative imaginary axis.
In the 4th quadrant, the frequencies move down the plane with increasing $n$ and across to the right with increasing $\ell$.
}
\label{fig:QNM freqs.}
\end{figure} 
Figure \ref{fig:QNM freqs.} shows the real and imaginary parts of the QNM frequencies $\wQNM$ for $n:0\to 7$ and $\ell: 0\to 100$. The plot demonstrates that the transition between the two asymptotic regimes is smooth. The imaginary part of the frequency remains negative for all $\ell$, $n$, indicating that all QNMs decay exponentially over time. As indicated by Eq.~(\ref{qn-freq-l}) and (\ref{qn-freq-n}), the damping rate increases with overtone number. It is natural to call the lowest overtones ($n=0$) the `fundamental' modes.

\begin{figure}
\begin{center}
\includegraphics[width=8.5cm]{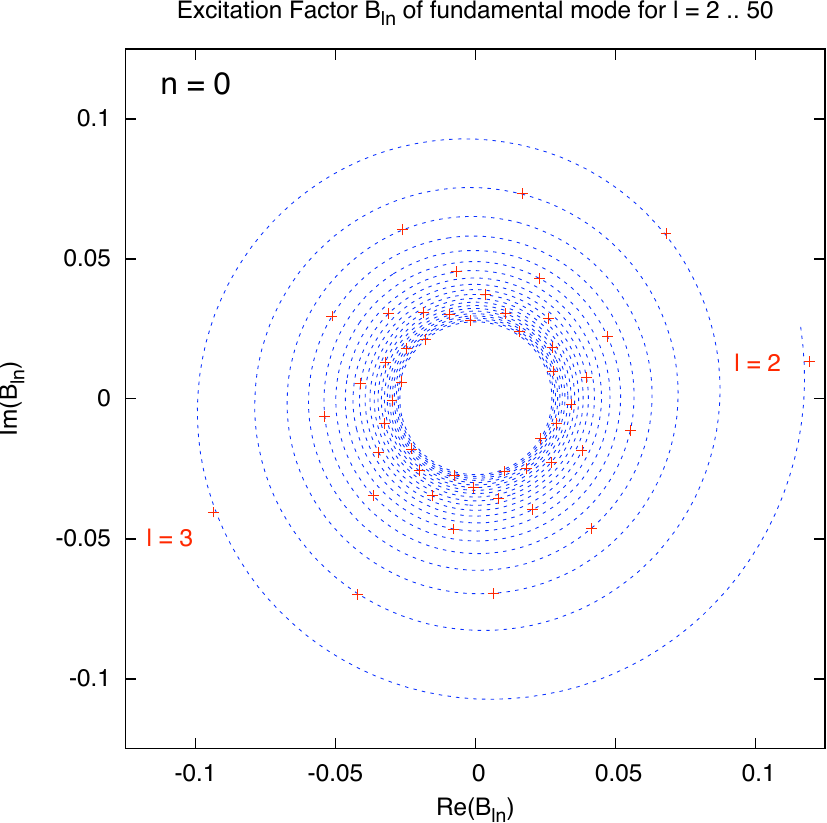}
\includegraphics[width=8.5cm]{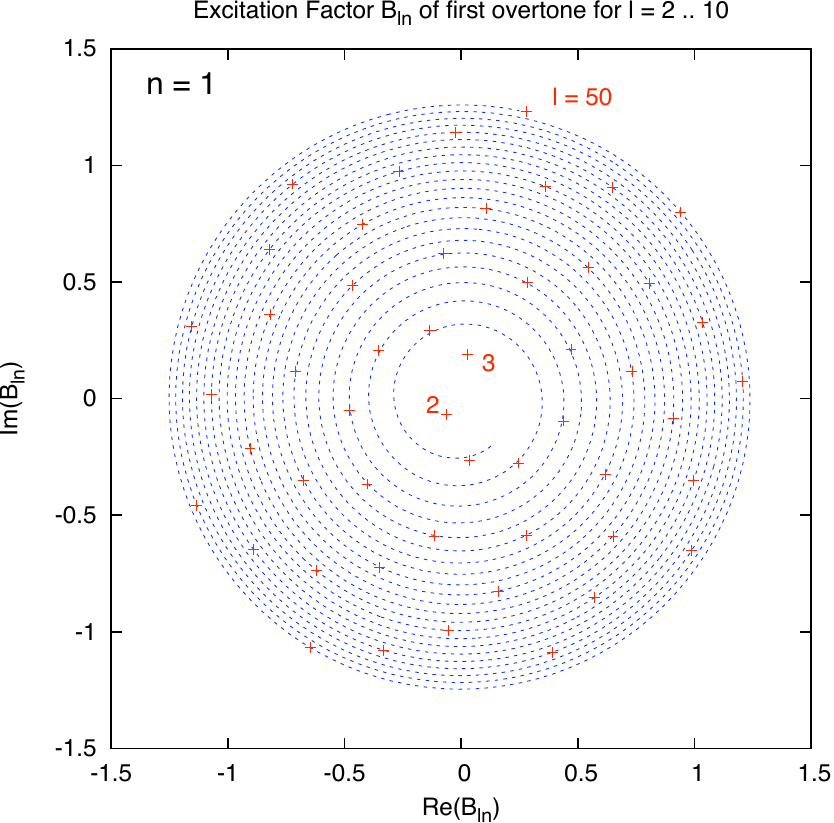}
\end{center}
\caption{The QNM excitation factors in the Argand plane. These plots show the real and imaginary parts of $\mathcal{B}_{\ell n}$ for the fundamental ($n=0$, left) and first overtone ($n=1$, right) of the scalar field on Schwarzschild spacetime, from $\ell = 2 \ldots 50$. The data points show values determined with the MST method (see text), and the dotted lines show the asymptotic approximation for large $\ell$ (Eq.~(31) in Ref.~\cite{Dolan:2011fh}). }
\label{fig:QNM exc.fac.}
\end{figure}
Figure \ref{fig:QNM exc.fac.} shows the
excitation
factors $\mathcal{B}_{\ell n}$ determined via the MST method for the fundamental and first-overtone modes, for multipoles up to $\ell = 50$. We have checked our values against the data in Tables III and IV of Ref.~\cite{Berti:2006wq}; against the asymptotic expressions in Ref.~\cite{Dolan:2011fh}; and against results from  Ref.~\cite{Zhang:2013ksa} where the MST method is also used. In Appendix \ref{sec:App} we present tables with the QNM frequencies $\wQNM$, the excitation factors $\mathcal{B}_{\ell  n}$ and the coefficients $A^{out}_{\ell,n}$ for the modes $n=0\to 5$ and $\ell:0\to 50$.


\subsubsection{Convergence of mode sums}
The sum over $n$ in Eq.~(\ref{eq:Gl QNM}) is taken over all QNMs in the fourth quadrant of the complex-$\omega$ plane. It is natural to ask whether this $n$-sum is convergent. Consider the case when $r$ and $r'$ are large, so $\f(r, \omega_{ln}) \sim \AoutQNM \exp(i \wQNM r_\ast)$ and we may write
\begin{equation}
\GlQNM(r,r';\Delta t) \sim 2 \, \text{Re} \sum_{n} C_{\ell,n} e^{-i \wQNM T} 
\end{equation}
where $C_{\ell,n}$ can be read off from Eq.(\ref{eq:Gl QNM})
and
$T \equiv \Delta t - r_\ast - r_\ast'$ is the `reflection time'. In the high overtone regime, the ratio of successive terms, $\sim |C_{\ell,n+1}/C_{\ell,n}| \exp (-\kappa T)$, is exponentially suppressed, via Eq.~(\ref{qn-freq-n}). It was demonstrated in Ref.~\cite{Andersson:1996cm} that for fixed $\ell$, $C_{\ell n}$ exhibits power-law dependence on $n$ at large $n$ and hence $T > 0$ is a sufficient condition for the absolute convergence of the $n$-sum. In the general case, where $r$ and $r'$ may not be large, the $n$-sum will still converge at sufficiently late times \cite{Casals:2011aa}. 

On the other hand, the sum over $\ell$ in Eq.~(\ref{eq:Green}) is not absolutely convergent, in \emph{any} regime. Asymptotic expressions for the quantities featuring in Eq.~(\ref{eq:Gl QNM}) in the large-$\ell$ regime ($\ell \gg 1$, $\ell \gg n$) were given in Ref.~\cite{Dolan:2011fh}, in Eqs.~(27)--(34) and (A41). In this regime, the magnitude of $\mathcal{B}_{\ell  n}$ scales in proportion to $L^{n-1/2}/n!$, whereas the complex phase varies linearly with $L$, where $L \equiv \ell + 1/2$.

It is is not a surprise to find that the series is not absolutely convergent, since we should recall that the GF is actually a distribution, which may exhibit infinitely-sharp features. The GF is singular on the light-cone. 
Conversely, we expect it to be well-defined and regular elsewhere. In \cite{Dolan:2011fh}, using the large-$\ell$ asymptotic results, it was shown that (i) the solution of a coherent phase condition ($\lim_{\ell \rightarrow \infty} \text{arg}( G^{QN}_{\ell+1} / \GlQNM )= 2 \pi k$ where $k \in \mathbb{Z}$) describes the light cone in spacetime, to within the expected error of the approximation; and (ii) the singularity structure of the GF follows a repeating four-fold pattern, as described in the Introduction:
$\delta(\tilde{\sigma})$, $\text{PV}\left(1/\pi\tilde\sigma\right)$, $-\delta(\tilde\sigma)$, $-\text{PV}\left(1/\pi\tilde\sigma\right)$, $\delta(\tilde\sigma)$, \dots,
where $\tilde{\sigma}$ is an approximation to Synge's world-function $\sigma$. 
We next describe a method for obtaining well-defined values of the QNM $\ell$-sum for points lying in-between null geodesic reintersections.

\subsubsection{Calculation of mode sums}

In order to perform the $\ell$-sum of the multipole modes $\FullGlQNM$ we use the ``smoothed sum method" of Sec.~VII.~D in Ref.~\cite{Casals:2009zh}, which is justified in Ref.~\cite{Hardy}. 
Our method relies on introducing a smoothing factor, $\tilde{S}(\ell)$, to attenuate the high-$\ell$ part of the mode sum.

As may be anticipated from Fourier theory, introducing a smooth filter in the $\ell$-mode decomposition is equivalent to convolving the GF with a narrow function which `smears out' the distributional features. To make this precise, consider a distribution $D(\cdot)$ and a smooth function $S(\cdot)$ with mode sum representations
\begin{equation}
D(\cos \theta) = \sum_\ell \left( \ell + \frac{1}{2} \right) \tilde{D}(\ell) P_{\ell}(\cos \theta) \quad \quad \text{and} \quad \quad
S(\cos \theta) = \sum_\ell \left( \ell + \frac{1}{2} \right) \tilde{S}(\ell) P_{\ell}(\cos \theta) .
\label{DSmodesum}
\end{equation}
The smoothed mode sum is equivalent to a convolution of these functions, i.e.
\begin{equation}
\sum_{\ell} \left(\ell + \frac{1}{2} \right) \tilde{D}(\ell) \tilde{S}(\ell) P_{\ell}(\cos \theta) = \frac{1}{2\pi} \int_{-1}^1 d(\cos \chi) \, \int_{-\pi}^\pi D( \cos \gamma ) S(\cos \chi) d \varphi  ,
\end{equation}
where $\cos \gamma = \cos \theta \cos \chi + \cos \varphi \sin \theta \sin \chi $. 

We make use of a smoothing factor $\tilde{S}(\ell)\equiv\exp(-\ell^2/2 \lcut^2)$.
 Via Eq.~(\ref{DSmodesum}), this smoothing factor corresponds to a smearing function $S(\cos \theta) \sim \Lcut^2 \exp(-\Lcut^2 \theta^2 / 2)$ for small $\theta$, where $\Lcut = \lcut + 1/2$. 
It is straightforward to show that, in the limit $\lcut \rightarrow \infty$, the smoothing factor smears a delta function, $\delta(\cos \theta - \cos \gamma)$, into a Gaussian $(\Lcut / \sqrt{2\pi \sin \theta \sin\gamma}) \, \exp\left( -\frac{1}{2} \Lcut^2 (\theta - \gamma)^2 \right)$ of angular width $1/\Lcut$. 

Such a smoothing factor can also be related to the numerical technique used in~\cite{Zenginoglu:2012xe}, where the four-dimensional
Dirac-$\delta$ source in the PDE obeyed by the GF is replaced by a `narrow' Gaussian distribution.
If in Eq.~(50) of Ref.~\cite{Dolan:2011fh} one introduces a smoothing factor $\tilde{S}(L)$ in the integrand, it is easy to show that the fourfold singularity structure of 
Eq.~(52)~\cite{Dolan:2011fh} becomes `smoothed out': instead of a Dirac-$\delta(\sigma)$ one obtains a Gaussian distribution and instead of the $\text{PV}\left(\frac{1}{\sigma}\right)$
type of singularity one obtains the Dawson integral of Eqs.~(9) and (10) in Ref.~\cite{Zenginoglu:2012xe} with the width of the four-dimensional Gaussian  being equal to $1/(M \ell_{\rm cut})$.
That is, the introduction of a smoothing factor  $\tilde{S}(L)$ in the $\ell$-sum is equivalent to a replacement of the four-dimensional Dirac-$\delta$ source by a Gaussian distribution. 


In the circular orbit case we calculated up to 180 $\ell$-modes and in the eccentric orbit case up to 157 $\ell$-modes.
For these values, we found it appropriate to take $\lcut = 25$.
We have checked that, for large values of $\ell$, 
the modes $\GlQNM$ calculated via the method of MST agree well with the large-$\ell$ asymptotics obtained in Ref.~\cite{Dolan:2011fh}. We find that the large-$\ell$ asymptotics are  accurate up to the expected level, i.e., up to terms of order $1/L$ in the complex phase. A small error in the phase can lead to a larger absolute error, particular for higher overtones, as $|\mathcal{B}_{\ell n}| \sim L^{n - 1/2}$. For this reason, we have used only the results from the MST method.

\subsubsection{Sample MST results}

\begin{figure}[htb!]
\begin{center}
\includegraphics[width=8cm]{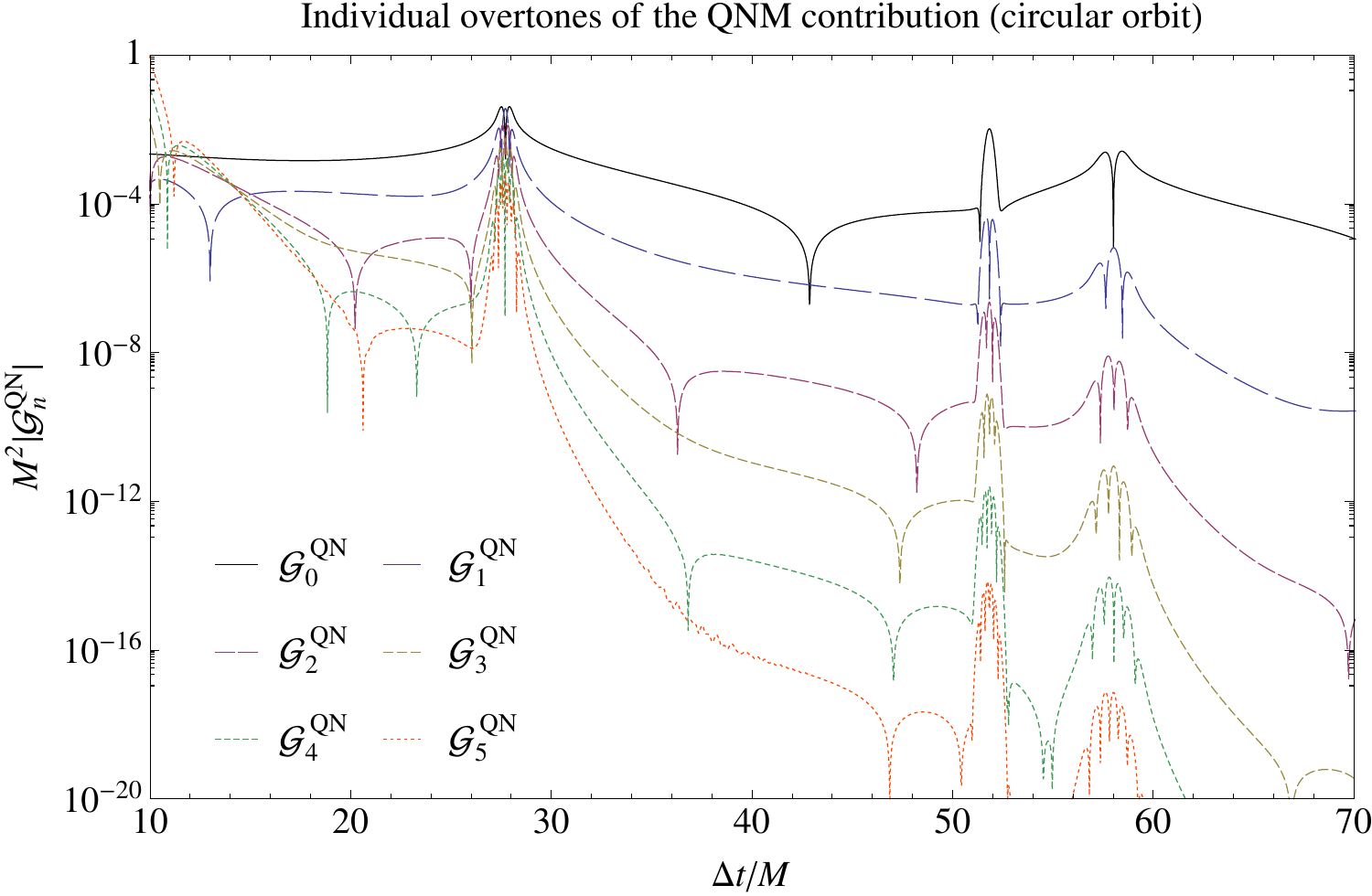}
\includegraphics[width=8cm]{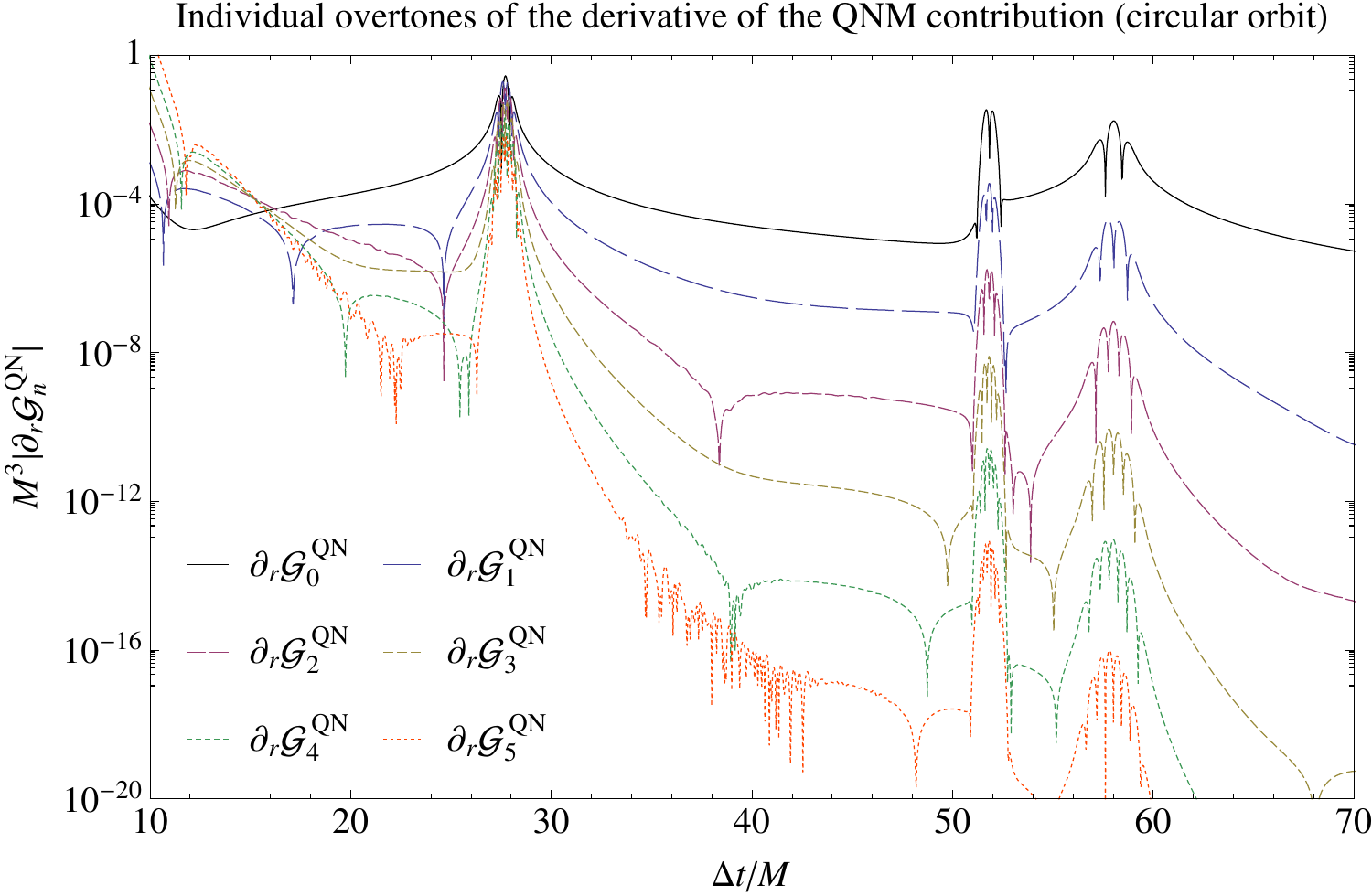}
\end{center}
\caption{
QNM contribution  for individual overtones $n=0,1,2,3,4,5$ to the GF (left)
and to the radial derivative of the GF (right) as a function of time $\Delta t$ along a circular
geodesic of radius $\rc=6M$.
Note that we use the symbol $\mathcal{G}_n^{QN}$ (with $\GQNM=\sum_{n=0}^{\infty}\mathcal{G}_n^{QN}$, in obvious notation)
 in this figure  only - not to be confused with  $\FullGlQNM$.
}
\label{fig:dGretdr Num QNM large-l circ r=6 n-mode} 
\end{figure}

Figure \ref{fig:dGretdr Num QNM large-l circ r=6 n-mode} shows the contribution of the lowest overtones to the GF, and its radial derivative, as a function of time along a circular geodesic. The plots shows that the higher overtones are negligible in comparison to the fundamental ($n=0$) modes for the majority of the time.
This figure illustrates that the singularities of the QNM contribution $\GQNM$ 
coincide precisely with the singularities of the full GF. That is, singular features in Fig.~\ref{fig:dGretdr Num QNM large-l circ r=6 n-mode} occur whenever there is a null geodesic connecting the points on the worldline.
These `light-crossing times'   (illustratively referred to as `caustic echos' in Ref.~\cite{Zenginoglu:2012xe})  may be found by solving the null geodesic equations, and
in the case of a circular geodesic at $\rc=6M$, the values are $\Delta t/M\approx 27.62,51.84, 58.05, 75.96, 100.09, 108.55, 124.21,\dots$ 



\subsection{Branch Cut} \label{sec:BC}

The BC integral $\GlBC$ can be calculated as~\cite{Casals:2011aa,Leung:2003ix}
\begin{equation}
\label{eq:GBC}
\GlBC(r,r';\Delta t)\equiv
-\frac{i}{2\pi}\int_{0}^{\infty} d\nu\ \DGw{r}{r'}{\nu}e^{-\nu \Delta t},
\end{equation}
where $\nu\equiv i\omega>0$ along the BC, with the 
BC modes
$\DGl$ expressed as
\begin{align} \label{eq:DeltaG in terms of Deltag}
\DGw{r}{r'}{\nu}=
-2i\nu \f(r,-i\nu)\f(r',-i\nu)
\frac{q(\nu)}{|W(-i\nu)|^2},
\quad 
\text{for}\quad \frac{r_*}{M},\frac{r'_*}{M}
\in \mathbb{R},
\end{align}
where the function 
\begin{equation} \label{eq:dg=qg}
q(\nu)\equiv -i\frac{ \lim_{\epsilon\to 0^+}\left(\g(r,\epsilon-i\nu)-\g(r,-\epsilon-i\nu)\right)}{\g(r,+i\nu)},
\end{equation}
is the so-called branch cut `strength'. 
In order to calculate the BC contribution to the GF we used the methods recently developed 
in~\cite{Casals:2012tb,Casals:Ottewill:2011smallBC,Casals:2012ng}.
We use a different method for 
three different frequency
regimes along the BC: `very small', `small' and `mid' frequency regimes.
In the two orbit cases studied in this paper and for the values of $\ell$ included in the BC contribution, these regimes
correspond to, respectively, $M\nu \in (0,0.05)$, $M\nu \in (0.05,0.225)$ and $M\nu \in (0.225,9)$.

It is clear from the exponential factor in Eq.~(\ref{eq:GBC}) that the regime for `very small' frequencies in the integral will only contribute significantly
to $\GlBC$ at `very late' times. Due to this exponential factor, we need to carry out an analytic -- rather than numeric -- integration in this `very small' frequency regime.
We use the 
MST method
in order to obtain  the  radial function $\f$ and the functions
$q(\nu)$ and $W(-i\nu)$.
In order to perform a small-$|M\omega|$ expansion of the radial function $\f$ we use the Barnes integral representation of
the hypergeometric functions~\cite{Casals:2012tb,Casals:Ottewill:2011smallBC}, which appear in the MST series representations for $\f$.
We calculate the small-$|M\omega|$ expansion of $\f$ up to three orders and of $q$ and $W$ up to fifteen orders (since the expansions for $q$ and $W$ are also used
in the `small' frequency regime).
We found that a Pad\'e approximation of the MST small-$|M\omega|$ expansion of the Wronskian $W$ significantly increases the validity of the expansion.
It is shown in~\cite{Casals:2012tb,Casals:Ottewill:2011smallBC} that this `very small' frequency regime yields the 
well-known~\cite{Price:1972pw,Price:1971fb,Leaver:1986}
power-law tail decay at late times of the GF plus a new logarithmic behaviour in time at higher order.

In the `small' frequency regime, we use the same small-$|M\omega|$ expansions of  $q$ and $W$ as those used in the `very small' frequency regime.
However, for the radial 
function $\f$ itself we choose to use the Jaff\'e series~\cite{Leaver:1986a,Casals:2012ng}
since it is easier to use and  in this regime we do not need to carry out an exact analytic integration in Eq.(\ref{eq:GBC}).

In the `mid' frequency regime we use the method developed in~\cite{Casals:2012ng}, since the convergence of the MST series representations
becomes too slow in this regime as well as the fact that the coefficients in these series (which are the same as the coefficients
in the Jaff\'e series) possess poles in this `mid' frequency regime on the BC. 
The calculation of the BC modes in this `mid' frequency regime enables us to see that, in the cases studied in this paper,
the contribution of this regime to the BC is rather small but is necessary in order to increase the accuracy of the final value obtained for the SF.

\begin{figure}[htb!$,]
\begin{center}
\includegraphics[width=8cm]{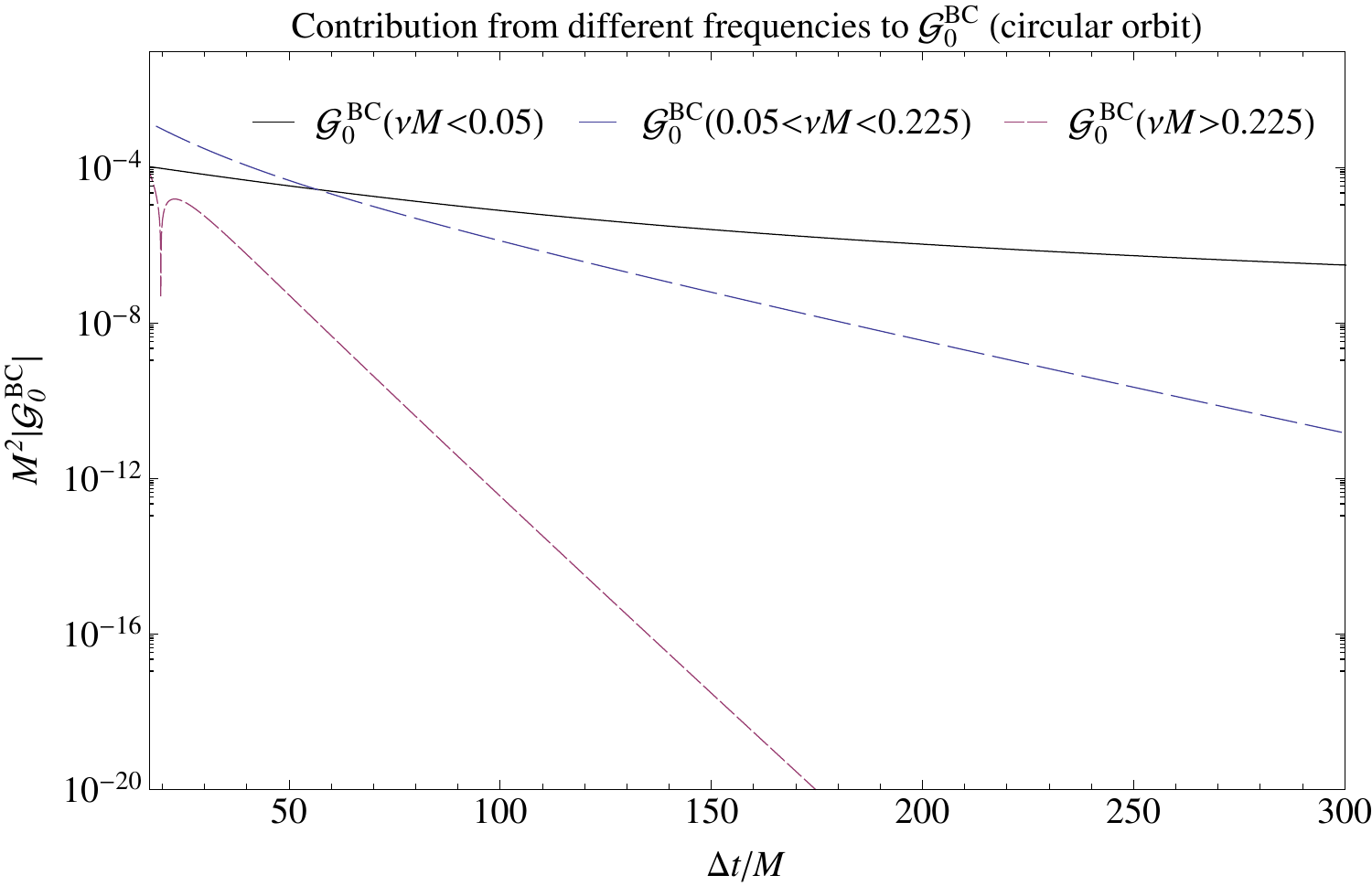}
\includegraphics[width=8cm]{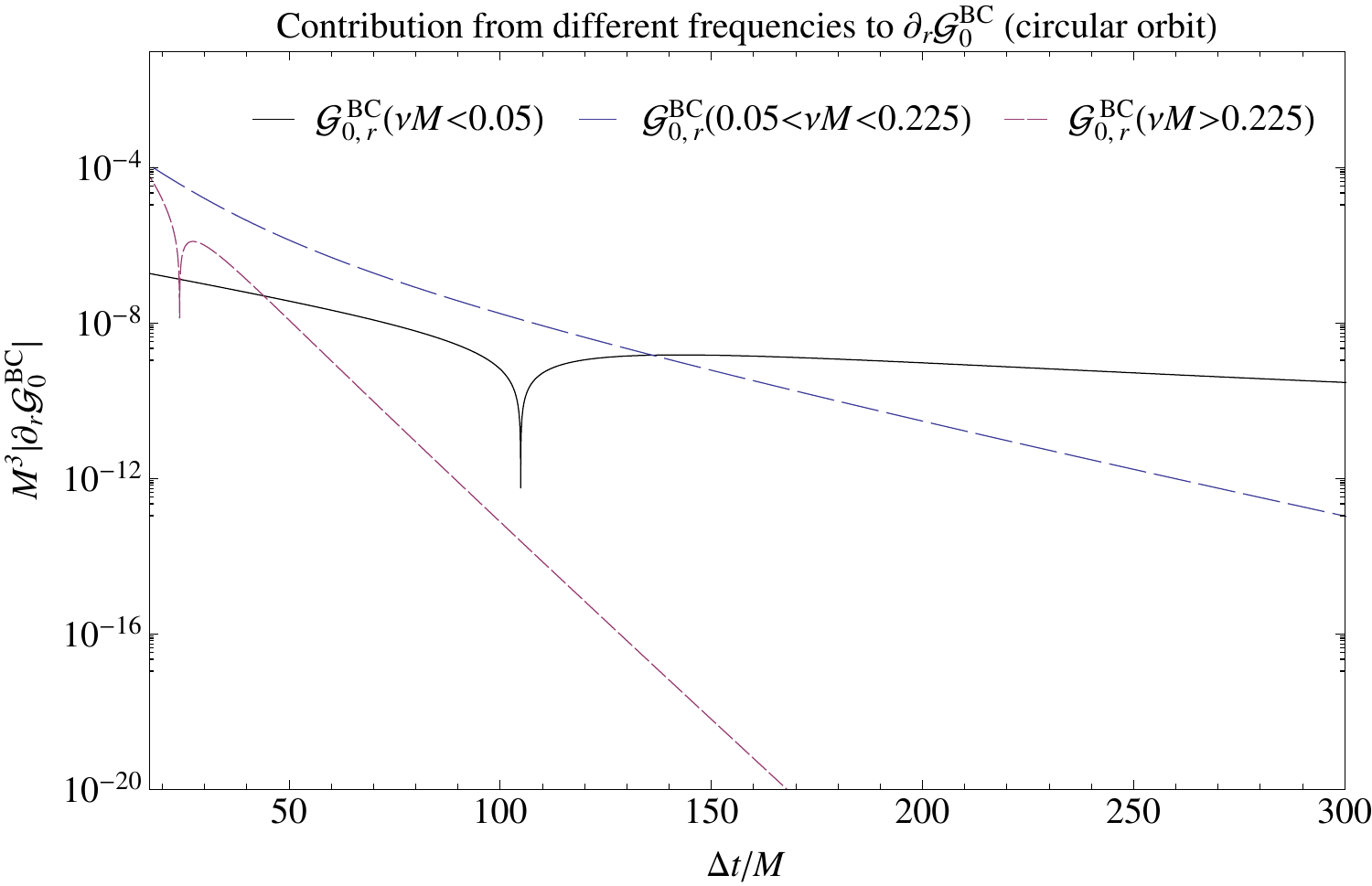}
\end{center}
\caption{Dominant BC mode $\FullGlBC{\ell=0}$ (left) and $\partial_r\FullGlBC{\ell=0}$
(right) 
along a circular
geodesic of radius $\rc=6M$ as a function of time.
We plot the various contributions from different integration intervals in 
Eq.~(\ref{eq:Gl}): black curve is integration over $M\nu: 0\to 0.05$
(`very small' frequency regime), dashed blue curve is over $M\nu: 0.05\to 0.225$ (`small' frequency
regime), dotted purple curve is over $M\nu: 0.225\to 9$ (`mid' frequency regime).
}
\label{fig:GBC s=l=0 r=6} 
\end{figure}

In Fig.~\ref{fig:GBC s=l=0 r=6} we show the contributions of these different frequency regimes along the BC to the radial derivative of $\FullGlBC{\ell=0}$.
Given how small the contribution from the `mid' frequency regime is in the DP, in the 
cases we studied here the contribution from larger frequencies (i.e., $M\nu>9$ in the cases studied in this paper) is negligible and so we do not need to evaluate it
(although asymptotics for `large' frequency were developed in~\cite{Casals:2011aa} which could be used if necessary in other cases).

\begin{figure}[htb!]
\begin{center}
\includegraphics[width=8cm]{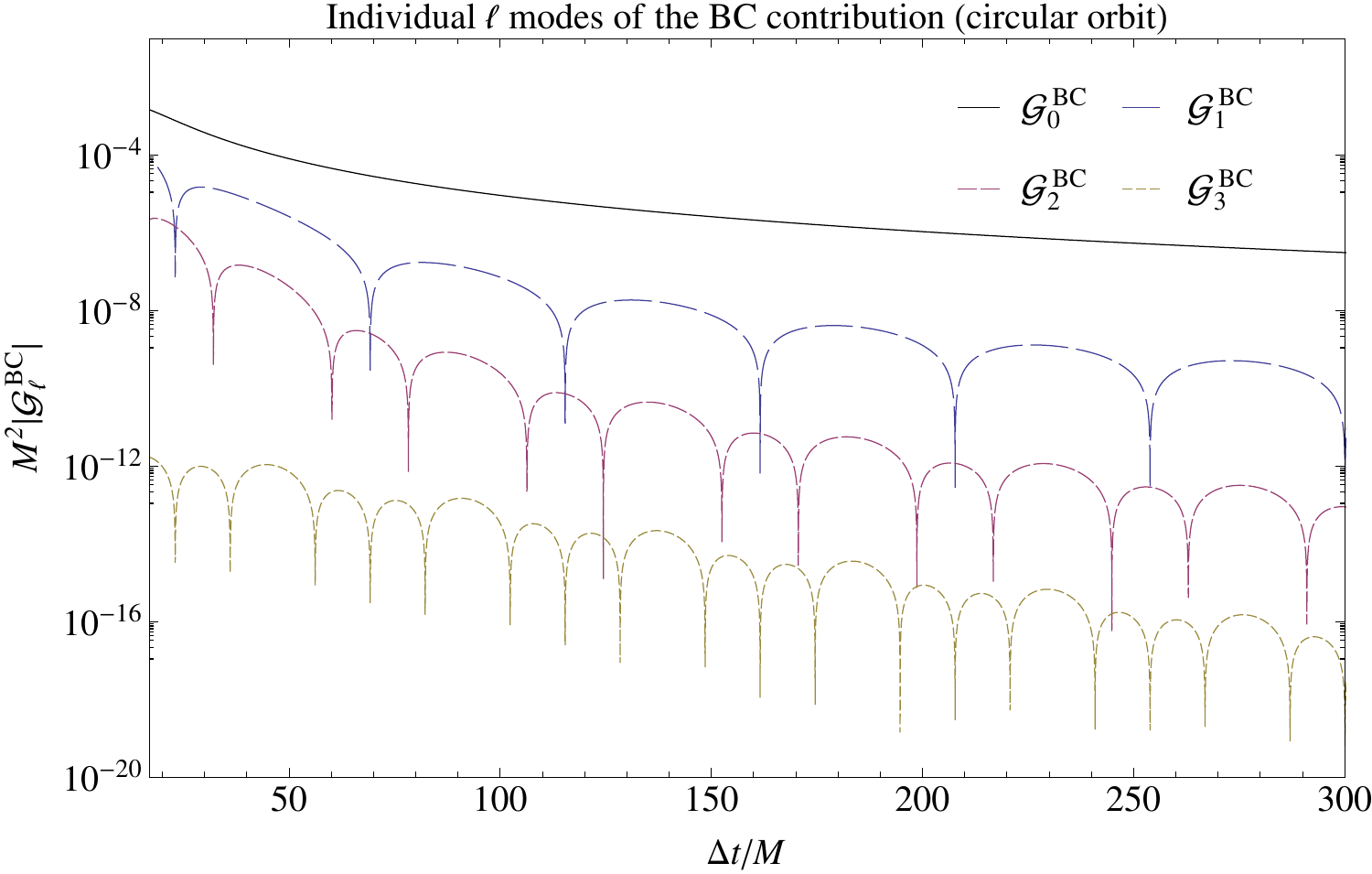}
\includegraphics[width=8cm]{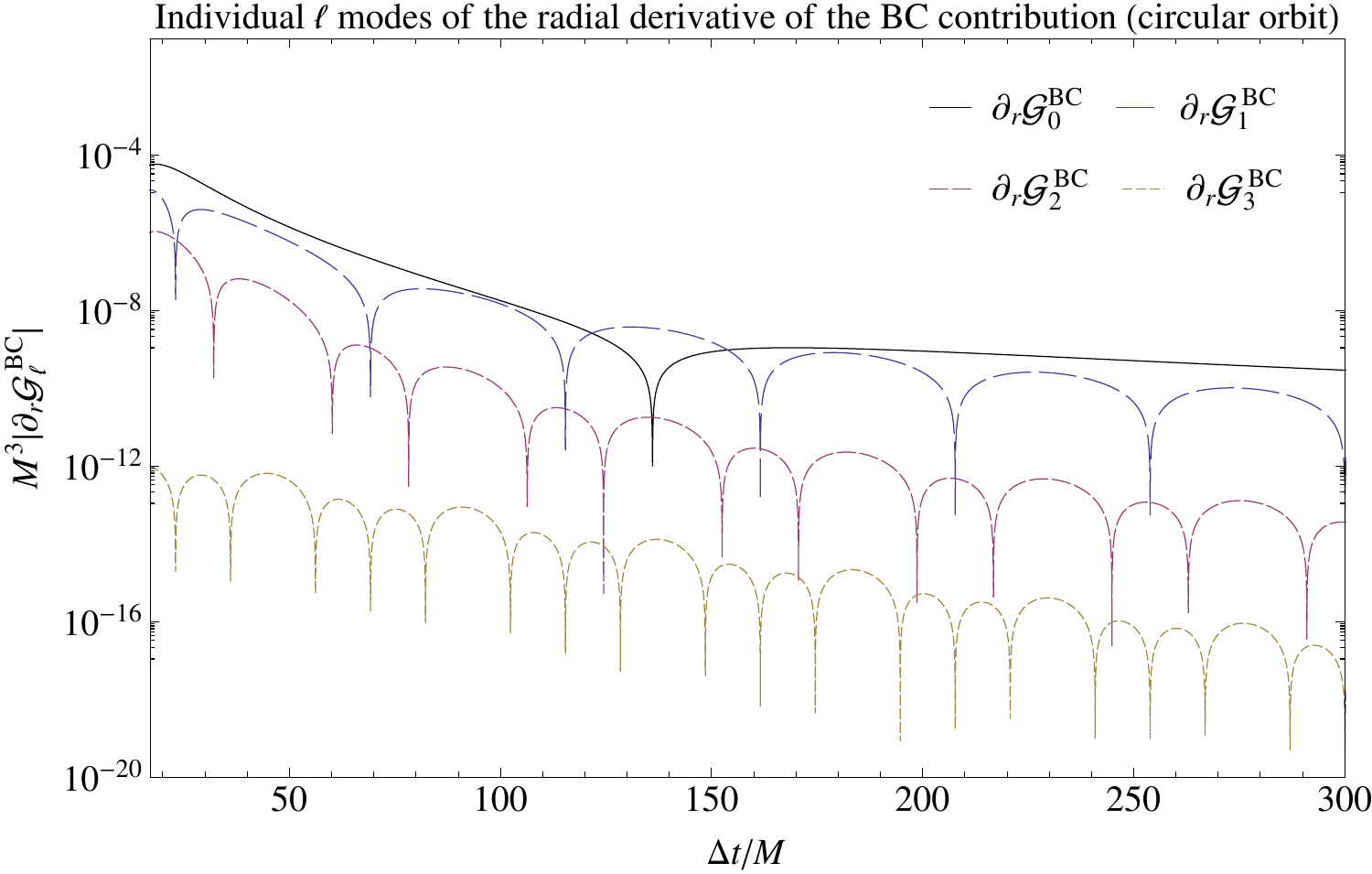}
\end{center}
\caption{
Individual $\ell$ modes of the BC contribution to the GF (left) and its radial
derivative (right) as a function of time along a circular geodesic of radius $\rc=6M$.
}
\label{fig:Gret Num Tail l-mode} 
\end{figure}

We calculated the BC modes $\FullGlBC{\ell}$ for $\ell=0$, $1$, $2$ and $3$.
Fig.~\ref{fig:Gret Num Tail l-mode} shows that the contribution to the GF from $\FullGlBC{\ell}$ decreases with increasing value of $\ell$.
This decrease is exponential at late-times as known from the power-law tail behaviour (e.g.,~\cite{Casals:2012tb}) and it appears
to continue to be exponential at any other times in the DP.
For $\ell=2$ we only needed to calculate the `very small'  and `small' frequency regimes and for $\ell=3$ only the `very small' frequency regime, 
since the contribution from larger frequencies for these modes
is negligible in the time region of interest (i.e., in the DP).
We have checked that the BC contribution $\GBC$, and particularly the mode $\FullGlBC{\ell=0}$, completely captures 
the late-time behaviour of the GF.
The behaviour with $\ell$ is similar for the radial derivative of the GF, however, higher-$\ell$ modes contribute more significantly up to
later times compared with the GF case.
The dependence on the angle $\gamma=\Omega \Delta t$ of the mode $\FullGlBC{\ell}$ obviously comes from the
Legendre polynomial $P_{\ell}(\cos\gamma)$ in Eq.~(\ref{eq:Green}).
Therefore, the dominant $\ell=0$ mode has no angular dependence, while the $\ell=1$ mode is expected to oscillate with time with the angular frequency 
$2\pi/\Omega$, the $\ell=2$ mode with twice that frequency, etc.
The panel on the right of Fig.~\ref{fig:Gret Num Tail l-mode} shows that the contribution of the $\ell=1$ mode to $\partial_r\GBC$ gives rise to its
 `wagging of the tail':  the oscillatory behaviour that it exhibits at late times (see, e.g., the panel on the right of Fig.~\ref{fig:QNM+BC}).

\subsection{Full distant past Green function}
\begin{figure}[htb!]
\begin{center}
\includegraphics[width=8cm]{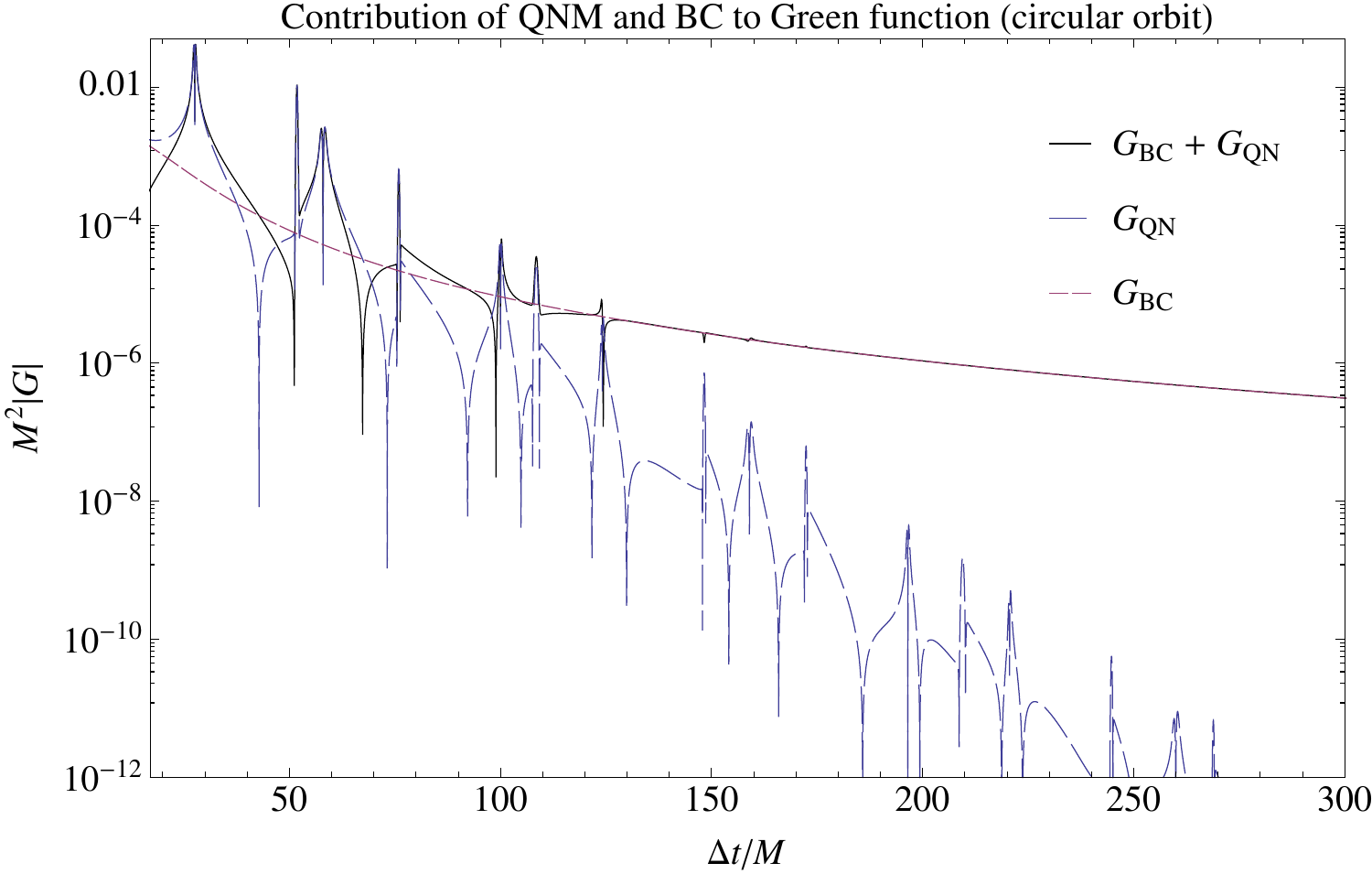}
\includegraphics[width=8cm]{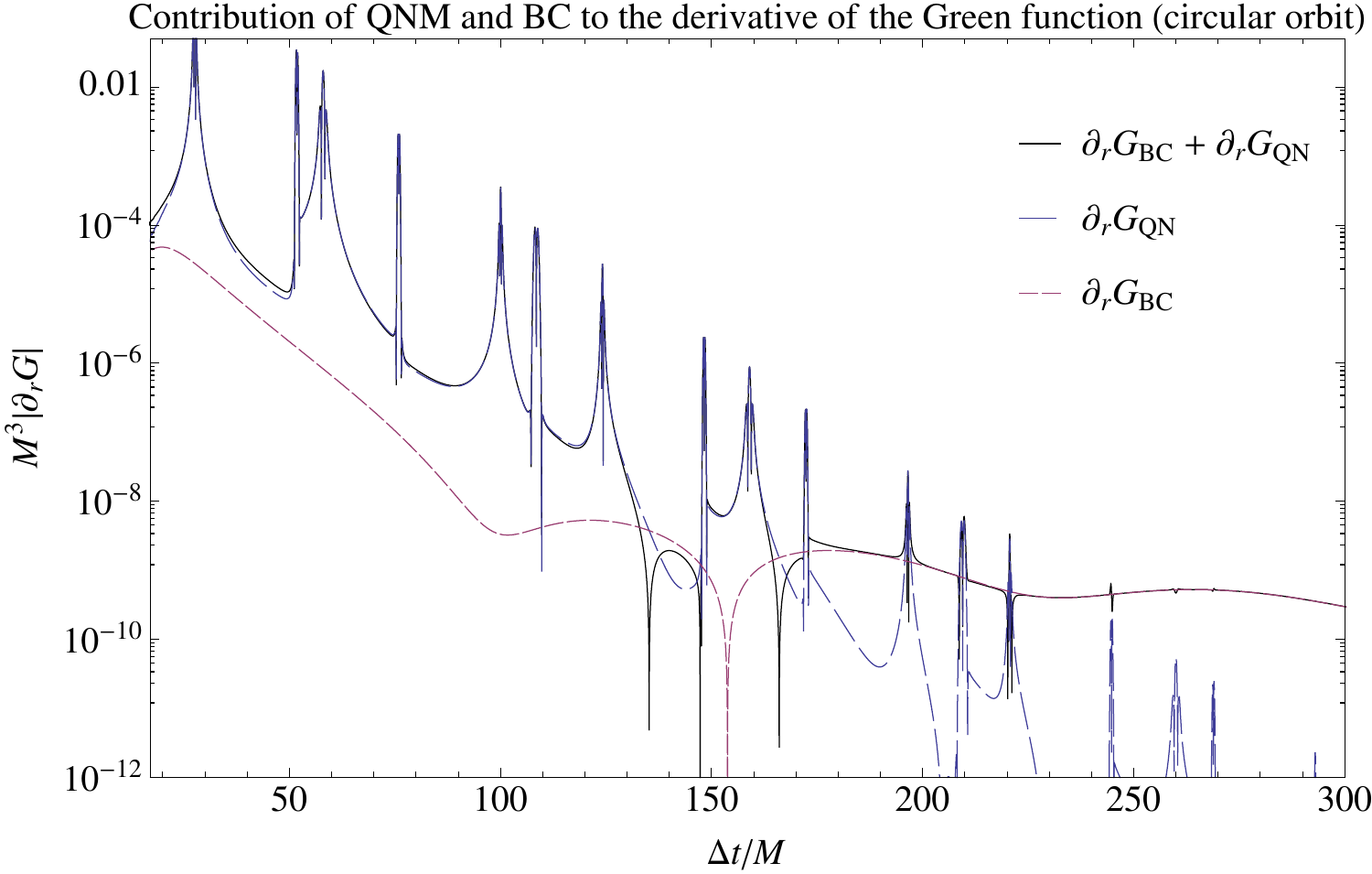}
\end{center}
\caption{
Combining QNM and BC contributions to compute the GF  (left) and radial derivative
of GF (right) in the DP  along a circular geodesic of radius $\rc=6M$.
}
\label{fig:QNM+BC} 
\end{figure}

By combining contributions from the QNM sum and BC integral, we obtain an expansion for the GF which is valid for all points $x$ and $x'$ sufficiently far apart, i.e., in the DP. 
Figure \ref{fig:QNM+BC} shows the QNM contribution, the BC contribution and the full QNM + BC Green function. As expected, the BC contribution yields the late-time tail while the QNM contribution yields the singular features at the light-crossing times. We note that the BC contribution to the GF dominates over the QNM contribution not only at late-times but also at certain `mid' time regimes and at `early' times.
In contrast, the BC contribution to the \emph{radial derivative} of the GF is less significant. 
This, we believe, is related to the fact that the leading-order coefficient in the small-$M\nu$ expansion of $\FullGlBC{\ell=0}$ (which is the dominating BC mode in the DP) is independent of $r$ and so it is eliminated from the radial derivative of the GF. Interference between the sub-dominant $\ell=0$ term and the $\ell=1$ part generates an oscillation in $\partial_r G$ at the orbital frequency.

In order to assess the
accuracy of the QNM+BC expansion in the DP, we validate it against the same GF computed using a variant of the numerical Gaussian method pioneered
in~\cite{Zenginoglu:2012xe}. A full description of our method --- together with the corresponding
numerical calculation of the SF --- will be presented in~\cite{CDGOWZ}. In brief, we numerically
evolve the $1+1$D wave equation in the time domain for each spherical harmonic mode and then sum
over $\ell$ to obtain the full $3+1$D solution. We use Gaussian initial data such that in the limit of
zero Gaussian width the numerical solution is the GF with one point fixed at the center
of the Gaussian. We choose a Gaussian pulse of width $ 0.1 M$, use a spatial resolution of
$\Delta r_\ast = 0.01M$ and include $200$ $\ell$-modes. To eliminate spurious large-$\ell$
oscillations in the final solution, we use a smooth-sum cut-off by multiplying each $\ell$-mode by
a smoothing factor  $\tilde{S}(\ell)$, with $\ell_{\rm cut} = 25$.

\begin{figure}[htb!]
\begin{center}
\includegraphics[width=8cm]{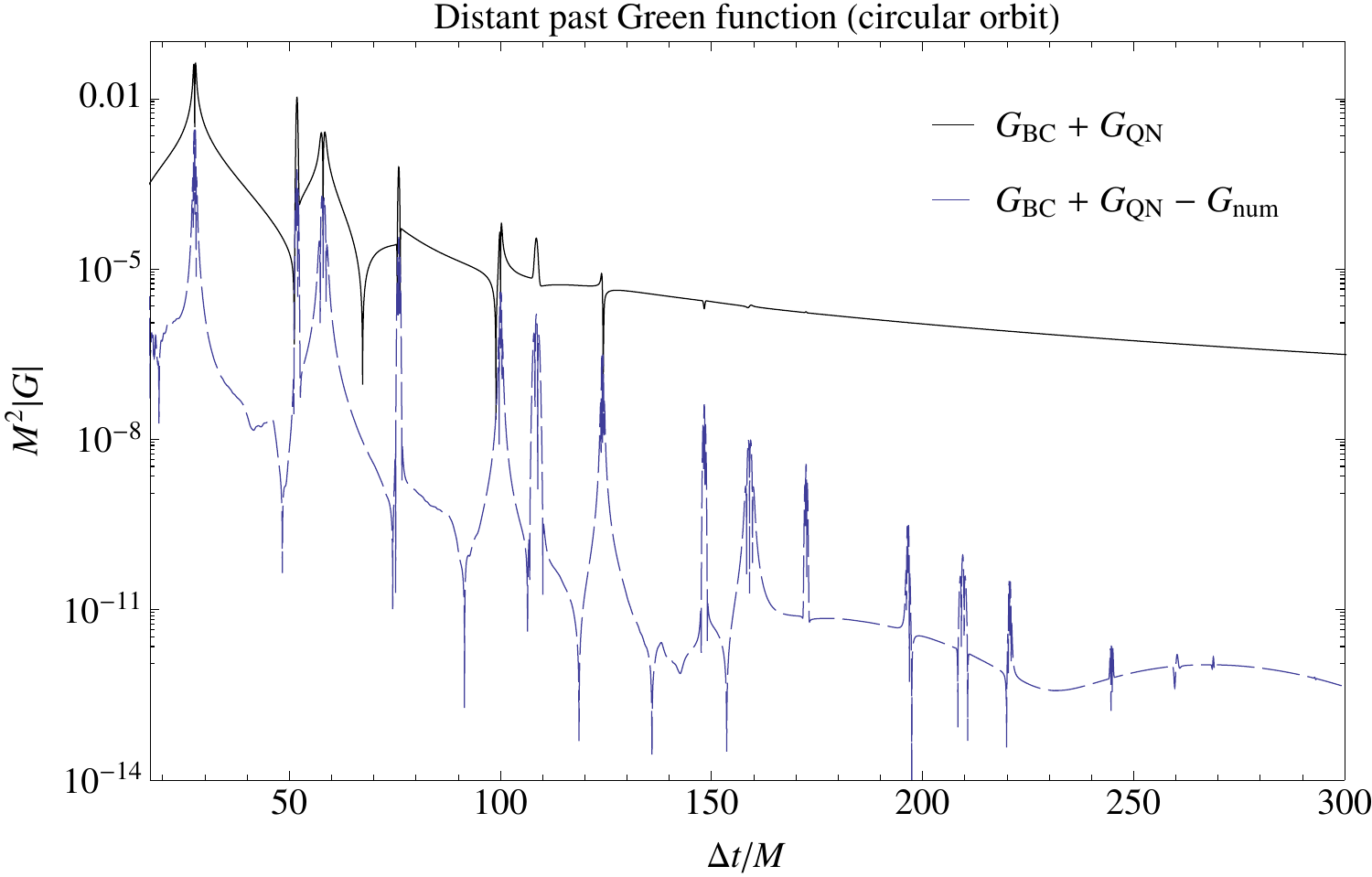}
\includegraphics[width=8cm]{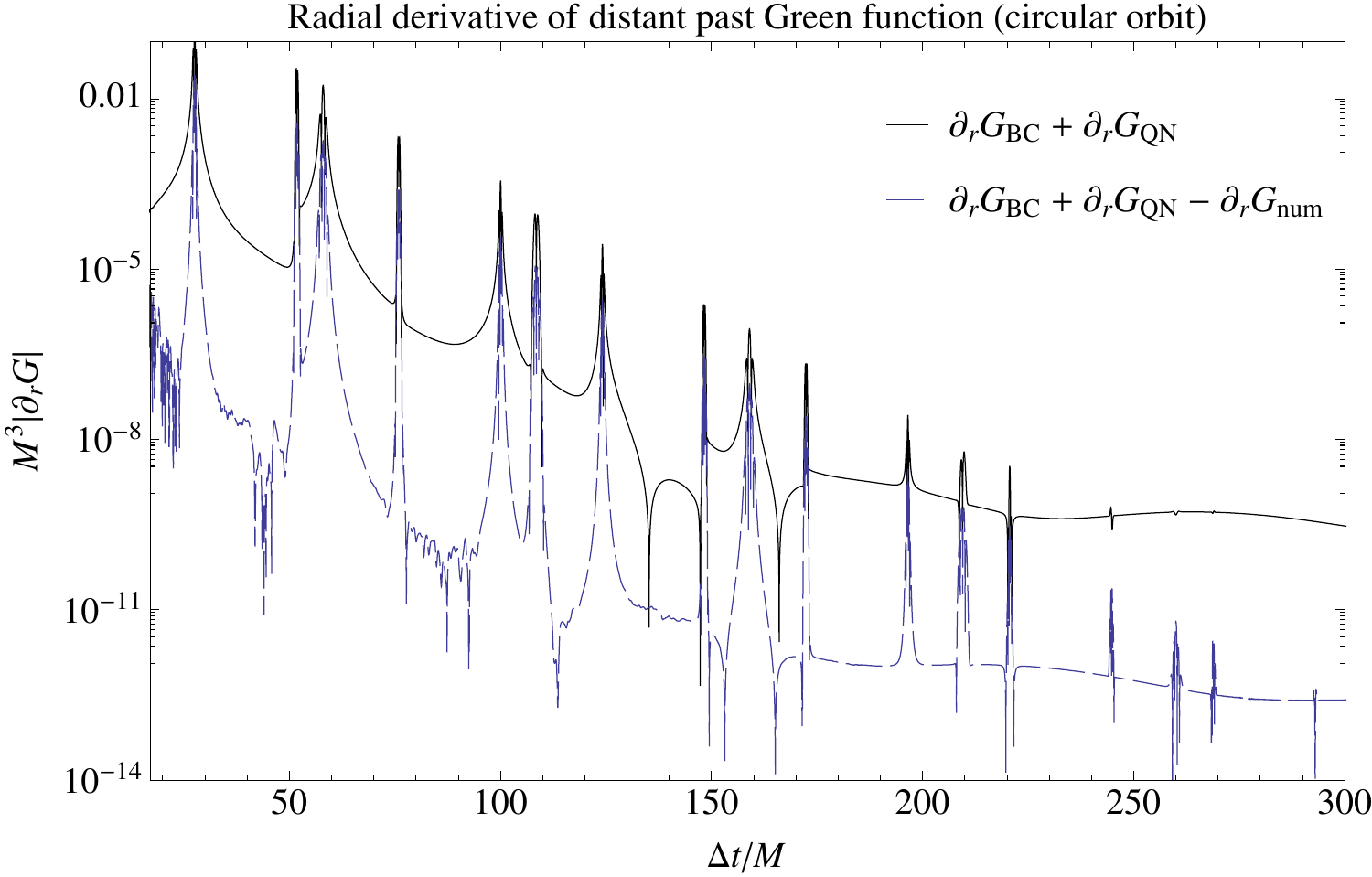}
\end{center}
\caption{
Validation of the GF (left) and its radial derivative (right) in the DP along a circular
geodesic of radius $\rc=6M$.
As a reference, we use a numerical time-domain
approximation to the GF~\cite{CDGOWZ}.
}
\label{fig:validation} 
\end{figure} 

In Fig.~\ref{fig:validation} we plot the full QNM + BC Green function together with the error relative to the numerical Gaussian GF. We see that
at almost all times in the DP ($t\gtrsim17M$, well within
the normal neighbourhood, $\mathcal{N}(z(\tau))$) the two completely independent calculations agree
very well. Indeed, at late times, the error is dominated by the numerical error in the numerical
Gaussian approach rather than in the QNM + BC calculation. This confirms that the HF contribution
is negligible in the DP, at least in the cases studied here. The notable
exceptions are the regions very close to the light-crossing times, where the smooth large-$\ell$
cuttoff means that we do not resolve the fine details in the singularities of GF.
Fortunately, the final value of the regularized SF is not sensitive to these singularities
and so our failure to resolve them does not significantly affect the computed SF.

\section{Computing the self-force}

\subsection{Matching}\label{sec:matching}

\begin{figure}[htb!]
\begin{center}
\includegraphics[width=8.5cm]{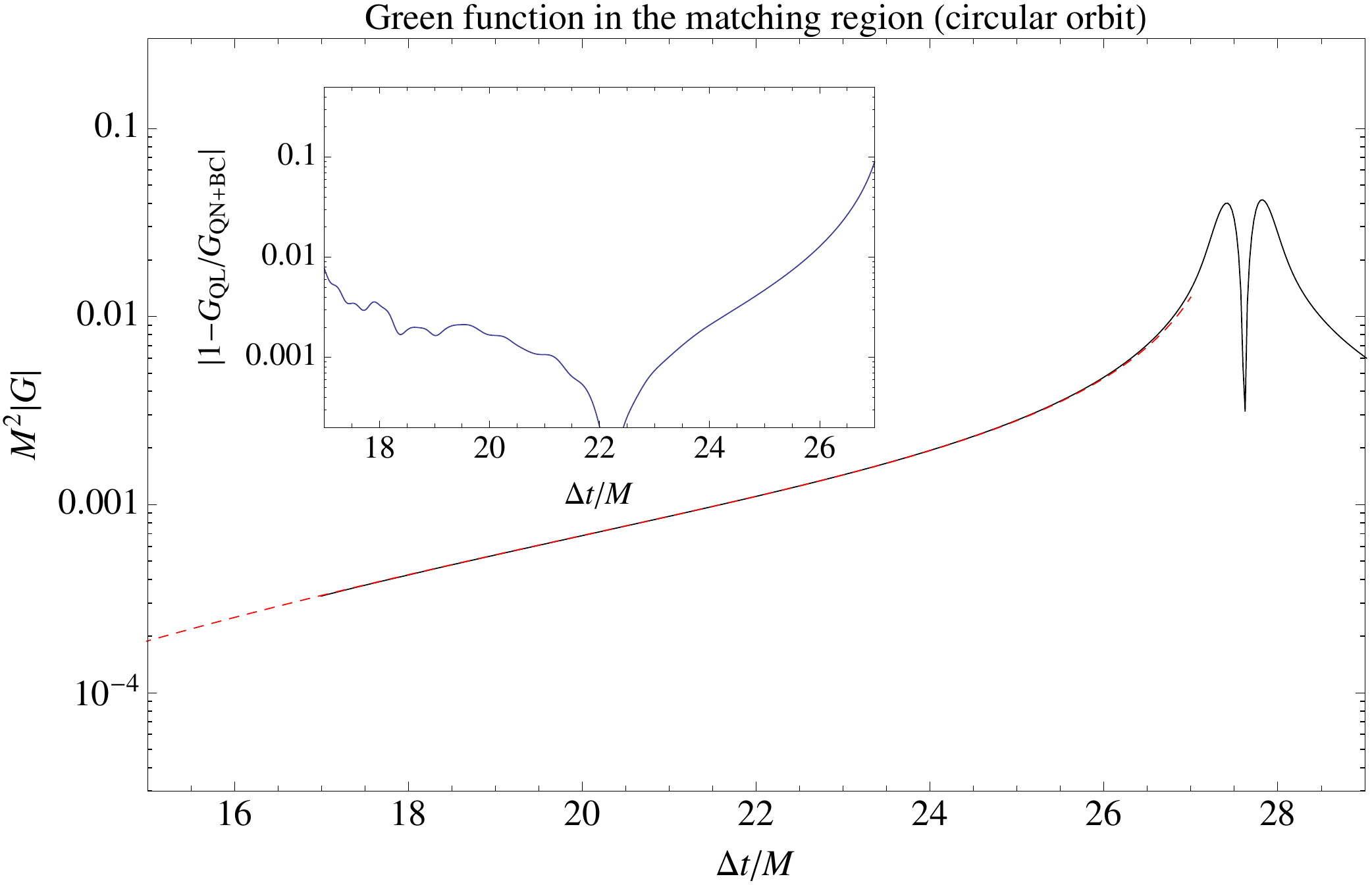}
\includegraphics[width=8.5cm]{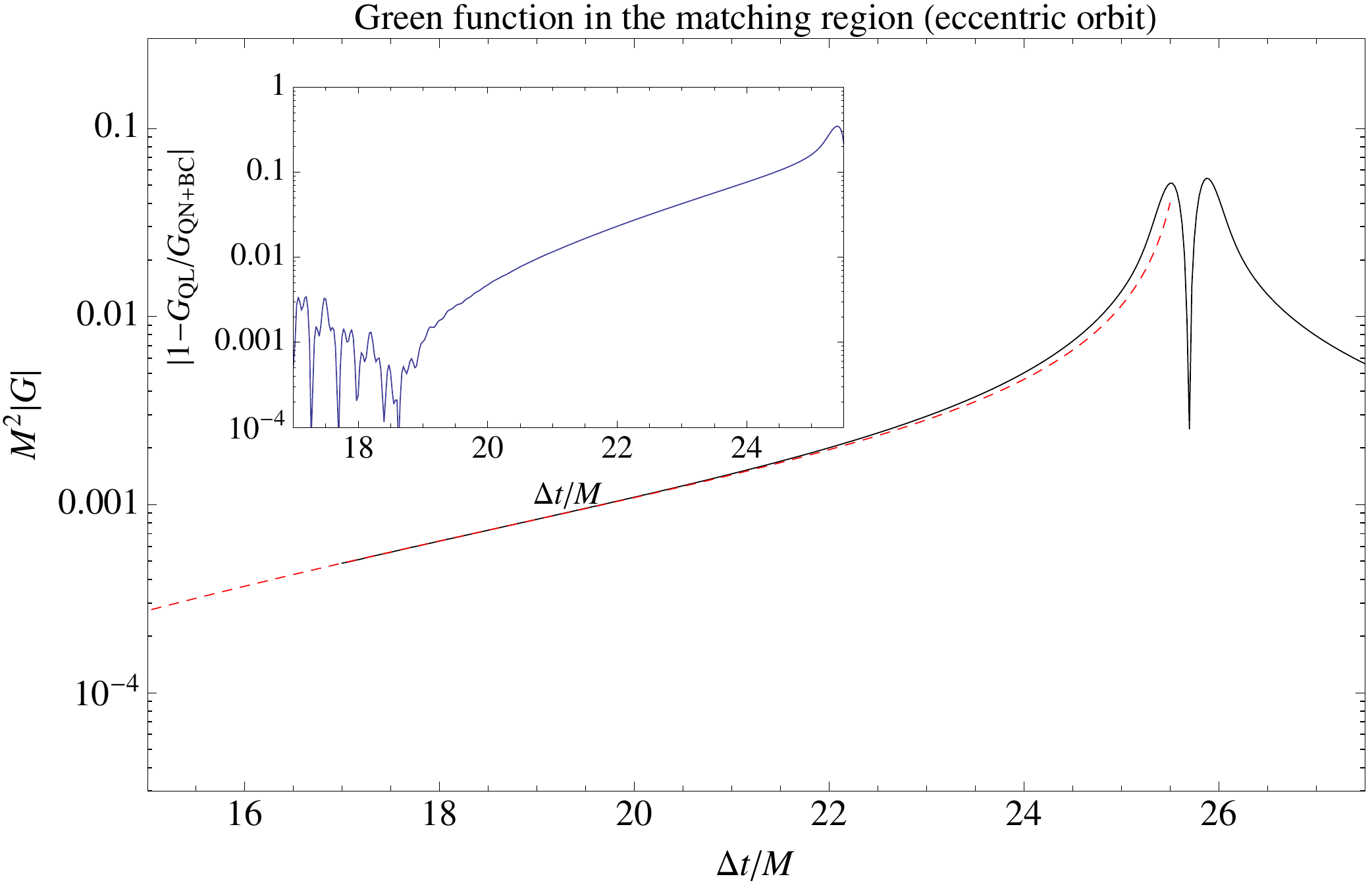}
\includegraphics[width=8.5cm]{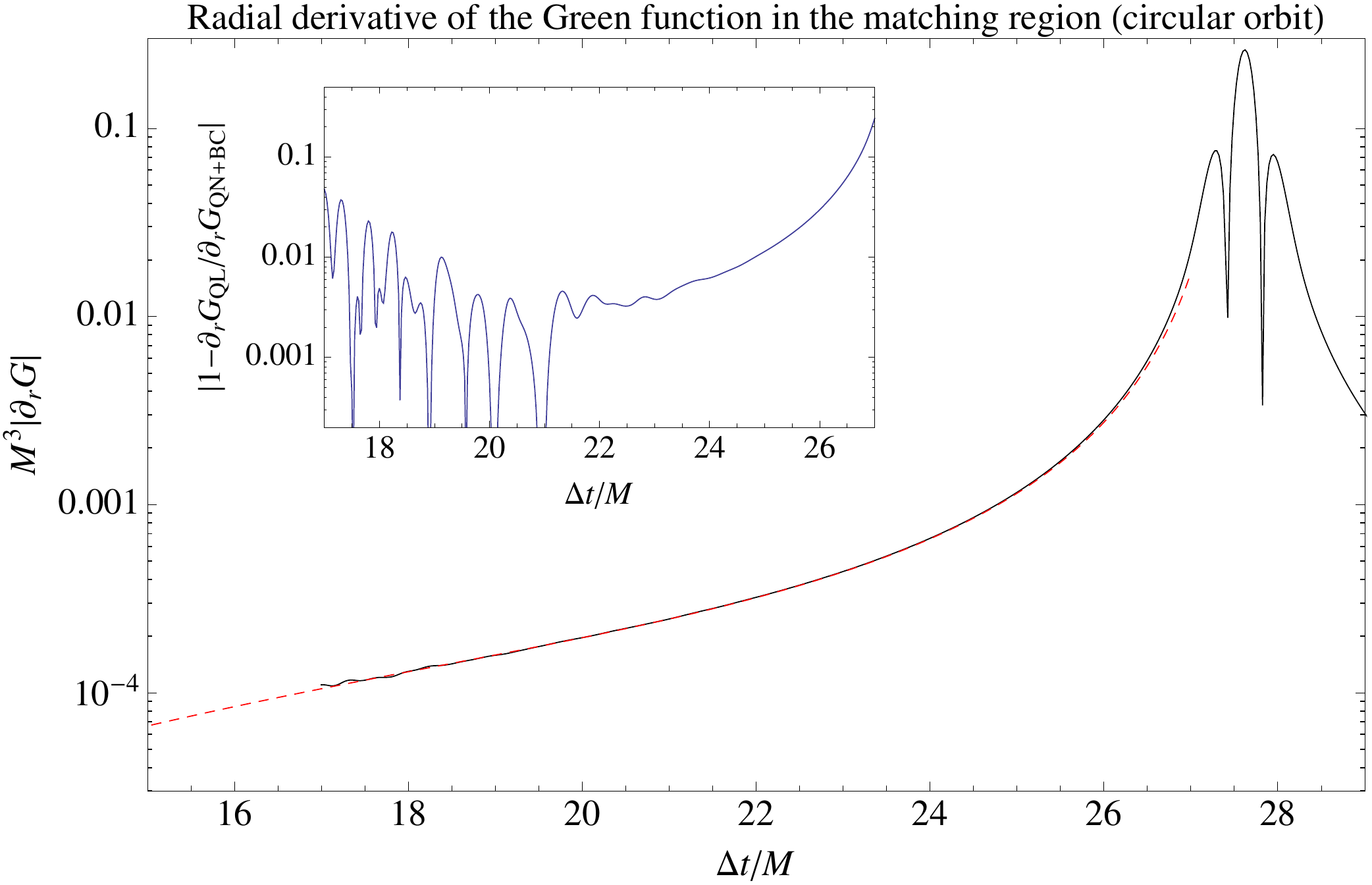}
\includegraphics[width=8.5cm]{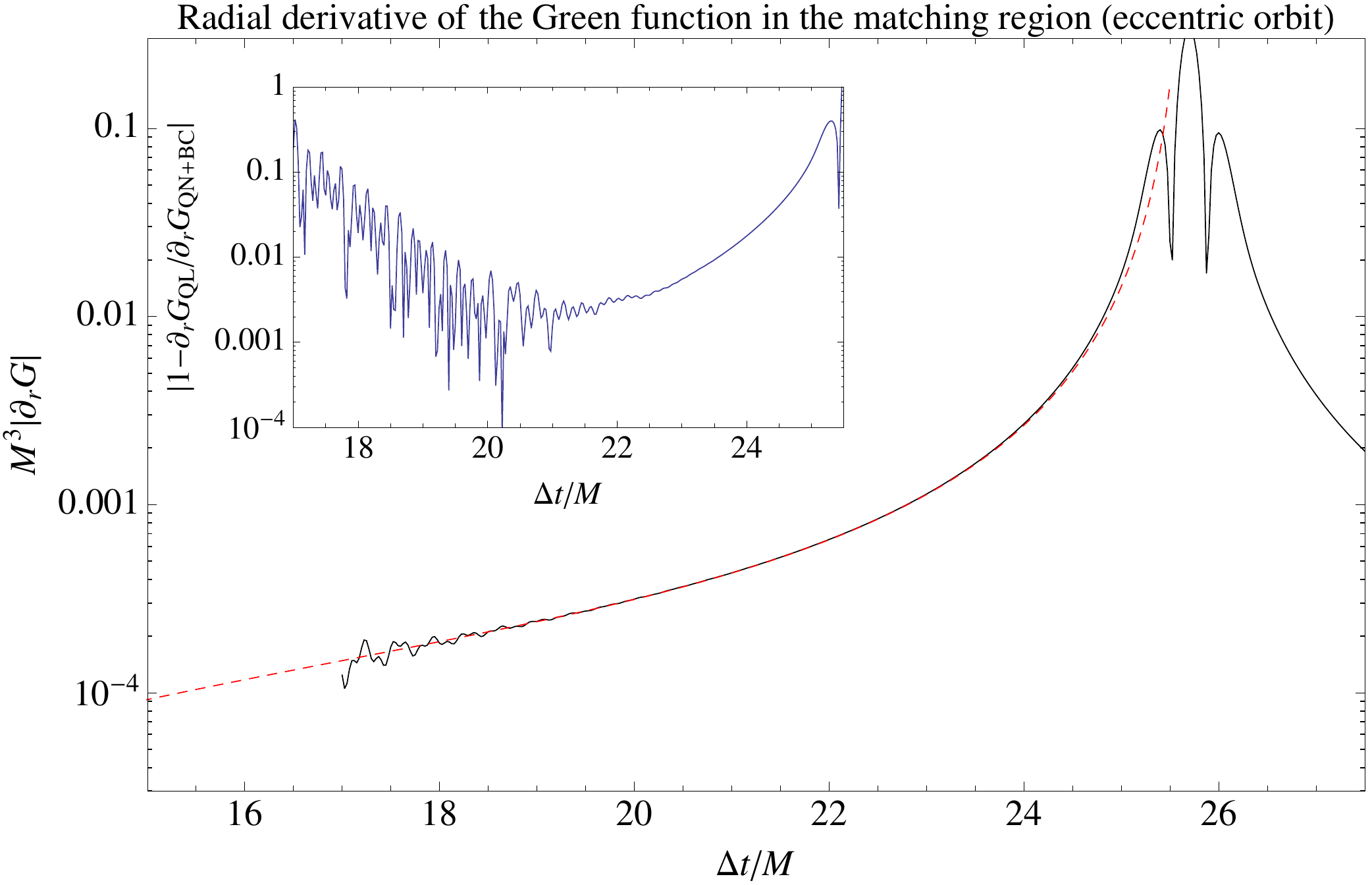}
\end{center}
\caption{
Matching between QL expansion (red dashed curve) and QNM+BC expansion in the DP (black curve) for GF (top) and its 
radial derivative (bottom) as a function of time along a circular
geodesic of radius $r_0=6M$ (left) and an eccentric orbit with $p=7.2$, $e=0.5$ instantaneously
moving outward ar $r_0=6M$. The inset shows the relative difference $|1-G^{\rm QL}/G^{\rm DP}|$
between the two expansions in the matching region.
}
\label{fig:matching} 
\end{figure}

In Fig.~\ref{fig:matching} we plot both QL and DP contributions to GF and its
radial derivative for circular and eccentric orbits. The QL expansion does impressively well to within a
short distance of the first singularity time, $\mathcal{N}(z(\tau))$. Importantly, the insets in
each plot also show that there is a reasonably large overlap in the regions of validity of the QL
and DP expansions. This was not guaranteed \emph{a priori} and is a crucial component of our matched
expansions calculation.

Comparing $\partial_r\Gret$ in the cases of circular and eccentric geodesics we see that
the singularity times --- which correspond to the light-crossing times --- occur at slightly
different times: the first two singularities occur slightly earlier in the eccentric case, the
third one slightly later, etc., as can be understood from Fig.~\ref{fig:intro}. We note that
this shift in singularity times is already significant for the first singularity and that the
difference in $\partial_r\Gret$ between circular and eccentric geodesics is already important
within the normal neighbourhood of $z(\tau)$.

\subsection{Self-Force}\label{sec:SF}

\begin{figure}[htb!]
\begin{center}
\includegraphics[width=8.5cm]{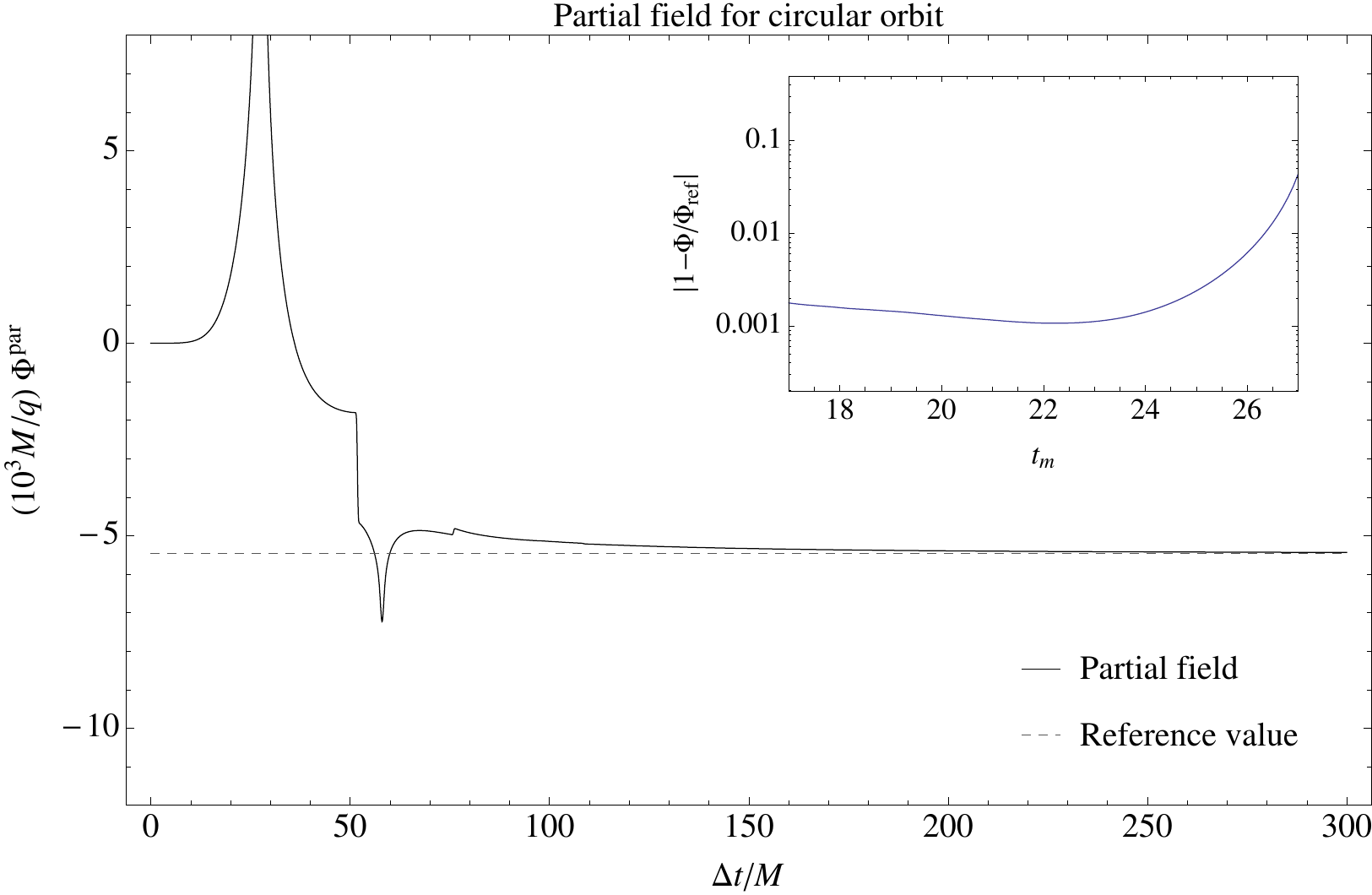}
\includegraphics[width=8.5cm]{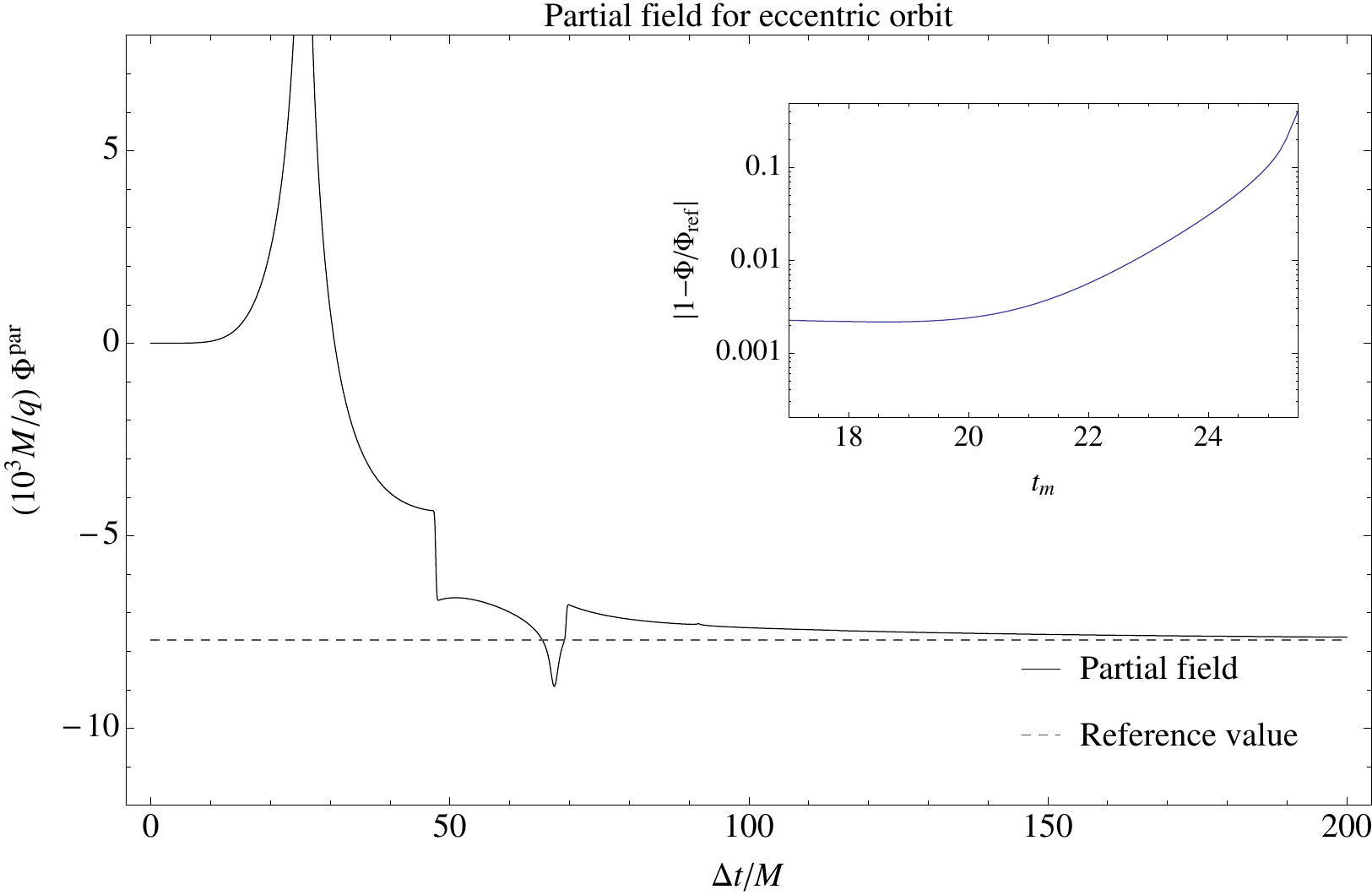}
\includegraphics[width=8.5cm]{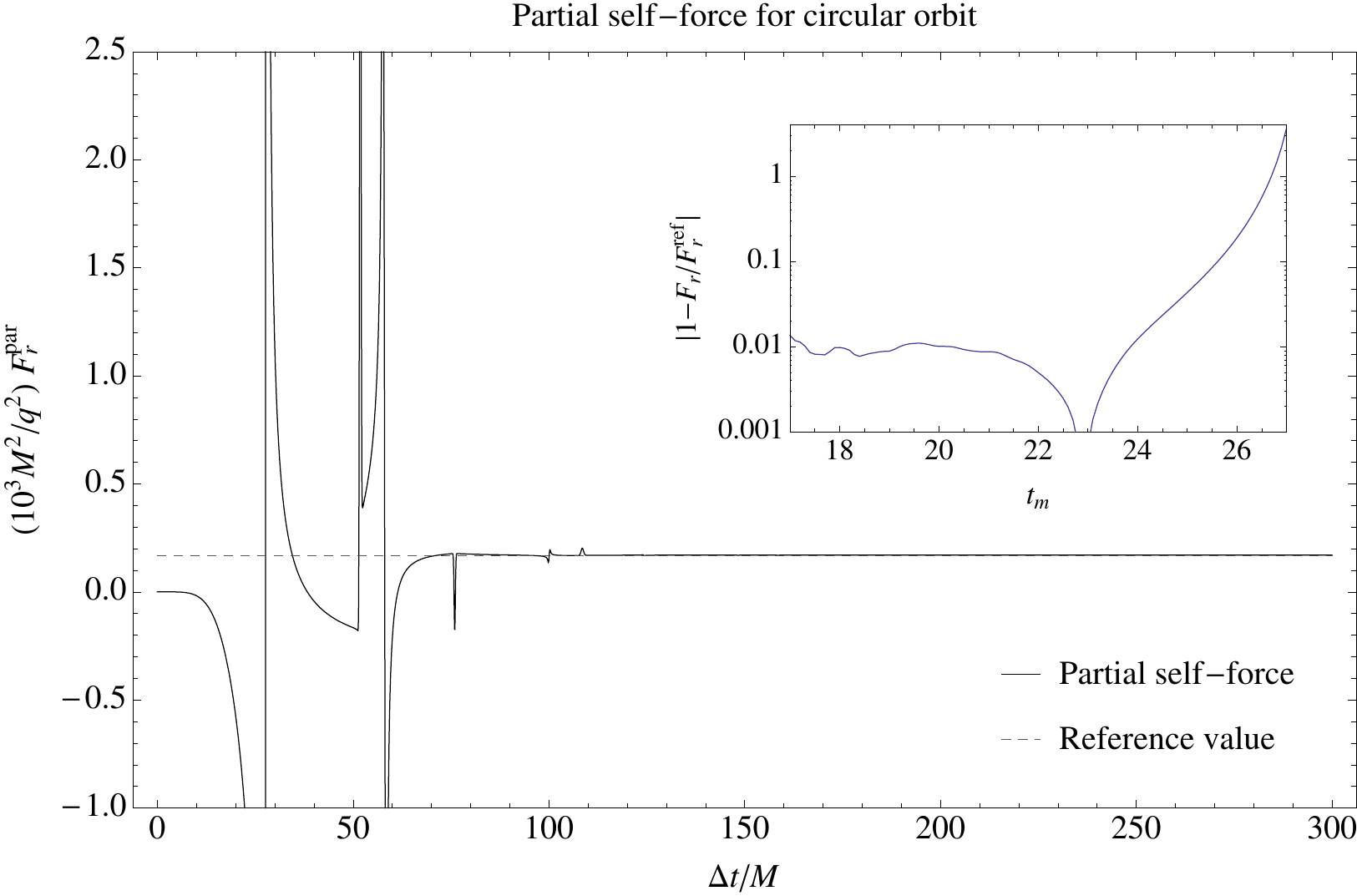}
\includegraphics[width=8.5cm]{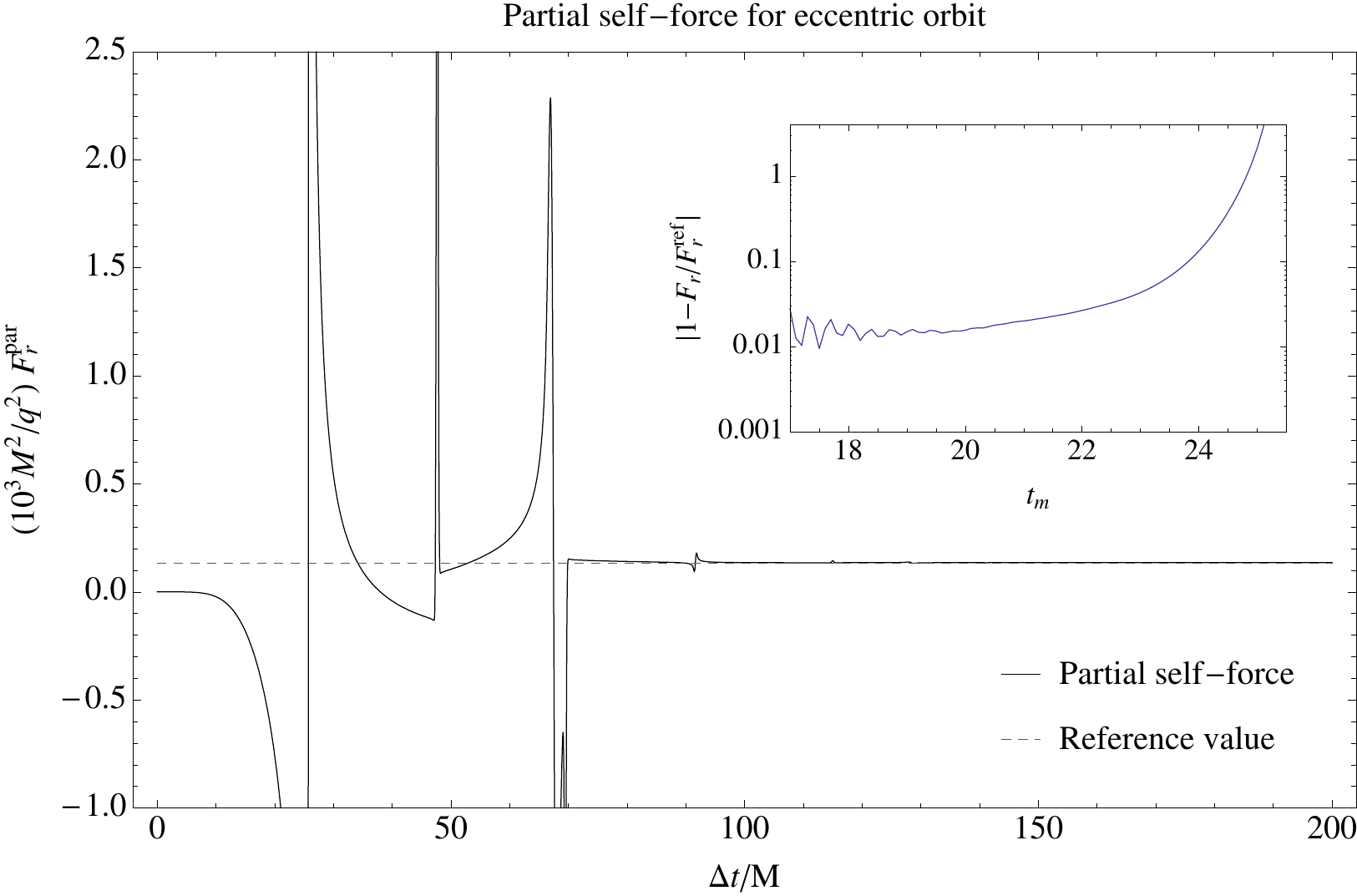}
\end{center}
\caption{
Partial field (top) and partial SF (bottom) as a function of integration time, $\Delta t$,
along a circular geodesic of radius $r_0=6M$ (left) and an eccentric orbit with $p=7.2$,
$e=0.5$, instantaneously moving outward at $r_0=6M$. The exact value computed using a
highly accurate frequency domain method \cite{Warburton:2011hp} is given by the black dashed
line. The insets show the sensitivity of the integration to the matching time. For these,
we consider the full field and self-force by integrating all the way to $t' = - \infty$, using
large-$t$ asymptotics where they are valid.
}
\label{fig:partial field} 
\end{figure}

Let us define the  `partial field $\Phi^{par}$ and the `partial SF'  $F_{\mu}^{par}$ as
\begin{align}\label{eq:partial S-F}
&
\Phi^{par}(\tau,\Delta\tau)\equiv q \int_{\tau-\Delta\tau}^{\tau^-}d\tau'\ \Gret(z(\tau),z(\tau')),
\nonumber \\
& F_{\mu}^{par}(\tau,\Delta\tau)\equiv q \nabla_{\mu}\Phi^{par}(\tau,\Delta\tau).
\end{align} 
Heuristically, the partial SF is the contribution to the SF from the worldline down to a proper time interval $\Delta\tau$ in the past.
We will denote by $\Delta t$ the coordinate time interval corresponding to the proper time interval $\Delta\tau$.
We plot the partial field and the partial SF in Fig.~\ref{fig:partial field} for both
circular and eccentric geodesics.
The plots show that it is not until just after the 3rd light-crossing time (i.e., $\Delta t\gtrapprox 65M$ in the circular case 
and $\Delta t\gtrapprox 70M$ in the eccentric case) 
that the value of the partial SF settles to a value `close' to the final value.

Finally, we calculate 
 the radial component of the scalar SF using the method of matched expansions 
  and compare our value with the value obtained using the mode-sum regularization method.
Our test cases are for a scalar charge at
$\rc=6M$ moving in (1) a circular geodesic and (2) an eccentric geodesic with $p=7.2$ and $e=0.5$.
Respectively, the reference values of the field and self-force obtained from the mode-sum regularization
method are $-5.454828 \times 10^{-3} q/M$ and
$1.67728\times 10^{-4}q^2M^{-2}$ in the circular case~\cite{DiazRivera:2004ik} and
$-7.70794\times 10^{-4}q^2M^{-2}$ and $1.31717\times 10^{-4}q^2M^{-2}$ in
the eccentric case~\cite{Warburton:2011hp}.
In the inset of Fig.~\ref{fig:partial field}  we plot the relative error in the field and in the SF as a function of the matching coordinate time $t_m$.
We see that the relative errors reach a plateau of at worst $1\%$ in the interval $t_m \approx (17M,23M)$.

\section{Discussion}
\label{sec:Discussion}

In the preceding sections, we have described the first complete implementation of the method of matched expansions for a self-force calculation on a black hole spacetime. We have shown that the direct evaluation of a tail integral leads to accurate values for SF which are in agreement with the results of mode-sum regularization and other methods.
Let us now take this opportunity to explore the possible advantages, limitations and extensions of this nascent method. 

\begin{figure}[htb!]
\begin{center}
\includegraphics[width=8.5cm]{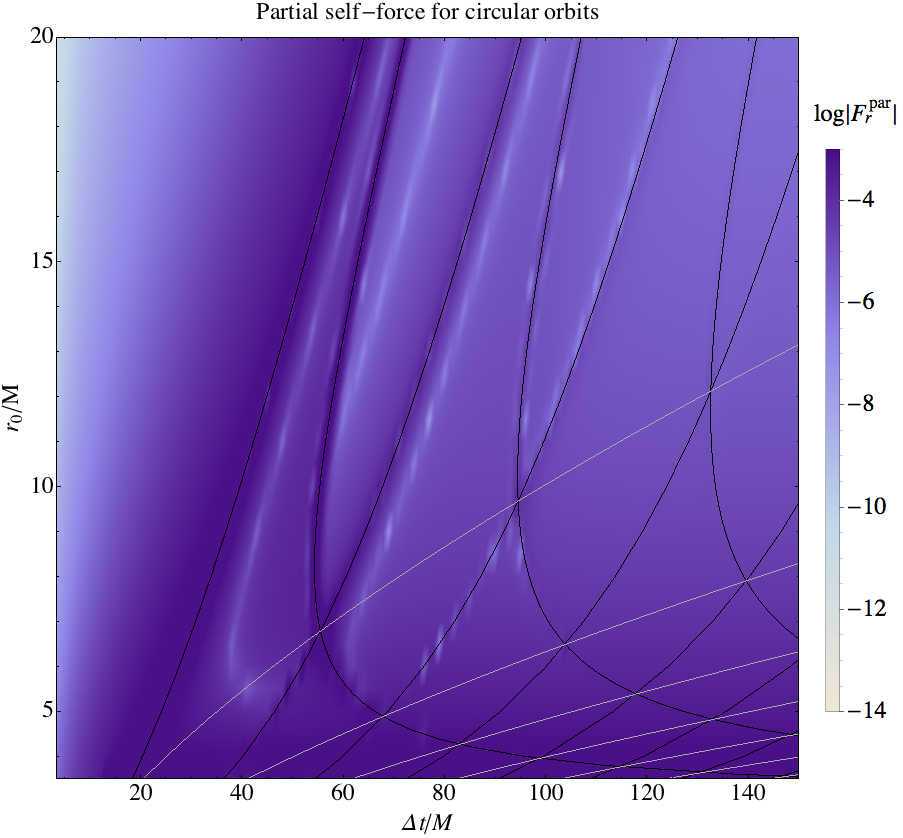}
\end{center}
\caption{
Partial self-force as a function of integration time, $\Delta t$,
along a range of circular geodesic orbits of radius $r_0$. The times when null geodesics intersect the
circular time-like geodesics are indicated by black lines. The light gray lines are contours of
constant orbital phase, $\gamma = \pi, 2\pi, 3\pi, \dots$
}
\label{fig:circular geodesics} 
\end{figure}

A key motivation for developing this method has been to shed more light on the physical nature of the self-force. For example, one may ask: what, if any, is the role of the null geodesics that begin and end on the worldline? 
Figure \ref{fig:partial field}, showing the partial field and force from integrating along the worldline, suggests that it is not straightforward to separate out the contribution of null rays from the remainder of the Green function, as the singular part associated with `odd' geodesics is extended in form. Nevertheless, the plots suggest that null rays play a significant role in the partial field. Perhaps more importantly, this approach provides information on the domain of dependence of the self-force. In other words, Fig.~\ref{fig:partial field} gives a clear indication of how far back along the worldline one must integrate in order to obtain an accurate value for the self-force.
Extending this picture to a range of circular geodesic orbits, we obtain Fig.~\ref{fig:circular geodesics},
which shows how the times at which features appear in the partial self-force is intimately linked
to the null geodesics intersecting the worldline. It also illustrates how, as the radius of
the orbit increases a larger portion of the past worldline (in terms of coordinate time) contributes
significantly to the self-force, while a smaller percentage of the orbit (in terms of orbital phase)
is important.

This suggests that an important application of this method will be in understanding the effect on the self-force of deviations in the past worldline. For instance, there is a concerted effort underway to evolve realistic Extreme Mass-Ratio Inspiral systems under the influence of the self-force. It is computationally efficient to use an ``instantaneously geodesic'' approximation for the past worldline~\cite{Warburton:2011fk}, whereas the ``true'' motion will actually be along an accelerated worldline (with respect to the background), which can only be found via a fully self-consistent approach~\cite{Diener:2011cc} which is computationally costly. It is important to quantify the error of using the instantaneous approximation, and it is here that information on the domain of dependence on the past worldline will be vital.

Existing methods are not ideally suited to computing the self-force along highly-eccentric, unbound, or plunging world-lines, particularly where the particle approaches the speed of light. In these cases, time-domain methods typically encounter difficulties due to initial 'junk' data propagating close to the particle's orbit, whereas
real-frequency-domain methods typically require a very large number of frequencies. 
Green function approaches offer a new way to try to circumvent these difficulties. Indeed, the use of the
method of matched expansions avoids any issues related to initial data.
As for the need for large frequencies, it remains to be investigated whether
a spectral decomposition provides any gains in efficiency relative to
straightforward integration along (just above) the real-frequency axis.

In this regard, there has been recent interest in (i) accelerated radial infall orbits, which may be used to test the cosmic censorship hypothesis~\cite{Zimmerman:2012zu} and to inform the potential detectability of gravitational wave signals from plunging Extreme Mass-Ratio Inspiral orbits~\cite{AmaroSeoane:2012cr}; (ii) orbits asymptotically close to the light-ring where the self-force exhibits `critical' behaviour \cite{Gundlach:Akcay:2012}; (iii) the marginally-bound zoom-whirl orbit, where accurate values of gravitational self-force may be used to help break a degeneracy within the parameter space of Effective One-Body theory \cite{Damour:2010}.

Now let us discuss some calculational aspects of the method of matched expansions. A crucial condition for the success (or otherwise) of the method is the existence of an `overlap' region between the QL and DP regimes, in which both expansions for the Green function are valid. We have shown that, though this overlap region exists, it is rather limited in extent. This is a list of  the challenges in trying to extend the overlap region:

\begin{itemize}
\item It is not realistic to extend the convergence of the spectral mode-sum decomposition
  to smaller values of $\Delta t$, as the QNM and BC -- and potentially
  also HF -- contributions diverge separately at early times, meaning that delicate cancellations between them are required.
\item Extending the overlap region closer to the first singularity would
  require the inclusion of many more $\ell$ modes in the QNM sum, which is computationally expensive.
 However, instead of MST data for the QNM sum one could make use of the already-known large-$\ell$ asymptotics for the QNMs or
one could try to obtain another analytic calculation of the behaviour of the GF close to the singularities at light-crossing times.
\item In the QL region, the calculation is based on the Hadamard form for the Green function,
  which is only valid within the normal neighbourhood which ends at the first
  singularity of the Green function. This puts a firm upper limit on the matching region. Furthermore, the QL   calculation uses an expansion
  about $\Delta t = 0$ which becomes increasingly poor as $\Delta t$ increases. We have been able
  to mitigate this problem by use of Pad\'{e} approximants in
  combination with knowledge of the singularity structure of the GF at the first light-crossing.
\end{itemize}

With regards to the efficiency/calculational cost, the QL contribution is extremely fast to
calculate for a particular orbit since its expression is just a power series which only needs to be
computed once and can then be used for all orbits. As for the spectral mode-sum decomposition in
the DP contribution, once the coordinate-independent  quantities 
(namely, $W(\omega)$
and $q(\nu)$ for the BC and $\mathcal{B}_{\ell n}$ and $\AoutQNM$ for the QNM) have been
obtained for the 
necessary modes
then essentially the only quantity left to calculate is the
radial function $\f(r,\omega)$.
That is, the calculation of the SF for {\it any} orbit has been reduced to the calculation of the QL expansion (which is immediate given
that we have already obtained the coefficients $v_{ijk}$)
and to solving a second-order ODE (namely the Regge-Wheeler equation).
Using the Jaff\'e series (or the MST method via Barnes integrals)
for calculating $\f(r,\omega)$ is too computationally-costly for
the method of matched expansions to be competitive with other methods for the calculation of
the SF. However, other, more efficient methods for calculating $\f$ could be used, such as a direct
numerical integration of the homogeneous version of the radial ODE (\ref{eq:radial ODE}). For
values of the frequency with $\text{Im}(\omega)<0$ as are required here it is not feasible to
impose the boundary conditions (\ref{eq:f,near hor}). However, one could calculate $\f(r,\omega)$
at a chosen negative value of $r_*$ which is not too large by using, e.g., the Jaff\'e series, and
then numerically integrating the radial ODE up to a value of $r_*$ as large as desired.
Furthermore, we note that if $\f$ is calculated for a broad general range of values of $r$ (and a
fixed range of values of the frequency both along the BC and on QNM values) then these values could
be easily interpolated in order to carry out a `fast' calculation of the SF for {\it any} desired
trajectory.

An alternative to the calculation of the GF in the DP via QNM and BC contributions is to calculate
the GF in the DP by integrating along the real frequencies, i.e., by using directly
Eq.~(\ref{eq:Green ell}). Indeed, this is the way that we initially attempted the calculation, but
we found that the GF presented many unphysical oscillations when plotted as a function of
time~\cite{Casals:Wardell:2008}. It would be interesting to repeat this calculation using an
integration along the real frequencies, but to additionally include a smooth-sum factor as used
here. 

Another alternative is to replace the DP expansion with a numerical time-domain scheme to obtain an approximation to the Green function. A description of this approach was given in \cite{CDGOWZ}.

The scenario of a scalar charge moving in Schwarzschild background spacetime provides an important testbed for any method for calculating the SF. However, the truly astrophysically-relevant case
is that of a small mass moving in the background of a Kerr black hole. Further technical challenges lie
ahead before we may apply the method of matched expansions in this case, although many of the techniques outlined here can be transferred. In particular, the techniques for
calculating the BC contribution are already in
place~\cite{Casals:2012tb,Casals:Ottewill:2011smallBC,Casals:2012ng}
for general spin -- including the gravitational case of spin-$2$ -- and we are currently generalizing
them to the Kerr case.
As for the QNM contribution in Kerr, significant progress  has been made in Ref.~\cite{Zhang:2013ksa} during the preparation of this paper.
The very high order QL expansions used here are
only valid for the Schwarzschild spacetime, spin-$0$ case, but lower order expansions are readily
available in the spin-$1$ and spin-$2$ cases for Kerr spacetime~\cite{Heffernan:2012vj}.
An outstanding issue, however, is the fact that while the
regularization procedure is well-established in the Lorenz gauge, there is ongoing discussion about 
how to carry out regularization in the radiation gauge (which is the one in which the metric
perturbations are obtained if using the techniques of~\cite{Leaver:1986a} for finding the
gravitational Green function in Kerr spacetime, i.e., for the Teukolsky equation). Recent progress \cite{Gralla:2011zr,Shah:2012gu,Merlin} suggests that this problem will be well understood in the near future.

\section*{Acknowledgements}
We thank Chad Galley and An{\i}l Zengino\u{g}lu for many fruitful discussions.
We are grateful to Niels Warburton for providing a highly accurate reference value for the SF
in the eccentric geodesic case.
We also thank the participants of the Capra meetings on Radiation Reaction for many illuminating conversations.
M.C. gratefully acknowledges support by a
IRCSET-Marie Curie International Mobility Fellowship in Science, Engineering and Technology.
 B.W. and A.C.O. gratefully
acknowledge support from Science Foundation Ireland under Grant No.
10/RFP/PHY2847. The authors wish to acknowledge the SFI/HEA Irish Centre for High-End Computing
(ICHEC) for the provision of computational facilities and support (project ndast005b).

\appendix
\section{QNM data} \label{sec:App}

In this appendix we present tables with the QNM data used in Sec.~\ref{sec:QNM} for a scalar field in Scharzschild spacetime.
Tables \ref{tb:freqs}, \ref{tb:exc fac} and \ref{tb:Aout} respectively show the frequencies $\wQNM$, the excitation factors $\mathcal{B}_{\ell  n}$
and the coefficients $A^{out}_{\ell,n}$ for the modes $n: 0\to 5$ and $\ell:0\to 50$.

\begingroup
\squeezetable
\begin{table*}
\begin{tabular}{||c||c|c|c|c|c|c||}
\hline
$\ell$ & $\omega_{\ell 0}$ & $\omega_{\ell 1}$ & $\omega_{\ell 2}$  & $\omega_{\ell 3}$  & $\omega_{\ell 4}$  & $\omega_{\ell 5}$ \\
\hline
 $0$ & $(0.220910,-0.209791)$ & $(0.172234,-0.696105)$ & $(0.151484,-1.20216)$ & $(0.140820,-1.70735)$ & $(0.134149,-2.21126)$ & $(0.129483,-2.71428)$ \\
 $1$ & $(0.585872,-0.195320)$ & $(0.528897,-0.612515)$ & $(0.459079,-1.08027)$ & $(0.406517,-1.57660)$ & $(0.370218,-2.08152)$ & $(0.344154,-2.58824)$ \\
 $2$ & $(0.967288,-0.193518)$ & $(0.927701,-0.591208)$ & $(0.861088,-1.01712)$ & $(0.787726,-1.47619)$ & $(0.722598,-1.95984)$ & $(0.669799,-2.45682)$ \\
 $3$ & $(1.35073,-0.192999)$ & $(1.32134,-0.584570)$ & $(1.26725,-0.992016)$ & $(1.19755,-1.42244)$ & $(1.12325,-1.87719)$ & $(1.05310,-2.35207)$ \\
 $4$ & $(1.73483,-0.192783)$ & $(1.71162,-0.581752)$ & $(1.66738,-0.980650)$ & $(1.60658,-1.39496)$ & $(1.53547,-1.82804)$ & $(1.46087,-2.28024)$ \\
 $5$ & $(2.11922,-0.192674)$ & $(2.10008,-0.580308)$ & $(2.06300,-0.974689)$ & $(2.01041,-1.37985)$ & $(1.94594,-1.79896)$ & $(1.87405,-2.23376)$ \\
 $6$ & $(2.50377,-0.192610)$ & $(2.48750,-0.579473)$ & $(2.45569,-0.971204)$ & $(2.40980,-1.37083)$ & $(2.35207,-1.78097)$ & $(2.28537,-2.20355)$ \\
 $7$ & $(2.88842,-0.192570)$ & $(2.87428,-0.578946)$ & $(2.84647,-0.968997)$ & $(2.80595,-1.36505)$ & $(2.75416,-1.76924)$ & $(2.69303,-2.18330)$ \\
 $8$ & $(3.27312,-0.192544)$ & $(3.26062,-0.578594)$ & $(3.23595,-0.967514)$ & $(3.19975,-1.36115)$ & $(3.15302,-1.76122)$ & $(3.09709,-2.16925)$ \\
 $9$ & $(3.65787,-0.192525)$ & $(3.64667,-0.578347)$ & $(3.62451,-0.966470)$ & $(3.59184,-1.35839)$ & $(3.54938,-1.75552)$ & $(3.49808,-2.15916)$ \\
 $10$ & $(4.04264,-0.192512)$ & $(4.03250,-0.578166)$ & $(4.01239,-0.965708)$ & $(3.98266,-1.35637)$ & $(3.94382,-1.75133)$ & $(3.89658,-2.15169)$ \\
 $11$ & $(4.42744,-0.192501)$ & $(4.41817,-0.578031)$ & $(4.39977,-0.965135)$ & $(4.37250,-1.35484)$ & $(4.33675,-1.74816)$ & $(4.29305,-2.14603)$ \\
 $12$ & $(4.81225,-0.192493)$ & $(4.80372,-0.577926)$ & $(4.78677,-0.964693)$ & $(4.76159,-1.35367)$ & $(4.72849,-1.74571)$ & $(4.68789,-2.14164)$ \\
 $13$ & $(5.19708,-0.192487)$ & $(5.18918,-0.577844)$ & $(5.17346,-0.964346)$ & $(5.15008,-1.35274)$ & $(5.11928,-1.74378)$ & $(5.08140,-2.13816)$ \\
 $14$ & $(5.58191,-0.192482)$ & $(5.57456,-0.577779)$ & $(5.55991,-0.964067)$ & $(5.53809,-1.35200)$ & $(5.50931,-1.74223)$ & $(5.47383,-2.13537)$ \\
 $15$ & $(5.96676,-0.192478)$ & $(5.95987,-0.577725)$ & $(5.94616,-0.963841)$ & $(5.92571,-1.35140)$ & $(5.89871,-1.74097)$ & $(5.86535,-2.13309)$ \\
 $16$ & $(6.35161,-0.192475)$ & $(6.34514,-0.577681)$ & $(6.33225,-0.963654)$ & $(6.31302,-1.35090)$ & $(6.28758,-1.73992)$ & $(6.25612,-2.13121)$ \\
 $17$ & $(6.73647,-0.192472)$ & $(6.73037,-0.577644)$ & $(6.71820,-0.963499)$ & $(6.70005,-1.35049)$ & $(6.67601,-1.73905)$ & $(6.64625,-2.12964)$ \\
 $18$ & $(7.12133,-0.192470)$ & $(7.11556,-0.577613)$ & $(7.10404,-0.963368)$ & $(7.08685,-1.35014)$ & $(7.06408,-1.73832)$ & $(7.03585,-2.12831)$ \\
 $19$ & $(7.50619,-0.192468)$ & $(7.50072,-0.577587)$ & $(7.48979,-0.963256)$ & $(7.47347,-1.34984)$ & $(7.45183,-1.73770)$ & $(7.42498,-2.12718)$ \\
 $20$ & $(7.89106,-0.192466)$ & $(7.88585,-0.577565)$ & $(7.87546,-0.963160)$ & $(7.85992,-1.34958)$ & $(7.83931,-1.73716)$ & $(7.81372,-2.12622)$ \\
 $21$ & $(8.27593,-0.192465)$ & $(8.27097,-0.577545)$ & $(8.26105,-0.963078)$ & $(8.24623,-1.34936)$ & $(8.22655,-1.73670)$ & $(8.20211,-2.12538)$ \\
 $22$ & $(8.66081,-0.192463)$ & $(8.65606,-0.577528)$ & $(8.64658,-0.963006)$ & $(8.63241,-1.34917)$ & $(8.61359,-1.73630)$ & $(8.59020,-2.12465)$ \\
 $23$ & $(9.04569,-0.192462)$ & $(9.04114,-0.577513)$ & $(9.03206,-0.962943)$ & $(9.01849,-1.34900)$ & $(9.00045,-1.73594)$ & $(8.97803,-2.12401)$ \\
 $24$ & $(9.43056,-0.192461)$ & $(9.42620,-0.577500)$ & $(9.41750,-0.962888)$ & $(9.40447,-1.34886)$ & $(9.38716,-1.73563)$ & $(9.36562,-2.12345)$ \\
 $25$ & $(9.81544,-0.192461)$ & $(9.81125,-0.577489)$ & $(9.80289,-0.962839)$ & $(9.79036,-1.34872)$ & $(9.77372,-1.73536)$ & $(9.75301,-2.12295)$ \\
 $26$ & $(10.2003,-0.192460)$ & $(10.1963,-0.577479)$ & $(10.1882,-0.962795)$ & $(10.1762,-1.34861)$ & $(10.1602,-1.73511)$ & $(10.1402,-2.12251)$ \\
 $27$ & $(10.5852,-0.192459)$ & $(10.5813,-0.577469)$ & $(10.5736,-0.962756)$ & $(10.5619,-1.34850)$ & $(10.5465,-1.73489)$ & $(10.5273,-2.12211)$ \\
 $28$ & $(10.9701,-0.192458)$ & $(10.9663,-0.577461)$ & $(10.9589,-0.962722)$ & $(10.9476,-1.34841)$ & $(10.9327,-1.73470)$ & $(10.9141,-2.12176)$ \\
 $29$ & $(11.3550,-0.192458)$ & $(11.3514,-0.577454)$ & $(11.3441,-0.962690)$ & $(11.3333,-1.34833)$ & $(11.3189,-1.73452)$ & $(11.3009,-2.12144)$ \\
 $30$ & $(11.7399,-0.192457)$ & $(11.7364,-0.577447)$ & $(11.7294,-0.962662)$ & $(11.7189,-1.34825)$ & $(11.7049,-1.73436)$ & $(11.6876,-2.12115)$ \\
 $31$ & $(12.1248,-0.192457)$ & $(12.1214,-0.577441)$ & $(12.1146,-0.962636)$ & $(12.1044,-1.34818)$ & $(12.0909,-1.73422)$ & $(12.0741,-2.12089)$ \\
 $32$ & $(12.5096,-0.192457)$ & $(12.5064,-0.577436)$ & $(12.4998,-0.962613)$ & $(12.4899,-1.34812)$ & $(12.4768,-1.73409)$ & $(12.4605,-2.12065)$ \\
 $33$ & $(12.8945,-0.192456)$ & $(12.8913,-0.577431)$ & $(12.8850,-0.962591)$ & $(12.8754,-1.34806)$ & $(12.8627,-1.73397)$ & $(12.8469,-2.12043)$ \\
 $34$ & $(13.2794,-0.192456)$ & $(13.2763,-0.577426)$ & $(13.2701,-0.962572)$ & $(13.2609,-1.34801)$ & $(13.2485,-1.73386)$ & $(13.2331,-2.12023)$ \\
 $35$ & $(13.6643,-0.192455)$ & $(13.6613,-0.577422)$ & $(13.6553,-0.962554)$ & $(13.6463,-1.34796)$ & $(13.6343,-1.73376)$ & $(13.6193,-2.12005)$ \\
 $36$ & $(14.0492,-0.192455)$ & $(14.0463,-0.577418)$ & $(14.0404,-0.962538)$ & $(14.0316,-1.34792)$ & $(14.0200,-1.73367)$ & $(14.0054,-2.11988)$ \\
 $37$ & $(14.4341,-0.192455)$ & $(14.4312,-0.577414)$ & $(14.4255,-0.962523)$ & $(14.4170,-1.34788)$ & $(14.4056,-1.73358)$ & $(14.3915,-2.11973)$ \\
 $38$ & $(14.8190,-0.192455)$ & $(14.8162,-0.577411)$ & $(14.8107,-0.962509)$ & $(14.8023,-1.34784)$ & $(14.7913,-1.73350)$ & $(14.7775,-2.11959)$ \\
 $39$ & $(15.2039,-0.192454)$ & $(15.2012,-0.577408)$ & $(15.1958,-0.962496)$ & $(15.1877,-1.34781)$ & $(15.1769,-1.73343)$ & $(15.1634,-2.11946)$ \\
 $40$ & $(15.5888,-0.192454)$ & $(15.5861,-0.577405)$ & $(15.5809,-0.962484)$ & $(15.5729,-1.34777)$ & $(15.5624,-1.73336)$ & $(15.5493,-2.11933)$ \\
 $41$ & $(15.9737,-0.192454)$ & $(15.9711,-0.577403)$ & $(15.9659,-0.962473)$ & $(15.9582,-1.34775)$ & $(15.9479,-1.73330)$ & $(15.9351,-2.11922)$ \\
 $42$ & $(16.3586,-0.192454)$ & $(16.3560,-0.577400)$ & $(16.3510,-0.962462)$ & $(16.3435,-1.34772)$ & $(16.3334,-1.73324)$ & $(16.3209,-2.11912)$ \\
 $43$ & $(16.7434,-0.192454)$ & $(16.7410,-0.577398)$ & $(16.7361,-0.962453)$ & $(16.7287,-1.34769)$ & $(16.7189,-1.73319)$ & $(16.7067,-2.11902)$ \\
 $44$ & $(17.1283,-0.192454)$ & $(17.1259,-0.577396)$ & $(17.1211,-0.962444)$ & $(17.1139,-1.34767)$ & $(17.1044,-1.73314)$ & $(17.0924,-2.11893)$ \\
 $45$ & $(17.5132,-0.192453)$ & $(17.5109,-0.577394)$ & $(17.5062,-0.962435)$ & $(17.4991,-1.34765)$ & $(17.4898,-1.73309)$ & $(17.4781,-2.11884)$ \\
 $46$ & $(17.8981,-0.192453)$ & $(17.8958,-0.577392)$ & $(17.8912,-0.962427)$ & $(17.8843,-1.34762)$ & $(17.8752,-1.73305)$ & $(17.8637,-2.11876)$ \\
 $47$ & $(18.2830,-0.192453)$ & $(18.2808,-0.577390)$ & $(18.2763,-0.962420)$ & $(18.2695,-1.34760)$ & $(18.2605,-1.73301)$ & $(18.2493,-2.11868)$ \\
 $48$ & $(18.6679,-0.192453)$ & $(18.6657,-0.577389)$ & $(18.6613,-0.962413)$ & $(18.6547,-1.34759)$ & $(18.6459,-1.73297)$ & $(18.6349,-2.11861)$ \\
 $49$ & $(19.0528,-0.192453)$ & $(19.0507,-0.577387)$ & $(19.0463,-0.962407)$ & $(19.0399,-1.34757)$ & $(19.0312,-1.73293)$ & $(19.0205,-2.11855)$ \\
 $50$ & $(19.4377,-0.192453)$ & $(19.4356,-0.577386)$ & $(19.4314,-0.962401)$ & $(19.4250,-1.34755)$ & $(19.4166,-1.73290)$ & $(19.4060,-2.11848)$ \\
\hline
\end{tabular}
\caption{QNM frequencies $\wQNM$ for a scalar field in Schwarzschild spacetime
for the modes $n: 0\to 5$ and $\ell:0\to 50$. The format is as $\left(\text{Re}\left(\wQNM\right),\text{Im}\left(\wQNM\right)\right)$
 and the units are such that $2M=1$.}
 \label{tb:freqs}
\end{table*}
\endgroup

\begingroup
\squeezetable
\begin{table*}
\begin{tabular}{||c||c|c|c|c|c|c||}
\hline
$\ell$ & $\mathcal{B}_{\ell 0}\times 100$ & $\mathcal{B}_{\ell 1} \times 10$ & $\mathcal{B}_{\ell 2}$  & $\mathcal{B}_{\ell 3}$  & $\mathcal{B}_{\ell 4}$  & $\mathcal{B}_{\ell 5}$ \\
\hline
 $0$ & $(21.235,-5.9276)$ & $(-0.63477,-0.66151)$ & $(0.02377,0.04351)$ & $(-0.01271,-0.03081)$ & $(0.008102,0.02362)$ & $(-0.005714,-0.01908)$ \\
 $1$ & $(-15.067,2.2814)$ & $(0.28973,1.8882)$ & $(0.09351,-0.08774)$ & $(-0.07810,0.01928)$ & $(0.05469,0.002346)$ & $(-0.03930,-0.008844)$ \\
 $2$ & $(11.935,1.3429)$ & $(0.35520,-2.6428)$ & $(-0.28608,0.04593)$ & $(0.16210,0.15711)$ & $(-0.02730,-0.15576)$ & $(-0.02720,0.10797)$ \\
 $3$ & $(-9.3638,-4.0471)$ & $(-1.3411,2.9416)$ & $(0.48787,0.12039)$ & $(-0.09265,-0.51733)$ & $(-0.29300,0.32710)$ & $(0.32705,-0.05070)$ \\
 $4$ & $(6.8044,5.9131)$ & $(2.4759,-2.7544)$ & $(-0.62107,-0.42124)$ & $(-0.28950,0.96779)$ & $(1.0201,-0.20510)$ & $(-0.69443,-0.58061)$ \\
 $5$ & $(-4.2150,-6.9900)$ & $(-3.5549,2.0749)$ & $(0.61058,0.82657)$ & $(1.0890,-1.3064)$ & $(-2.0197,-0.65225)$ & $(0.47548,2.1172)$ \\
 $6$ & $(1.6909,7.3201)$ & $(4.3853,-0.95604)$ & $(-0.39862,-1.2698)$ & $(-2.2942,1.2584)$ & $(2.8081,2.6470)$ & $(1.4454,-4.3672)$ \\
 $7$ & $(0.63105,-6.9682)$ & $(-4.8058,-0.48790)$ & $(-0.04070,1.6576)$ & $(3.7373,-0.54056)$ & $(-2.5599,-5.9087)$ & $(-6.2905,6.1838)$ \\
 $8$ & $(-2.6129,6.0323)$ & $(4.7048,2.0922)$ & $(0.69098,-1.8848)$ & $(-5.0882,-1.0568)$ & $(0.25452,10.064)$ & $(14.708,-5.2578)$ \\
 $9$ & $(4.1381,-4.6429)$ & $(-4.0356,-3.6604)$ & $(-1.4897,1.8533)$ & $(5.8887,3.5864)$ & $(5.0410,-14.084)$ & $(-25.906,-1.6518)$ \\
 $10$ & $(-5.1254,2.9555)$ & $(2.8238,4.9878)$ & $(2.3321,-1.4911)$ & $(-5.6269,-6.8773)$ & $(-13.797,16.287)$ & $(36.860,17.922)$ \\
 $11$ & $(5.5380,-1.1394)$ & $(-1.1667,-5.8868)$ & $(-3.0826,0.76920)$ & $(3.8428,10.504)$ & $(25.634,-14.538)$ & $(-41.938,-45.727)$ \\
 $12$ & $(-5.3861,-0.63610)$ & $(-0.77565,6.2110)$ & $(3.5928,0.28584)$ & $(-0.24737,-13.804)$ & $(-39.022,6.6593)$ & $(33.270,84.369)$ \\
 $13$ & $(4.7258,2.2148)$ & $(2.7977,-5.8745)$ & $(-3.7242,-1.5878)$ & $(-5.1644,15.947)$ & $(51.177,8.9910)$ & $(-2.0838,-128.67)$ \\
 $14$ & $(-3.6533,-3.4664)$ & $(-4.6709,4.8660)$ & $(3.3712,2.9953)$ & $(12.032,-16.060)$ & $(-58.237,-32.865)$ & $(-58.986,167.98)$ \\
 $15$ & $(2.2950,4.2966)$ & $(6.1701,-3.2529)$ & $(-2.4829,-4.3248)$ & $(-19.602,13.383)$ & $(55.757,63.629)$ & $(152.86,-186.27)$ \\
 $16$ & $(-0.79608,-4.6538)$ & $(-7.1007,1.1774)$ & $(1.0784,5.3725)$ & $(26.760,-7.4391)$ & $(-39.512,-97.770)$ & $(-274.80,163.84)$ \\
 $17$ & $(-0.69352,4.5318)$ & $(7.3227,1.1567)$ & $(0.74559,-5.9409)$ & $(-32.150,-1.8046)$ & $(6.5060,129.57)$ & $(409.81,-80.743)$ \\
 $18$ & $(2.0322,-3.9690)$ & $(-6.7696,-3.5052)$ & $(-2.8181,5.8684)$ & $(34.345,13.826)$ & $(43.958,-151.50)$ & $(-531.46,-78.243)$ \\
 $19$ & $(-3.0995,3.0426)$ & $(5.4591,5.6094)$ & $(4.9079,-5.0567)$ & $(-32.075,-27.507)$ & $(-109.35,155.18)$ & $(603.12,317.74)$ \\
 $20$ & $(3.8058,-1.8603)$ & $(-3.4949,-7.2253)$ & $(-6.7472,3.4924)$ & $(24.472,41.182)$ & $(183.35,-132.69)$ & $(-581.93,-626.05)$ \\
 $21$ & $(-4.0999,0.54890)$ & $(1.0588,8.1517)$ & $(8.0640,-1.2600)$ & $(-11.294,-52.788)$ & $(-255.87,78.242)$ & $(425.80,970.86)$ \\
 $22$ & $(3.9719,0.75774)$ & $(1.6069,-8.2548)$ & $(-8.6169,-1.4577)$ & $(-6.9020,60.107)$ & $(313.74,10.120)$ & $(-102.72,-1297.9)$ \\
 $23$ & $(-3.4528,-1.9321)$ & $(-4.2239,7.4855)$ & $(8.2314,4.3919)$ & $(28.688,-61.069)$ & $(-342.17,-129.05)$ & $(-398.61,1533.6)$ \\
 $24$ & $(2.6100,2.8643)$ & $(6.5075,-5.8888)$ & $(-6.8298,-7.2123)$ & $(-51.804,54.095)$ & $(326.88,268.97)$ & $(1058.8,-1593.0)$ \\
 $25$ & $(-1.5393,-3.4725)$ & $(-8.1978,3.6016)$ & $(4.4519,9.5630)$ & $(73.341,-38.427)$ & $(-256.65,-414.05)$ & $(-1820.5,1391.7)$ \\
 $26$ & $(0.35540,3.7091)$ & $(9.0894,-0.84170)$ & $(-1.2631,-11.105)$ & $(-90.034,14.381)$ & $(125.96,543.13)$ & $(2586.0,-862.51)$ \\
 $27$ & $(0.82045,-3.5645)$ & $(-9.0558,-2.1137)$ & $(-2.4533,11.558)$ & $(98.671,16.498)$ & $(62.594,-631.84)$ & $(-3222.0,-26.386)$ \\
 $28$ & $(-1.8719,3.0668)$ & $(8.0661,4.9563)$ & $(6.3191,-10.747)$ & $(-96.552,-51.450)$ & $(-296.78,655.66)$ & $(3571.8,1254.4)$ \\
 $29$ & $(2.6989,-2.2777)$ & $(-6.1912,-7.3787)$ & $(-9.8982,8.6271)$ & $(81.945,86.667)$ & $(554.28,-593.72)$ & $(-3475.7,-2735.3)$ \\
 $30$ & $(-3.2265,1.2856)$ & $(3.5999,9.1086)$ & $(12.744,-5.3063)$ & $(-54.488,-117.65)$ & $(-803.83,432.87)$ & $(2797.7,4312.7)$ \\
 $31$ & $(3.4120,-0.19579)$ & $(-0.54353,-9.9388)$ & $(-14.452,1.0455)$ & $(15.450,139.71)$ & $(1007.9,-171.41)$ & $(-1455.6,-5764.9)$ \\
 $32$ & $(-3.2476,-0.87973)$ & $(-2.6689,9.7520)$ & $(14.714,3.7585)$ & $(32.186,-148.56)$ & $(-1126.7,-178.17)$ & $(-550.01,6823.6)$ \\
 $33$ & $(2.7604,1.8339)$ & $(5.7020,-8.5359)$ & $(-13.360,-8.6084)$ & $(-83.856,140.93)$ & $(1124.1,588.00)$ & $(3112.9,-7204.4)$ \\
 $34$ & $(-2.0087,-2.5748)$ & $(-8.2293,6.3876)$ & $(10.394,12.957)$ & $(133.78,-115.16)$ & $(-972.89,-1015.6)$ & $(-6008.9,6647.9)$ \\
 $35$ & $(1.0755,3.0344)$ & $(9.9687,-3.5066)$ & $(-6.0080,-16.265)$ & $(-175.55,71.640)$ & $(660.96,1407.0)$ & $(8900.8,-4967.3)$ \\
 $36$ & $(-0.05946,-3.1744)$ & $(-10.714,0.17582)$ & $(0.57074,18.068)$ & $(202.84,-13.054)$ & $(-195.84,-1702.3)$ & $(-11361.,2096.4)$ \\
 $37$ & $(-0.93520,2.9895)$ & $(10.359,3.2661)$ & $(5.3956,-18.036)$ & $(-210.24,-55.671)$ & $(-392.58,1842.4)$ & $(12915.,1870.6)$ \\
 $38$ & $(1.8090,-2.5069)$ & $(-8.9101,-6.4597)$ & $(-11.266,16.020)$ & $(194.06,127.68)$ & $(1052.0,-1777.9)$ & $(-13100.,-6653.3)$ \\
 $39$ & $(-2.4773,1.7827)$ & $(6.4918,9.0622)$ & $(16.378,-12.084)$ & $(-153.00,-194.78)$ & $(-1710.6,1477.0)$ & $(11540.,11783.)$ \\
 $40$ & $(2.8779,-0.89603)$ & $(-3.3330,-10.785)$ & $(-20.102,6.5150)$ & $(88.643,248.33)$ & $(2283.4,-933.64)$ & $(-8017.5,-16628.)$ \\
 $41$ & $(-2.9770,-0.05986)$ & $(-0.25303,11.424)$ & $(21.919,0.19311)$ & $(-5.5843,-280.21)$ & $(-2681.5,172.13)$ & $(2543.7,20447.)$ \\
 $42$ & $(2.7719,0.98694)$ & $(3.9003,-10.886)$ & $(-21.484,-7.3834)$ & $(-88.753,283.93)$ & $(2822.8,750.43)$ & $(4592.7,-22473.)$ \\
 $43$ & $(-2.2904,-1.7924)$ & $(-7.2276,9.1991)$ & $(18.681,14.297)$ & $(184.69,-255.54)$ & $(-2644.2,-1746.3)$ & $(-12806.,22018.)$ \\
 $44$ & $(1.5875,2.3976)$ & $(9.8784,-6.5135)$ & $(-13.651,-20.153)$ & $(-271.21,194.50)$ & $(2112.9,2703.8)$ & $(21234.,-18584.)$ \\
 $45$ & $(-0.73895,-2.7461)$ & $(-11.560,3.0875)$ & $(6.7894,24.233)$ & $(337.14,-104.05)$ & $(-1235.7,-3497.4)$ & $(-28799.,11974.)$ \\
 $46$ & $(-0.16616,2.8081)$ & $(12.073,0.73647)$ & $(1.2742,-25.970)$ & $(-372.45,-8.7134)$ & $(64.083,4002.3)$ & $(34316.,-2378.9)$ \\
 $47$ & $(1.0351,-2.5836)$ & $(-11.340,-4.5670)$ & $(-9.7379,25.016)$ & $(369.62,133.30)$ & $(1305.5,-4110.2)$ & $(-36630.,-9567.1)$ \\
 $48$ & $(-1.7809,2.1013)$ & $(9.4099,8.0034)$ & $(17.703,-21.302)$ & $(-324.76,-256.67)$ & $(-2737.0,3745.6)$ & $(34782.,22791.)$ \\
 $49$ & $(2.3304,-1.4152)$ & $(-6.4597,-10.678)$ & $(-24.269,15.054)$ & $(238.48,364.55)$ & $(4066.0,-2880.2)$ & $(-28175.,-35838.)$ \\
 $50$ & $(-2.6320,0.59900)$ & $(2.7767,12.296)$ & $(28.634,-6.7965)$ & $(-116.25,-442.97)$ & $(-5116.7,1543.2)$ & $(16724.,46998.)$ \\
\hline
\end{tabular}
\caption{QNM excitation factors $\mathcal{B}_{\ell  n}$ for a scalar field in Schwarzschild spacetime
for the modes $n: 0\to 5$ and $\ell:0\to 50$. The format is as $\left(\text{Re}\left(\mathcal{B}_{\ell  n}\right),\text{Im}\left(\mathcal{B}_{\ell  n}\right)\right)$.}
 \label{tb:exc fac}
\end{table*}
\endgroup

\begingroup
\squeezetable
\begin{table*}
\begin{tabular}{||c||c|c|c|c|c|c||}
\hline
$\ell$ & $A^{out}_{\ell 0}$ & $A^{out}_{\ell 1}$ & $A^{out}_{\ell 2}$  & $A^{out}_{\ell 3}$  & $A^{out}_{\ell 4}$  & $A^{out}_{\ell 5}$ \\
\hline
 $0$ & $(1.47452,0.0888637)$ & $(-4.32470,-0.114684)$ & $(11.9546,0.668667)$ & $(-32.4450,-2.48657)$ & $(87.7319,7.95094)$ & $(-237.111,-23.8583)$ \\
 $1$ & $(1.47605,-0.624535)$ & $(-3.92222,1.97912)$ & $(11.2588,-6.05607)$ & $(-32.7896,16.8678)$ & $(94.0354,-44.6667)$ & $(-265.548,115.791)$ \\
 $2$ & $(1.14717,-1.14263)$ & $(-2.94032,3.18388)$ & $(7.76027,-9.29836)$ & $(-21.6237,27.3282)$ & $(62.6096,-78.6495)$ & $(-183.181,220.720)$ \\
 $3$ & $(0.672428,-1.47811)$ & $(-1.67535,3.98725)$ & $(4.07260,-11.0897)$ & $(-10.0780,31.6447)$ & $(26.4662,-91.4468)$ & $(-74.8255,264.019)$ \\
 $4$ & $(0.121568,-1.62130)$ & $(-0.226788,4.31695)$ & $(0.158910,-11.6778)$ & $(1.02034,32.2244)$ & $(-6.54401,-90.6517)$ & $(25.7781,258.736)$ \\
 $5$ & $(-0.439168,-1.56646)$ & $(1.24207,4.14000)$ & $(-3.70220,-11.0029)$ & $(11.3617,29.5722)$ & $(-35.1458,-80.6845)$ & $(107.941,223.832)$ \\
 $6$ & $(-0.944776,-1.32512)$ & $(2.56252,3.48053)$ & $(-7.12276,-9.11612)$ & $(20.2274,23.9358)$ & $(-58.3825,-63.3166)$ & $(170.258,169.500)$ \\
 $7$ & $(-1.33732,-0.928074)$ & $(3.58292,2.41716)$ & $(-9.72868,-6.21346)$ & $(26.7903,15.8294)$ & $(-74.7697,-40.1421)$ & $(211.155,101.842)$ \\
 $8$ & $(-1.57216,-0.422905)$ & $(4.18639,1.07421)$ & $(-11.2263,-2.61461)$ & $(30.3665,6.07180)$ & $(-82.9283,-13.3056)$ & $(228.710,26.8602)$ \\
 $9$ & $(-1.62294,0.131003)$ & $(4.30416,-0.392348)$ & $(-11.4441,1.27579)$ & $(30.5607,-4.29796)$ & $(-82.0835,14.5385)$ & $(221.989,-48.6246)$ \\
 $10$ & $(-1.48452,0.669101)$ & $(3.92347,-1.81275)$ & $(-10.3555,5.01617)$ & $(27.3418,-14.1445)$ & $(-72.3276,40.5044)$ & $(191.954,-117.356)$ \\
 $11$ & $(-1.17354,1.12899)$ & $(3.08898,-3.02290)$ & $(-8.08337,8.17979)$ & $(21.0600,-22.3721)$ & $(-54.6990,61.8195)$ & $(141.804,-172.437)$ \\
 $12$ & $(-0.726448,1.45751)$ & $(1.89755,-3.88318)$ & $(-4.88680,10.4046)$ & $(12.4145,-28.0559)$ & $(-31.1102,76.1628)$ & $(76.8532,-208.171)$ \\
 $13$ & $(-0.195280,1.61682)$ & $(0.487069,-4.29448)$ & $(-1.13138,11.4354)$ & $(2.37729,-30.5519)$ & $(-4.15511,81.9504)$ & $(4.06743,-220.796)$ \\
 $14$ & $(0.358337,1.58870)$ & $(-0.979432,-4.20958)$ & $(2.75251,11.1535)$ & $(-7.91680,-29.5743)$ & $(23.1737,78.5345)$ & $(-68.6585,-208.980)$ \\
 $15$ & $(0.870304,1.37662)$ & $(-2.33261,-3.63852)$ & $(6.31921,9.59093)$ & $(-17.2990,-25.2316)$ & $(47.8211,66.2923)$ & $(-133.361,-174.044)$ \\
 $16$ & $(1.28143,1.00526)$ & $(-3.41630,-2.64742)$ & $(9.15908,6.92689)$ & $(-24.7011,-18.0160)$ & $(67.0172,46.5953)$ & $(-182.903,-119.862)$ \\
 $17$ & $(1.54425,0.517678)$ & $(-4.10553,-1.35079)$ & $(10.9457,3.46725)$ & $(-29.2780,-8.74854)$ & $(78.5956,21.6616)$ & $(-211.789,-52.4664)$ \\
 $18$ & $(1.62846,-0.0296820)$ & $(-4.32084,0.101628)$ & $(11.4737,-0.390459)$ & $(-30.5061,1.51341)$ & $(81.2446,-5.69090)$ & $(-216.800,20.6127)$ \\
 $19$ & $(1.52442,-0.573538)$ & $(-4.03752,1.54222)$ & $(10.6820,-4.20275)$ & $(-28.2439,11.5973)$ & $(74.6614,-32.3624)$ & $(-197.384,91.1800)$ \\
 $20$ & $(1.24425,-1.05106)$ & $(-3.28835,2.80479)$ & $(8.66178,-7.53116)$ & $(-22.7491,20.3492)$ & $(59.5915,-55.3236)$ & $(-155.729,151.298)$ \\
 $21$ & $(0.820349,-1.40710)$ & $(-2.15983,3.74369)$ & $(5.64518,-9.99279)$ & $(-14.6501,26.7669)$ & $(37.7475,-71.9619)$ & $(-96.5397,194.189)$ \\
 $22$ & $(0.301737,-1.60059)$ & $(-0.782198,4.25068)$ & $(1.97925,-11.3043)$ & $(-4.87428,30.1145)$ & $(11.6163,-80.3809)$ & $(-26.5085,215.007)$ \\
 $23$ & $(-0.251681,-1.60924)$ & $(0.685608,4.26732)$ & $(-1.91419,-11.3148)$ & $(5.45794,30.0076)$ & $(-15.8228,-79.6192)$ & $(46.4309,211.396)$ \\
 $24$ & $(-0.776004,-1.43207)$ & $(2.07428,3.79176)$ & $(-5.58706,-10.0230)$ & $(15.1616,26.4580)$ & $(-41.4378,-69.7625)$ & $(114.003,183.770)$ \\
 $25$ & $(-1.21071,-1.08959)$ & $(3.22368,2.87889)$ & $(-8.61657,-7.57746)$ & $(23.1231,19.8724)$ & $(-62.3021,-51.9358)$ & $(168.533,135.269)$ \\
 $26$ & $(-1.50565,-0.621363)$ & $(4.00126,1.63402)$ & $(-10.6540,-4.25970)$ & $(28.4284,11.0065)$ & $(-76.0296,-28.1765)$ & $(203.818,71.4108)$ \\
 $27$ & $(-1.62679,-0.0814377)$ & $(4.31739,0.200743)$ & $(-11.4647,-0.451668)$ & $(30.4681,0.877945)$ & $(-81.0493,-1.20198)$ & $(215.841,-0.531215)$ \\
 $28$ & $(-1.56019,0.467864)$ & $(4.13564,-1.25566)$ & $(-10.9553,3.40822)$ & $(29.0074,-9.35001)$ & $(-76.7857,25.9001)$ & $(203.234,-72.3570)$ \\
 $29$ & $(-1.31356,0.963157)$ & $(3.47699,-2.56723)$ & $(-9.18438,6.87552)$ & $(24.2141,-18.5023)$ & $(-63.7264,50.0258)$ & $(167.437,-135.873)$ \\
 $30$ & $(-0.915376,1.34730)$ & $(2.41743,-3.58276)$ & $(-6.35593,9.55094)$ & $(16.6385,-25.5273)$ & $(-43.3672,68.4107)$ & $(112.536,-183.830)$ \\
 $31$ & $(-0.411589,1.57599)$ & $(1.07915,-4.18517)$ & $(-2.79560,11.1264)$ & $(7.15114,-29.6172)$ & $(-18.0413,78.9469)$ & $(44.8036,-210.749)$ \\
 $32$ & $(0.139667,1.62285)$ & $(-0.383549,-4.30501)$ & $(1.08660,11.4204)$ & $(-3.15741,-30.3018)$ & $(9.34762,80.4259)$ & $(-28.0174,-213.554)$ \\
 $33$ & $(0.674798,1.48250)$ & $(-1.80200,-3.92848)$ & $(4.84355,10.3991)$ & $(-13.1020,-27.5023)$ & $(35.6580,72.6776)$ & $(-97.5991,-191.926)$ \\
 $34$ & $(1.13208,1.17115)$ & $(-3.01268,-3.09903)$ & $(8.04256,8.18003)$ & $(-21.5392,-21.5402)$ & $(57.8709,56.5905)$ & $(-155.980,-148.339)$ \\
 $35$ & $(1.45876,0.724705)$ & $(-3.87600,-1.91228)$ & $(10.3151,5.01885)$ & $(-27.4984,-13.1011)$ & $(73.4369,34.0106)$ & $(-196.478,-87.7835)$ \\
 $36$ & $(1.61719,0.194684)$ & $(-4.29245,-0.505085)$ & $(11.3995,1.27963)$ & $(-30.2942,-3.15567)$ & $(80.5687,7.52970)$ & $(-214.456,-17.1945)$ \\
 $37$ & $(1.58909,-0.357782)$ & $(-4.21402,0.960332)$ & $(11.1708,-2.60693)$ & $(-29.6049,7.15221)$ & $(78.4472,-19.8117)$ & $(-207.855,55.3417)$ \\
 $38$ & $(1.37771,-0.868975)$ & $(-3.64978,2.31502)$ & $(9.65530,-6.19312)$ & $(-25.5096,16.6367)$ & $(67.3156,-44.8736)$ & $(-177.431,121.513)$ \\
 $39$ & $(1.00745,-1.27994)$ & $(-2.66478,3.40282)$ & $(7.02757,-9.06585)$ & $(-18.4793,24.2064)$ & $(48.4524,-64.7771)$ & $(-126.673,173.734)$ \\
 $40$ & $(0.521016,-1.54330)$ & $(-1.37258,4.09832)$ & $(3.59031,-10.8942)$ & $(-9.32301,28.9904)$ & $(24.0245,-77.2354)$ & $(-61.4008,206.017)$ \\
 $41$ & $(-0.0255001,-1.62867)$ & $(0.0778388,4.32136)$ & $(-0.260509,-11.4675)$ & $(0.905787,30.4380)$ & $(-3.16140,-80.8168)$ & $(10.8992,214.660)$ \\
 $42$ & $(-0.569069,-1.52624)$ & $(1.51928,4.04624)$ & $(-4.08127,-10.7197)$ & $(11.0299,28.3825)$ & $(-29.9810,-75.1094)$ & $(81.9327,198.672)$ \\
 $43$ & $(-1.04701,-1.24780)$ & $(2.78557,3.30467)$ & $(-7.43181,-8.73691)$ & $(19.8842,23.0605)$ & $(-53.3518,-60.7691)$ & $(143.549,159.888)$ \\
 $44$ & $(-1.40421,-0.825487)$ & $(3.73074,2.18216)$ & $(-9.92612,-5.74763)$ & $(26.4494,15.0843)$ & $(-70.5871,-39.4440)$ & $(188.676,102.758)$ \\
 $45$ & $(-1.59950,-0.307989)$ & $(4.24585,0.808100)$ & $(-11.2768,-2.09620)$ & $(29.9699,5.37210)$ & $(-79.7053,-13.5859)$ & $(212.135,33.8421)$ \\
 $46$ & $(-1.61035,0.245017)$ & $(4.27151,-0.659103)$ & $(-11.3283,1.79670)$ & $(30.0404,-4.95822)$ & $(-79.6578,13.8324)$ & $(211.231,-38.9484)$ \\
 $47$ & $(-1.43552,0.769766)$ & $(3.80479,-2.05032)$ & $(-10.0746,5.48257)$ & $(26.6525,-14.7174)$ & $(-70.4500,39.6578)$ & $(186.070,-107.254)$ \\
 $48$ & $(-1.09518,1.20576)$ & $(2.89948,-3.20519)$ & $(-7.66025,8.53674)$ & $(20.1964,-22.7821)$ & $(-53.1406,60.9206)$ & $(139.540,-163.231)$ \\
 $49$ & $(-0.628563,1.50272)$ & $(1.65995,-3.99059)$ & $(-4.36329,10.6073)$ & $(11.4152,-28.2235)$ & $(-29.7203,75.1750)$ & $(76.9875,-200.449)$ \\
 $50$ & $(-0.0894812,1.62643)$ & $(0.229077,-4.31599)$ & $(-0.563627,11.4558)$ & $(1.31999,-30.4154)$ & $(-2.88278,80.7814)$ & $(5.59782,-214.633)$ \\
\hline
\end{tabular}
\caption{QNM coefficients  $A^{out}_{\ell,n}$ for a scalar field in Schwarzschild spacetime
for the modes $n:0\to 5$ and $\ell:0\to 50$. The format is as 
$\left(\text{Re}\left(A^{out}_{\ell,n}\right),\text{Im}\left(A^{out}_{\ell,n}\right)\right)$ and the units are such that $2M=1$.
}
\label{tb:Aout}
\end{table*}
\endgroup

\bibliography{references}

\end{document}